\documentclass{jfm}
\usepackage{graphicx}
\usepackage{epstopdf, epsfig}
\usepackage{comment}
\usepackage{subfig}
\usepackage{amsmath}
\usepackage{adjustbox}
\usepackage{multirow}
\usepackage{array}
\usepackage{makecell}
\usepackage{booktabs}
\usepackage{bbding}%cehcmark
\usepackage{natbib}
\usepackage{tikz}
\usetikzlibrary{positioning}

\AppendGraphicsExtensions{-tiff-converted-to.png}

\shorttitle{Guidelines for authors}
\shortauthor{P. Nibourel, C. Leclercq, F. Demourant, E. Garnier and D. Sipp}

\title{Reactive control of $2^{nd}$ Mack mode in a supersonic boundary layer with freestream velocity/density variations}

\author{Pierre Nibourel\aff{1}
  \corresp{\email{pierre.nibourel@onera.fr}},
  Colin Leclercq\aff{1}, Fabrice Demourant\aff{2}, Eric Garnier\aff{1}
 \and Denis Sipp\aff{1}}

\affiliation{\aff{1}ONERA-DAAA, Université Paris-Saclay, 8 Rue des Vertugadins, 92190 Meudon, France
\aff{2}ONERA-DTIS, 2 Avenue Edouard Belin, 31000 Toulouse, France}

\begin{document}

\maketitle

\begin{abstract}
We consider closed-loop control of a two-dimensional supersonic boundary layer at $M=4.5$ that aims at reducing the linear growth of second Mack mode instabilities. These instabilities are first characterized with local spatial and global resolvent analyses, which allow to refine the control strategy and to select appropriate actuators and sensors. After linear input-output reduced order models have been identified, multi-criteria structured mixed \(H_{2}\)/\(H_{\infty}\) synthesis allows to fix beforehand the controller structure and to minimize appropriate norms of various transfer functions: the $H_{2}$ norm to guarantee performance (reduction of perturbation amplification in nominal condition) and the $H_{\infty}$ norm to maintain performance robustness (with respect to sensor noise) and stability robustness (with respect to uncertain free-stream velocity/density variations). Both feedforward and feedback setups, i.e. with estimation sensor placed respectively upstream/downstream of the actuator, allow to maintain the local perturbation energy below a given threshold over a significant distance downstream of the actuator, even in the case of noisy estimation sensors or free-stream density variations. However, the feedforward setup becomes completely ineffective when convective time-delays are altered by free-stream velocity variations of $\pm 5\%$, which highlights the strong relevance of the feedback setup for performance robustness in convectively unstable flows. 
\end{abstract}

\begin{keywords}
\end{keywords}

\section{Introduction}

%Due to its ability to limit undesirable fluid effects which limit the effectiveness of many industrial applications, flow control have been an active area of research over the past decades in research topics such as noise suppression, lift enhancement, mixing augmentation, separation or laminar/turbulent transition delay \citep{gad_flow_book}.

%\subsection{Challenges for reactive control of supersonic boundary layer transition}

Transition to turbulence in a boundary layer results in increased wall friction, penalizing aircraft drag. At high speeds, the generated heat is significant and becomes a major concern for the design of supersonic/hypersonic vehicles \citep{Juliano_art}. The transition to turbulence in boundary layers is initiated by amplification of external disturbances of various kinds (roughness, sound waves, freestream turbulence, etc.) and several paths to transition are possible depending on the nature and intensity of incoming disturbances \citep{Morkovin_art}. With low levels of disturbances, their growth is described by linear stability theory. The stability of a supersonic boundary layer has been widely studied in the literature \citep[and many others]{mack_1984_art,malick_89_art,Ma_Zhong_art,Bugeat_art}. For sufficiently high Mach numbers, this configuration is characterised by the presence of two distinct inviscid instability mechanisms: a generalised inflection point for the first Mack mode \citep{mack_1984_art} and a region where the streamwise base-flow velocity relative to the disturbance phase velocity is supersonic for the second Mack mode implying that acoustic noise is trapped in this region \citep{mack_1984_art,Fedorov_art}. A classical approach for relating instability to transition is precisely based on this linear framework and is called the \(N\)-factor method \citep{smith1956transition}, wherein transition is assumed to occur when a perturbation has been amplified by a factor of \(e^{N}\), which defines an energy threshold depending on the disturbance environment.

Numerous studies addressed the problem of transition delay in the supersonic boundary layer flow using active control: \citet{Gaponov_art} injected heavy gas through porous wall to reduce surface friction and heat transfer, \citet{Sharma_art} resorted to the generation of streaks to counter transient instabilities, \citet{Yao_wall_osci} investigated the impact of spanwise wall oscillation on the drag of a supersonic turbulent boundary layer and \citet{Jahanbakhshi_control_art} took advantage of the sensitivity of the Mack modes to temperature to delay transition to turbulence. However, all the aforementioned studies employed predetermined active strategies which do not exploit any real-time measurement and may therefore be less cost effective and robust to changes in operating conditions than a \textit{reactive} control strategy \citep{gad_flow_book}. To the best of our knowledge, reactive control of convective instabilities in the supersonic boundary layer flow has not yet been considered.

Contrary to oscillator flows \citep{Barbagallo_citation_art,Sipp_control_art} which are, by definition, linearly globally unstable \citep{Huerre_art} and have intrinsic dynamics, noise-amplifier flows like the supersonic boundary layer are extremely sensitive to external disturbances, which are amplified downstream as they are convected by the flow (hence the name convective instabilities). In this context, the purpose of reactive control is to cancel out noise-induced perturbations \citep{Bagheri_art,Barbagallo_art} by producing destructive interference with an actuator. This task is difficult for mainly two reasons: a) the detection of the time-delay associated with the convection of perturbations which may trigger out of phase action with respect to the incoming perturbations, b) the wide spatially evolving range of amplified frequencies along the plate, from higher frequencies upstream to lower ones downstream.

\subsection{Historical dominance of feedforward/LQG for the control of noise-amplifier flows}

Controller synthesis is only feasible for models of small dimensions, of the order of $10^2$ degrees of freedom at most, because of the computational cost and storage requirements of currently available tools \citep{Riccati_complexity}. Therefore, most fluidic control problems require the identification of reduced order models (ROMs), using for instance the eigensystem realization algorithm (ERA) on impulse response data. This popular tool, introduced by \citet{juang_pappa_art}, has already been used in many control studies for noise-amplifier flows \citep[and many others]{Belson_art,Dadfar2D_art,Sasaki3D_art}. Once ROMs are obtained, the control law is built with classical tools of control theory which are mathematically well-established in a linear framework and thus perfectly suited for controlling the linear growth of small perturbations. 

In noise amplifier flows, there is no synchronization of the dynamics at a global scale, perturbations from an actuator $u$ are rapidly damped in the upstream direction, hence the control setup changes fundamentally depending on the position of the estimation sensor $y$ relative to $u$. When $y$ is placed upstream, actuator-induced perturbations are not observable and the configuration is termed "feedforward" \citep{Bagheri_art,Semeraro_art,Herve_art,Juillet_art,Morra_art}. On the other hand, when $y$ is placed downstream, the sensor measures the superposition of noise-induced and actuator-induced perturbations, hence the term "feedback" \citep{Barbagallo_art,Belson_art,Semeraro_fb_art,Vemuri_fb,Tol_fbff_art}. In this case though, there may be a significant time-delay before the effect of actuation may be seen by the sensor, because perturbations are convected at a finite rate by the underlying base flow: the farther downstream $y$ is, the longer the delay.

The literature on noise-amplifier control is dominated by the linear-quadratic-Gaussian (LQG) synthesis \citep[and many others]{Semeraro_art,Barbagallo_art,Juillet_art,Sasaki3D_wavecancelling_art,Tol_fbff_art}, a synthesis method dating back to the 1960s \citep{Kalman64}. Despite being theoretically optimal with respect to a performance criterion, this method comes with no guarantees on stability margins \citep{DoyleLQG_art}. In other words, tiny errors in the model may end up in an unstable feedback loop when $y$ is placed downstream of $u$ (feedback setup), which represents a major drawback for practical applications. Using the loop-transfer-recovery (LTR) method, it is in some cases possible to overcome this lack of stability robustness by overwhelming the control signal entering the estimator \citep{LTR_Kwakernaak,LTR_Doyle}. This procedure has for example been successfully used by \citet{Sipp_Important_art} to improve the stability robustness of their controller in the case of a flow over an open square cavity (oscillator flow). The recovery procedure works by inverting the plant dynamics in order to obtain ultra-fast estimators. This procedure leads to an unstable closed-loop in the case of systems with time-delays, because they possess right-half plane zeros which are converted into right-half plane poles \citep{LTR_Delay,Skogestad_art,Sipp_Important_art}. As a result, this method is not suitable for noise-amplifier flows in general, and in particular, the supersonic boundary layer flow. Contrary to the feedback structure, the feedforward design is unconditionally stable and its implementation via LQG is not a problem. Therefore, feedforward configurations combined with LQG syntheses dominate the noise-amplifier flow control literature, particularly in the incompressible boundary layer control studies \citep{Bagheri_art,Semeraro_art,Semeraro2_art,Semeraro_fb_art,Dadfar2D_art,Dadfar_centralized_art,Sasaki3D_wavecancelling_art,Sasaki3D_art,Morra_art,Freire_art}. 

\subsection{Feedforward "Achilles heel": performance robustness}

However, the use of a feedforward setup raises the problem of robustness to performance, which can be defined as the control law's ability to remain efficient in terms of perturbation amplitude reduction despite modelling errors or free-stream condition variations around the reference case. This problem has been little addressed in the boundary layer control literature, despite the advent of robust synthesis, introduced by \citet{Doyle2_art}. So far, these modern methods have been mainly used in the case of oscillator flows \citep{Flinois_art,Leclercq_art,Shaqarin_art} to have some stability guarantees, because using a feedback setup is mandatory to stabilize a globally unstable flow. %due to the intrinsic feedback nature of such flows (and so to have some stability guarantees), but remains rarely employed in the case of noise-amplifier flows. 

To improve performance robustness compared to a simple fixed-structure LQG feedforward controller, \citet{erdmann2011active,Fabbiane_review_art,Fabbiane_JFMrapids_art} used an adaptive feedforward method for boundary layer control, based on the filtered-X least-mean-squares (FXLMS) algorithm, where the controller structure is adjusted according to the variations of the flow conditions through real-time measurements. However, this method is not robust to abrupt changes in inflow conditions because the controller coefficients are adjusted in a quasi-static fashion. Due to its natural ability to be robust to unknown disturbances or uncertainties on the model \citep{Skogestad_art}, feedback design appears to be a promising alternative for performance robustness on short time scales. \citet{Barbagallo_art} employed a feedback structure combined with an LQG synthesis to control instabilities over a backward-facing step and emphasized the importance of placing the estimation sensor close to the actuator to obtain a reasonable performance. Doing so increases the controllable bandwidth indeed, as it is limited in feedback setup by the convection delay of the disturbances from the actuator to the estimation sensor. However, some of their feedback controllers turned out to be unstable on the real plant (the full linearized Navier--Stokes equations), because of the poor stability robustness of LQG to tiny erros in the ROM. \citet{Tol_fbff_art} also obtained some unstable controllers when trying to control Tollmien--Schlichting (TS) waves in an incompressible two-dimensional boundary layer using LQG on a feedback setup. \citet{Belson_art} are among the first to demonstrate the feasibility of a feedback setup with stability and performance robustness for the same flow, using a simple proportional-integral (PI) controller that was tuned by hand. However, the simple structure of the PI controller did not allow to obtain a satisfactory performance for the chosen actuator/sensor pair, forcing the authors to change it, despite the good performance obtained with LQG on the ROMs with the same actuator/sensor pair. A similar approach was used by \citet{Vemuri_fb}, in order to cancel out TS waves in an experimental setup. The authors tuned a proportional controller by hand to optimise the controller gain in closed loop while ensuring robust stability of their feedback configuration. Such loop-shaping approaches provide guarantees on stability robustness but are far from optimal from a performance viewpoint. And perhaps more importantly, they are very limited in the sense that they cannot be applied to more complex controller structures in a systematic way.

\subsection{Designing robust controllers: structured mixed \(H_{2}\)/\(H_{\infty}\) synthesis techniques}

In contrast, modern tools for robust multi-criteria synthesis, such as the structured mixed \(H_{2}\)/\(H_{\infty}\) synthesis \citep{apkarian_systune_art}, allow to optimize complex control laws. The structured mixed \(H_{2}\)/\(H_{\infty}\) synthesis is able to treat different kinds of mathematical criteria simultaneously, contrary to the LQG method which minimizes a single quadratic criterion based on performance and cost. Furthermore, structured synthesis \citep{apkarian_hinfstruct} has the advantage to limit the controller order and to impose its structure beforehand (\textit{e.g.} state-space model of order 10, PID controller, etc.), unlike methods that solve Riccati equations, such as LQG \citep{Freire_art}, \(H_{\infty}\) \citep{Flinois_art} or \(H_{2}\) \citep{Tol_channel} optimal controls, which lead to high-order controllers (of the same order as the plant augmented by weighting functions). These are often too expensive to use in real-time applications and require reducing the controller order in a post-processing step. Performing this reduction optimally while maintaining stability and performance guarantees on the closed-loop remains an open problem \citep{Reduction_Chen,Reduction_perfo_Goddard}. The possibility of working with both \(H_{2}\) (an integrated gain over all frequencies) and \(H_\infty\) (the maximum gain over all frequencies) criteria ensures performance, robustness to stability and robustness to performance \citep{apkarian_systune_art2}. Indeed, the use of \(H_{\infty}\) criteria on some transfer functions allows to respect stability margins on the feedback design (what was missing within the LQG synthesis) despite modelling errors and to desensitize the controller on certain frequency ranges, allowing optimal performance to be maintained despite the presence of, for example, noise on the estimation sensor. The use of \(H_{2}\) criteria makes it possible to have a performance objective of disturbance rejection during the synthesis (which was sometimes lacking in previous feedback studies).

\subsection{Objective and outline of the paper}

In the present paper, we will consider a supersonic boundary layer at $ M=4.5$ and focus on two-dimensional, i.e. spanwise invariant, and linear perturbations. We will not be dealing with oblique modes or finite-amplitude perturbations, even if they often do play a significant role in transition in practice. Hence the present work is only a first step in learning how to design robust control laws for the problem of transition in the supersonic boundary layer. One key question we wish to address before introducing more physical complexity is how do the feedforward and feedback setups compare on this noise-amplifier flow, using modern robust synthesis tools? With the help of multi-criteria structured \(H_{2}\)/\(H_{\infty}\) controller synthesis, can we design a feedback setup which outperforms the often-used feedforward/LQG with regards to performance robustness to realistic changes in operating conditions, i.e. velocity and density variations? 

The paper is organized as follows. Sections \S \ref{sec:config_mathmodel} and \S \ref{sec:Numerical_methods} provide a description of the flow configuration and numerical methods. In \S \ref{sec:amplifier_behaviour}, local and global linear stability tools are used to define appropriate closed-loop specifications, i.e. determining the actuators, sensors and performance criterion to be optimized. Section \S \ref{sec:indentif_synthesis} is devoted to ROM identification from impulse responses using the Eigensystem Realization Algorithm (ERA), with special emphasis on the problem of time-delays in such noise-amplifier flows. Next, we formally introduce the multi-criteria structured mixed \(H_{2}\)/\(H_{\infty}\) synthesis and the associated constraint minimization problem we wish to solve. In \S \ref{sec:Result_ff_vs_fb} we compare the results obtained on and off-design (noisy sensors, density and velocity variations) for the feedforward and feedback setups. Conclusions are drawn in \S \ref{sec:conclusions}. 

\section{Flow configuration}
\label{sec:config_mathmodel}

A two-dimensional compressible ideal gas flowing over a flat plate is considered. The flow is governed by the Navier--Stokes equations:
\begin{subeqnarray}
\label{eq:Navier_Stokes}
\frac{\partial \rho }{\partial t}+\boldsymbol{\nabla}\cdot(\rho\mathbf{u})&=&0,\\
\frac{\partial \rho\mathbf{u} }{\partial t}+\boldsymbol{\nabla}\cdot(\rho\mathbf{u}\otimes\mathbf{u})&=&-\boldsymbol{\nabla} p + \boldsymbol{\nabla}\cdot\boldsymbol{\tau},\\
\frac{\partial \rho E}{\partial t}+\boldsymbol{\nabla}\cdot(\rho E\mathbf{u})&=&\boldsymbol{\nabla}\cdot(- p\mathbf{u}+\boldsymbol{\tau }\cdot\mathbf{u}-\boldsymbol{\theta}),
\end{subeqnarray}
where \(\rho\) is the fluid density, \(\mathbf{u}\) the velocity vector, \(p\) the static pressure, \(E=\frac{p}{\rho(\gamma-1)}+\frac{\mathbf{u}\cdot\mathbf{u}}{2}\) the total energy, \(\boldsymbol{\tau}\) the viscous stress tensor and \(\boldsymbol{\theta}\) the heat flux vector. The viscous stress tensor and the heat flux vector are given by:
\begin{equation}
\label{eq:stress_tensor}
 \boldsymbol{\tau}=\mu(\boldsymbol{\nabla \otimes \mathbf{u}}+(\boldsymbol{\nabla \otimes \mathbf{u})}^{T}-\frac{2}{3}(\boldsymbol{\nabla} \cdot\boldsymbol{\mathbf{u}})\mathcal{I}),\\
\end{equation}
\begin{equation}
\label{eq:heat_flux}
 \boldsymbol{\theta}=-k\boldsymbol{\nabla}T,
\end{equation}
with \(\mathcal{I}\) the identity tensor, \(k\) the thermal conductivity
and \(\mu\) the dynamic viscosity which is deduced from the local temperature \(T\) via Sutherland's law,
\begin{equation}
    \label{eq:Sutherland}
    \mu=\mu_{\mathrm{ref}}\left(\frac{T}{T_{\mathrm{ref}}}\right)^{\frac{3}{2}}\frac{T_{\mathrm{ref}}+S}{T+S}.
\end{equation}
The parameters of Sutherland's law are taken as: \(\mu_{\mathrm{ref}}=1.716 \times 10^{-5}\ \mathrm{Pa.s}\), \(T_{\mathrm{ref}}=273.15\ \mathrm{K}\) and \(S=110.4\ \mathrm{K}\). The gas considered being air, we have \(\gamma=1.4\), \(r=287\ \mathrm{J.K^{-1}.kg^{-1}}\) and \(Pr=\frac{\mu \gamma r}{k(\gamma-1)}=0.725\). The free-stream flow conditions are very close to those used experimentally by \citet{Kendall_art} and in the simulations of \citet{Ma_Zhong_art}, \textit{i.e.}: \(T_{\infty}=65.149 \ \mathrm{K}\), \( U_{\infty}=728.191\ \mathrm{m.s^{-1}}\) and \(p_{\infty}=728.312\ \mathrm{Pa}\).  Thus, the free-stream Mach number of the simulation is \(M_{\infty}=\frac{U_{\infty}}{\sqrt{\gamma rT{\infty}}}=4.5\). %These values are also not far from those used for the HIFiRE (Hypersonic International Flight Research Experimentation) cone experiments, for which one of the goals is to provide high-quality and flight data for boundary layer stability analysis to predict transition location \citep{HIFIRE_art}.

\begin{figure}
\centering
\subfloat[\label{fig:numerical_domain}]{\includegraphics[height=0.32\textwidth,width=0.64\textwidth]{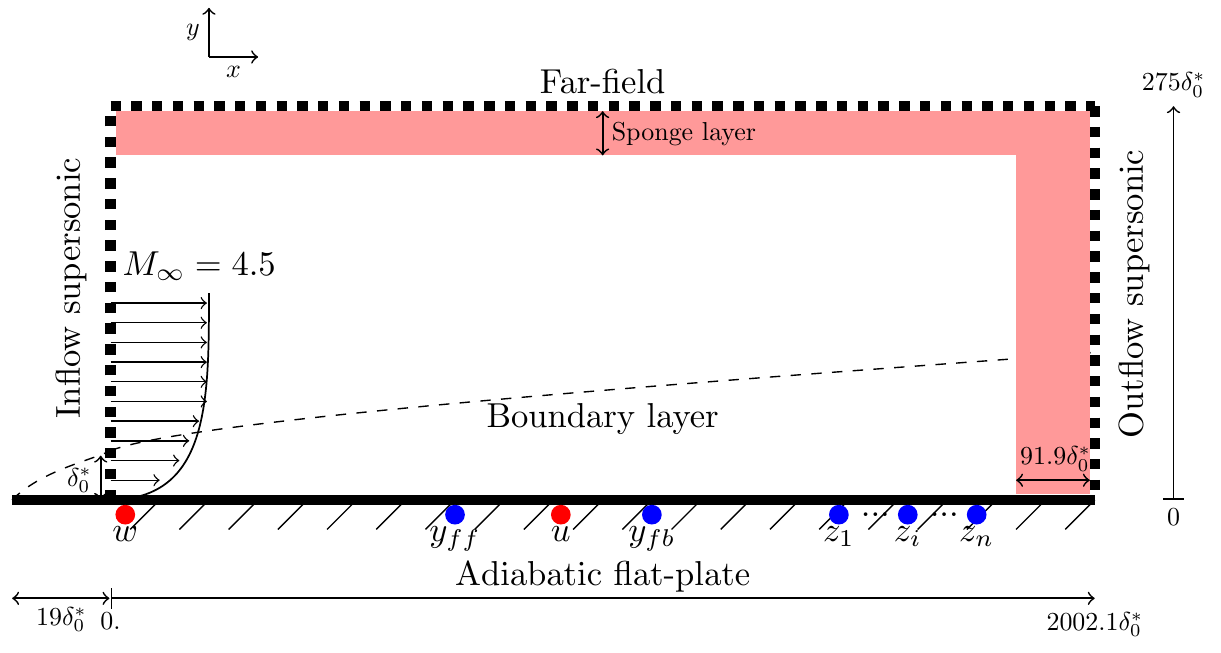}}
\adjustbox{raise=1.5pc}{\subfloat[\label{fig:inlet_profile}]{\includegraphics[height=0.2335\linewidth,width=0.291875\linewidth]{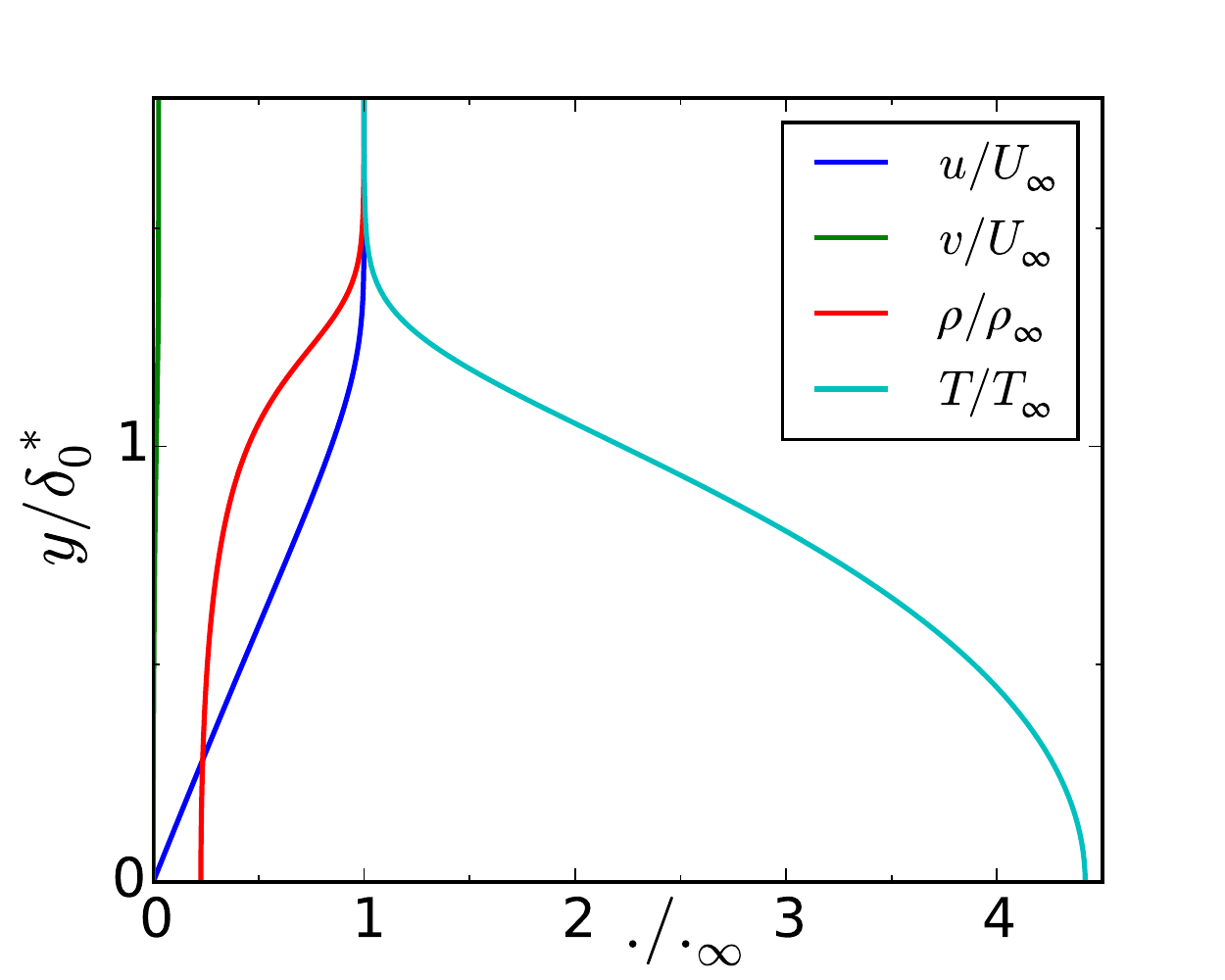}}}
\caption{a) Diagram of the computational domain. Inputs and outputs of the control problem are in red and blue, respectively. b) Boundary layer profile used for the inlet condition.}
\label{fig:numerical_domain_tot}
\end{figure}

The computational domain is represented in figure \ref{fig:numerical_domain}: it consists of a rectangular domain where the lower boundary is an adiabatic flat plate of length \(L_{x}=2002.1\delta_{0}^{*}\), with \(\delta_{0}^{*}=3.2656 \times 10^{-4}\ \mathrm{m}\) the compressible displacement thickness at the inlet of the domain (defined as \(\delta_{0}^{*}=\int^{\infty}_{0} (1-\frac{\rho u}{\rho_{\infty}u_{\infty}})dy\)), which results in $ Re_{\delta_0^*}=\frac{\rho_{\infty}U_{\infty}\delta_0^*}{\mu_{\infty}}\simeq2121$.

A far-field and a supersonic exit conditions are respectively applied at the top (\(y=275\delta_{0}^{*}\)) and at the outlet of the computational domain. Furthermore, a sponge area is used downstream and in the upper part of the domain to minimize reflections. This sponge area consists of stretching the mesh in the longitudinal direction for the downstream boundary (\(L_{\mathrm{sponge}}=91.9\delta_{0}^{*}\) and 30 cells in the streamwise direction) and adding a source term in (\ref{eq:Navier_Stokes}) on the last 10 cells to bring the flow back to its equilibrium point. A supersonic inlet condition is imposed at the upstream boundary where the complete state is prescribed and matches a zero-pressure gradient laminar boundary layer profile (see figure \ref{fig:inlet_profile}) computed with the ONERA boundary layer code CLICET (see for instance \citet{Clicet_art}). It corresponds to a profile taken at a distance of \(19\delta_{0}^{*}\) from the leading edge. The beginning of the numerical domain has been chosen to be in a stable area for all frequencies according to local linear stability theory (see \S \ref{sec:Result_local_global}). The boundary layer thickness (denoted  $\delta$) at the end of the domain of interest leads to $Re_{\delta}\simeq35081$. Overall, the useful numerical domain (i.e. not counting the length of the sponge area) extends from \(4\times10^{4}<Re_{x}=\frac{\rho_{\infty}U_{\infty}x}{\mu_{\infty}}<4.1\times10^{6}\).

\section{Base-flow and spatial stability analyses}
\label{sec:Numerical_methods}
\subsection{Base flow and linearized DNS}

Direct numerical simulations (DNS) are performed using the finite volume code \textit{elsA} \citep{Elsa_art}. An upwind AUSM + up scheme \citep{AUSM_art} associated with a third order MUSCL extrapolation method \citep{MUSCL_art} is used for the spatial discretization of the convective fluxes. The viscous fluxes are obtained by a second-order centered scheme.
The semi-discretized Navier-Stokes equations then read:
\begin{equation}
    \label{eq:NS_disc}
    \frac{\partial \mathbf{q}}{\partial t}=\mathcal{N}(\mathbf{q})+\mathcal{P}\mathbf{f},
\end{equation}
where \(\mathbf{q}=[\rho, \rho\mathbf{u}, \rho E]^{T}\) and \(\mathcal{N}(\mathbf{q})\) is the discretized compressible Navier--Stokes equations (including the boundary conditions). The momentum forcing \(\mathbf{f}\) may either represent a noise source or the effect of an actuator. The matrix $ \mathcal{P} $ represents the prolongation operator that transforms the momentum forcing into a full state-vector forcing by adding zero components. 
The laminar base-flow $ \mathbf{\bar{q}}$, defined as
\begin{equation}
    \label{eq:BF}
    \mathcal{N}(\mathbf{\bar{q}})=0,
\end{equation}
is obtained by time-stepping the unforced unsteady equations \eqref{eq:NS_disc} with an implicit time-stepping method based on a local time step, up to convergence of the residuals.
The unsteady simulations for the development of instabilities are performed with an implicit second-order Gear scheme \citep{gear_book} with 4 sub-iterations and a time-step \(dt\) ensuring a CFL number lower than \(1.4\) in the whole domain. For these unsteady simulations, the amplitude of the forcing $\mathbf{{f}} $ is chosen sufficiently small to ensure that the induced perturbation $\mathbf{{q}'}=\mathbf{{q}}-\mathbf{\bar{q}}$ remains in the linear regime until the end of the computational domain. The time step and the number of sub-iterations of the temporal method have been validated by comparing transfer functions from the linearized DNS and those determined from the frequency-domain resolvent approach (defined in \S \ref{sec:Numerical_global_analysis}).

A resolution of 3200\(\times\)220 cells for the useful domain was chosen. The mesh is uniform in the \(x\) direction while a geometric law is used in the \(y\) direction to resolve strong gradients near the wall. The base-flow and linear growth rates have been verified against linearized DNS results of \citet{Ma_Zhong_art}, allowing to validate the resolution and the numerical schemes. %(this was done with a different fluid configuration than the one presented in \S \ref{sec:config_mathmodel}, in order to be in agreement with theirs); 

%The resolution and the numerical scheme have been validated using the fluid configuration of \citet{Ma_Zhong_art} and comparing the base flow and the linear growth rates coming from their linearized DNS.  

\subsection{Global resolvent analysis}
\label{sec:Numerical_global_analysis}

For purposes of controlling instabilities, the choice of the type and position of the actuator/sensors will play an essential role. This choice is guided by resolvent analysis, which characterizes the noise-amplifier behaviour from an input-output viewpoint. The method is briefly detailed in this section.  

The purpose of control is to reduce the amplitude of disturbances which naturally develop in the boundary layer, and thus to maintain the flow as close as possible to its equilibrium \(\mathbf{\bar{q}}\). By injecting the ansatz $ \mathbf{{q}}=\mathbf{\bar{q}}+\mathbf{{q}'}$ into (\ref{eq:NS_disc}) and by considering only small-amplitude forcing \(\mathbf{f}\), we obtain after linearization:
\begin{equation}
    \label{eq:NS_perturb_1}
    \frac{\partial \mathbf{q'}}{\partial t}=\mathcal{A}\mathbf{q'}+\mathcal{P}\mathbf{f},
\end{equation}
where $ \mathcal{A} $ is the Jacobian matrix defined as \(\mathcal{A}=\frac{d  \mathcal{N}}{d \mathbf{q}}\big|_{\mathbf{\bar{q}}}\). In our configuration, all the eigenvalues of \(\mathcal{A}\) have a negative real part and the flow is therefore globally stable.  Switching to the frequency domain, a direct relation between the spatial structure of a harmonic forcing \(\mathbf{f}(x,y,t)=\widetilde{\mathbf{f}}(x,y)e^{i\omega t}\) and its flow response \(\mathbf{q'}(x,y,t)=\widetilde{\mathbf{q}}(x,y)e^{i\omega t}\) is established: 
\begin{equation}
    \label{eq:resolvent_equation}
    \widetilde{\mathbf{q}}=\mathcal{R}\widetilde{\mathbf{f}},
\end{equation}
where \(\mathcal{R}=(i\omega\mathcal{I} -\mathcal{A})^{-1}\mathcal{P}\) is the resolvent operator and \(\omega=2\pi f \in \mathbb{R}\) is the angular frequency. For a given frequency and among all the possible forcings, we examine the one which maximizes the gain
\(\widetilde{g}^{2}(\omega)=\underset{\widetilde{\mathbf{f}} \neq 0}{\sup}\frac{||\widetilde{\mathbf{q}}||^{2}_{E}}{||\widetilde{\mathbf{f}}||^{2}_{F}}\) where \(||.||^{2}_{E}\) and \(||.||^{2}_{F}\) respectively denote the Chu energy norm and the energy of the momentum forcing \citep{Bugeat_art}. The Chu energy is defined as
\begin{equation}
 E_{\mathrm{Chu}}=\frac{1}{2}\int_{\mathcal{V}} (\overbrace{\underbrace{\bar{\rho}(|{u'}|^{2}+|{v'}|^{2}}_{E_{u'}})+\underbrace{r\frac{\bar{T}}{\bar{\rho}}|\rho'|^{2}}_{E_{\rho'}}+\underbrace{\frac{r}{\gamma-1}\frac{\bar{\rho}}{\bar{T}}|T'|^{2}}_{E_{T'}}}^{E_{tot}})dV,
\end{equation}
it contains terms relative to thermodynamic perturbations in addition to the kinetic one, and is therefore commonly used to study the global behaviour of compressible flows \citep{Hanifi_art,Bugeat_art}. For a given frequency, the fields \(\widetilde{\mathbf{f}}\) and \(\widetilde{\mathbf{q}}\) corresponding to the optimal gain \(\widetilde{g}\) are respectively called optimal forcing and response modes. Determining the optimal gain amounts to computing the largest eigenvalue of a positive generalised eigenvalue problem with the Arnoldi algorithm \citep[ARPACK library,][]{ARPACK_art} using a sparse LU solver \citep[MUMPS library,][]{MUMPS_art} for linear system solution. This global analysis tool developed in previous work \citep{Beneddine_art} was validated on the supersonic boundary layer results of \citet{Bugeat_art}. In our study, the domains involved in the definition of \(||.||^{2}_{E}\) and \(||.||^{2}_{F}\) correspond both to \(x\in[0;1910.2\delta_{0}^{*}]\) and \(y\in[0;92\delta_{0}^{*}]\).

\subsection{Local stability analysis}
\label{sec:Numerical_local_analysis}

The primary aim of the local linear stability theory (LLST) for the present study is to classify the mechanisms involved in our DNS and resolvent analysis by associating local modal mechanisms from the LLST with those observed in our purely non-modal DNS and global resolvent study. Indeed, the flow being globally stable, the growth of disturbances is only due to non-modal phenomena. These non-modal effects are a consequence of the non-normality of \(\mathcal{A}\) \citep{Schmidt_nonmodal_art}. The non-normal effects can be cast in two categories for open-flows: the \textit{component-type non-normality} and the \textit{convective-type non-normality} \citep{Sipp_Marquet_art}. Component-type non-normality is characterized by a component-wise transfer of energy between the forcing and response fields like in the Orr or lift-up mechanisms \citep{Bugeat_art} (but note the latter is absent here since lift-up is three-dimensional). Convective-type non-normality is caused by modal amplification on the local scale and is characterized by a separation of the spatial supports of the forcing and response fields. %This effect can be seen as the equivalent of a local modal instability but in a non-modal framework.

In LLST, we consider perturbations which are evolving very rapidly in the $x$ direction compared to the base flow. At each streamwise position, the base flow is considered frozen with respect to the perturbations \(\phi'=[\rho',u',v',T']\), therefore the latter can be sought in the form 
\begin{equation}
\phi'=\widetilde{\phi}(y)e^{i(\alpha x-\omega t)},\label{eq:ansatz}
\end{equation}
where in general the wavenumber $\alpha$ and the frequency $\omega$ are complex numbers. Plugging this ansatz in the linearized Navier--Stokes equations with frozen base flow profile leads to a different dispersion relation $D(\alpha,\omega;x)=0$ for each value of $x$. In the spatial stability framework, we consider real angular frequencies $\omega$ and solve for the complex wavenumber $\alpha=\alpha_r+i\alpha_i$, where $\alpha_r$ is the wavenumber and $-\alpha_i$ is the spatial growth rate along $x$. All perturbations are assumed to vanish at the free-stream boundary $y\to\infty$ while on the flat plate $y=0$, $\tilde{u}=\tilde{v}=0$ and $\mathrm{d}\tilde{\rho}/\mathrm{d}y=\mathrm{d}\tilde{T}/\mathrm{d}y=0$ (adiabatic plate). Equations are discretized along the wall-normal direction $y$ using a Chebyshev collocation method. For each values of $x$ and $\omega$, an eigenvalue problem is solved, using the LAPACK library, in order to determine the complex eigenvalue $\alpha$ and corresponding eigenvector $\widetilde{\phi}=[\tilde{\rho},\tilde{u},\tilde{v},\tilde{T}]$. The analysis is performed using an in-house code fully detailed in \citet{thesis_saintjames} and validated here on the linear local growth rates of the supersonic boundary layer \citet{Ma_Zhong_art}.

\section{Noise-amplifier behaviour and control setup}
\label{sec:amplifier_behaviour}

\subsection{Characterisation of instabilities}
\label{sec:Result_local_global}

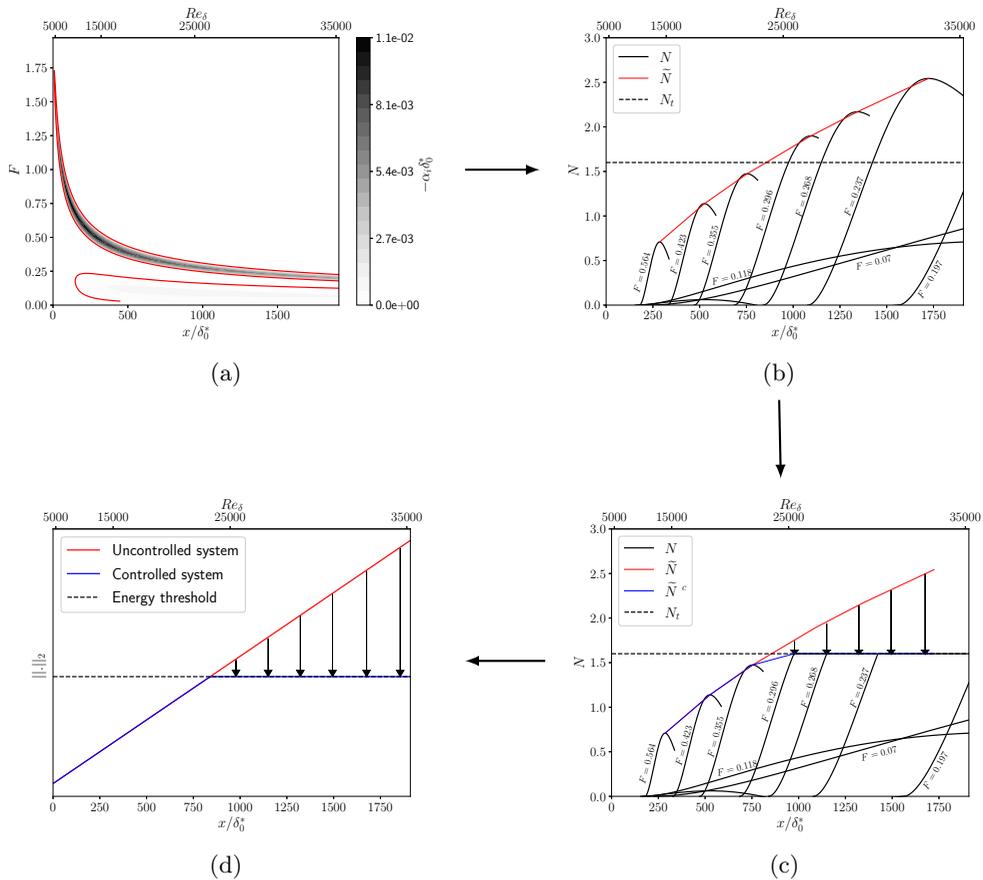
\begin{figure}
\center
\subfloat[\label{fig:neutral_curve}]{\tikz[remember
picture]{\node(1AL){\includegraphics[scale=0.3]{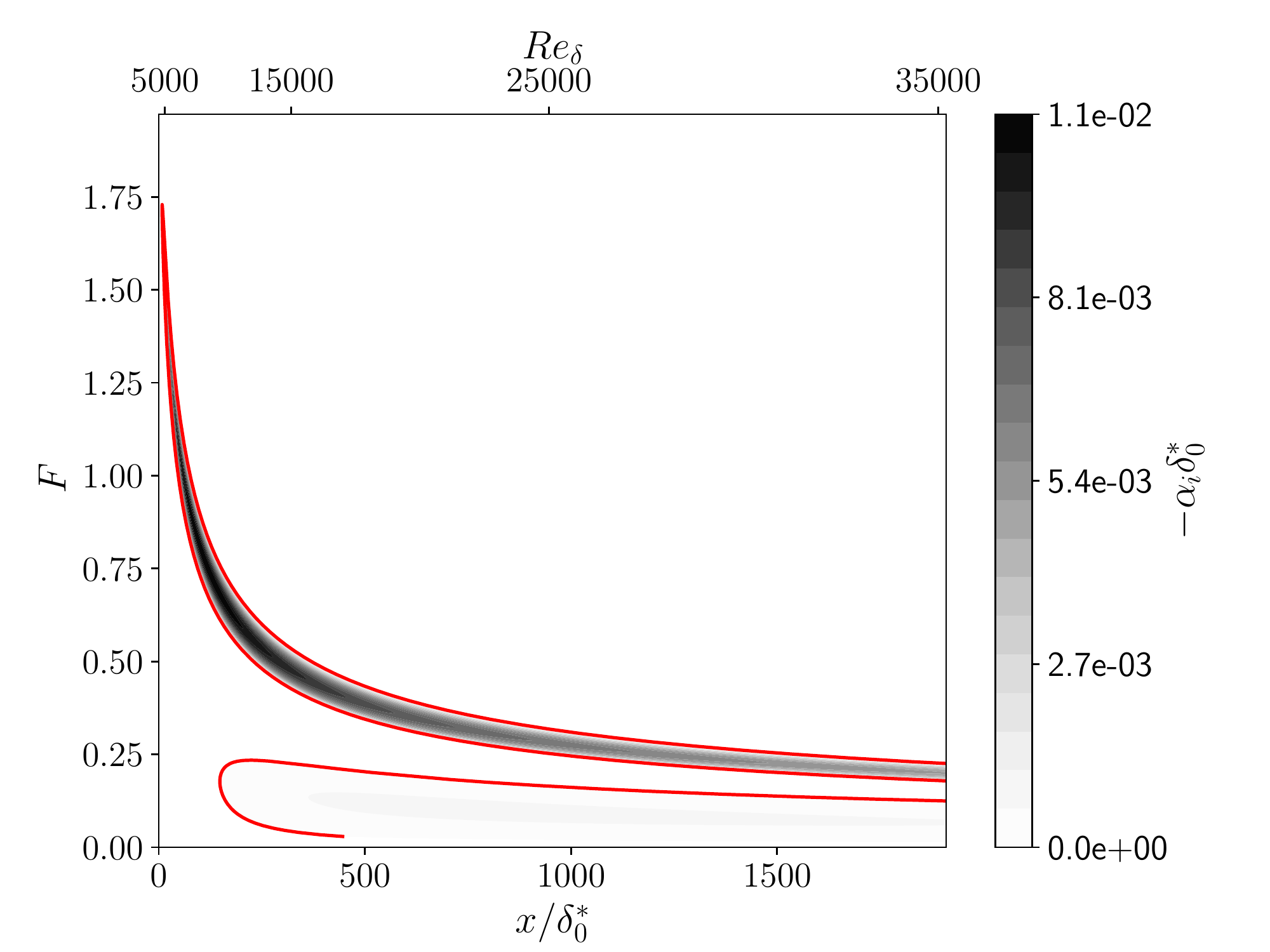}};}}%
\hspace*{1.cm}% Espacement entre les figures
\subfloat[\label{fig:Nfactor}]{\tikz[remember picture]{\node(1AR){\includegraphics[scale=0.3]{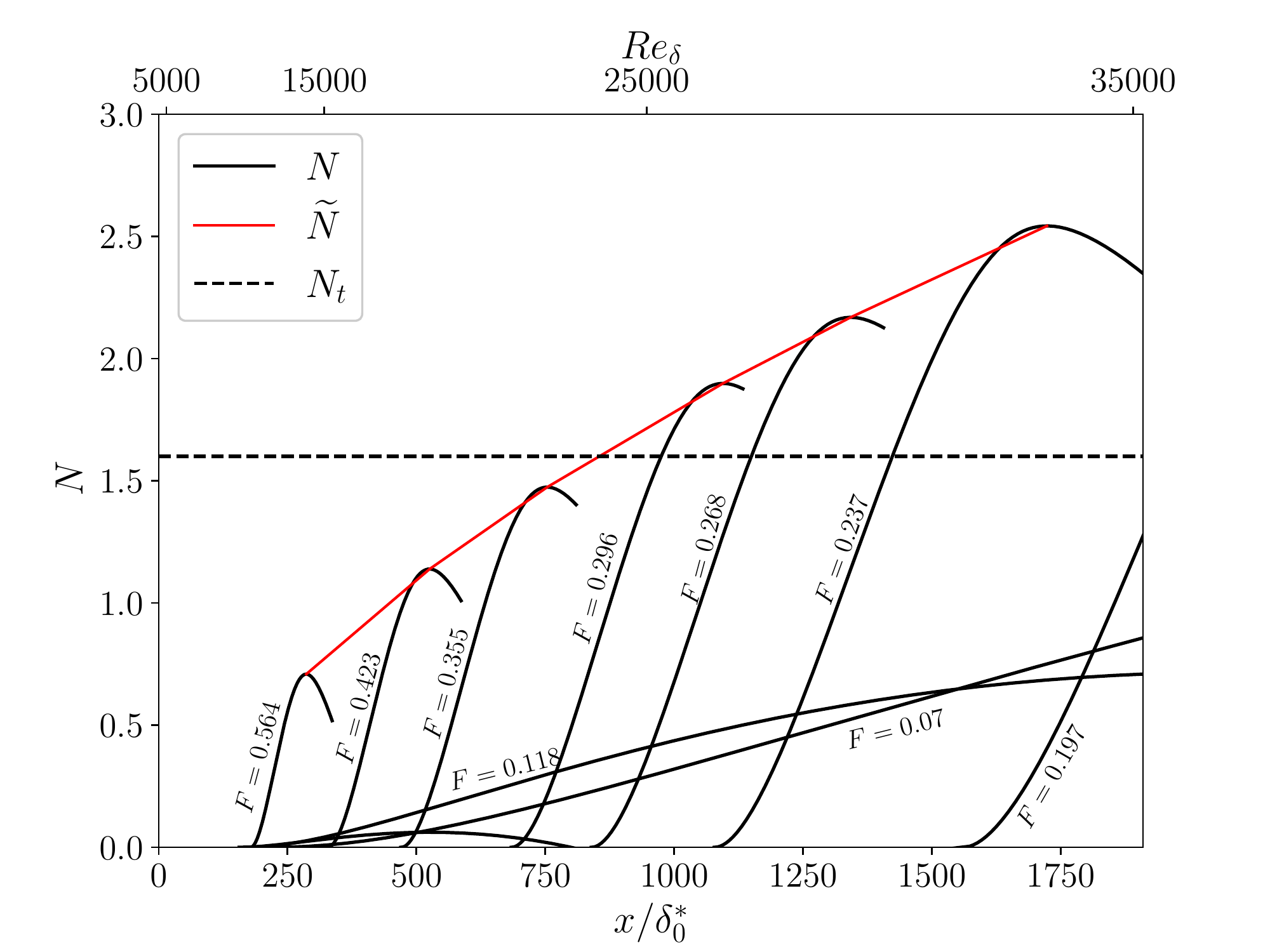}};}}\\
\vspace*{0.8cm}
\setcounter{subfigure}{3}
\subfloat[\label{fig:Nfactor_H2}]{\tikz[remember picture]{\node(2AL){\includegraphics[scale=0.3]{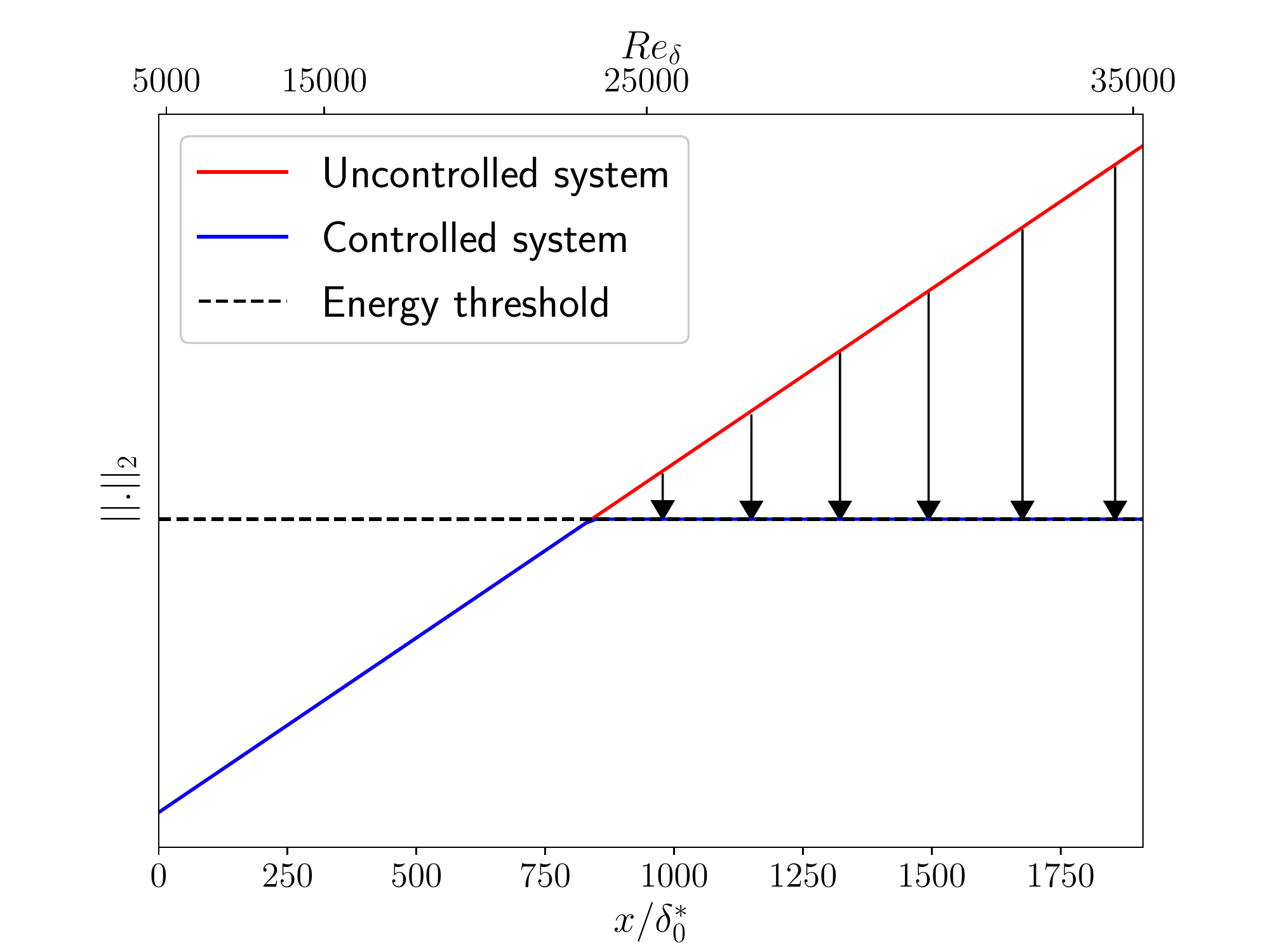}};}}
\hspace*{1.cm}% Espacement entre les figures
\setcounter{subfigure}{2}
\subfloat[\label{fig:Nfactor_controlled}]{\tikz[remember picture]{\node(2AR){\includegraphics[scale=0.3]{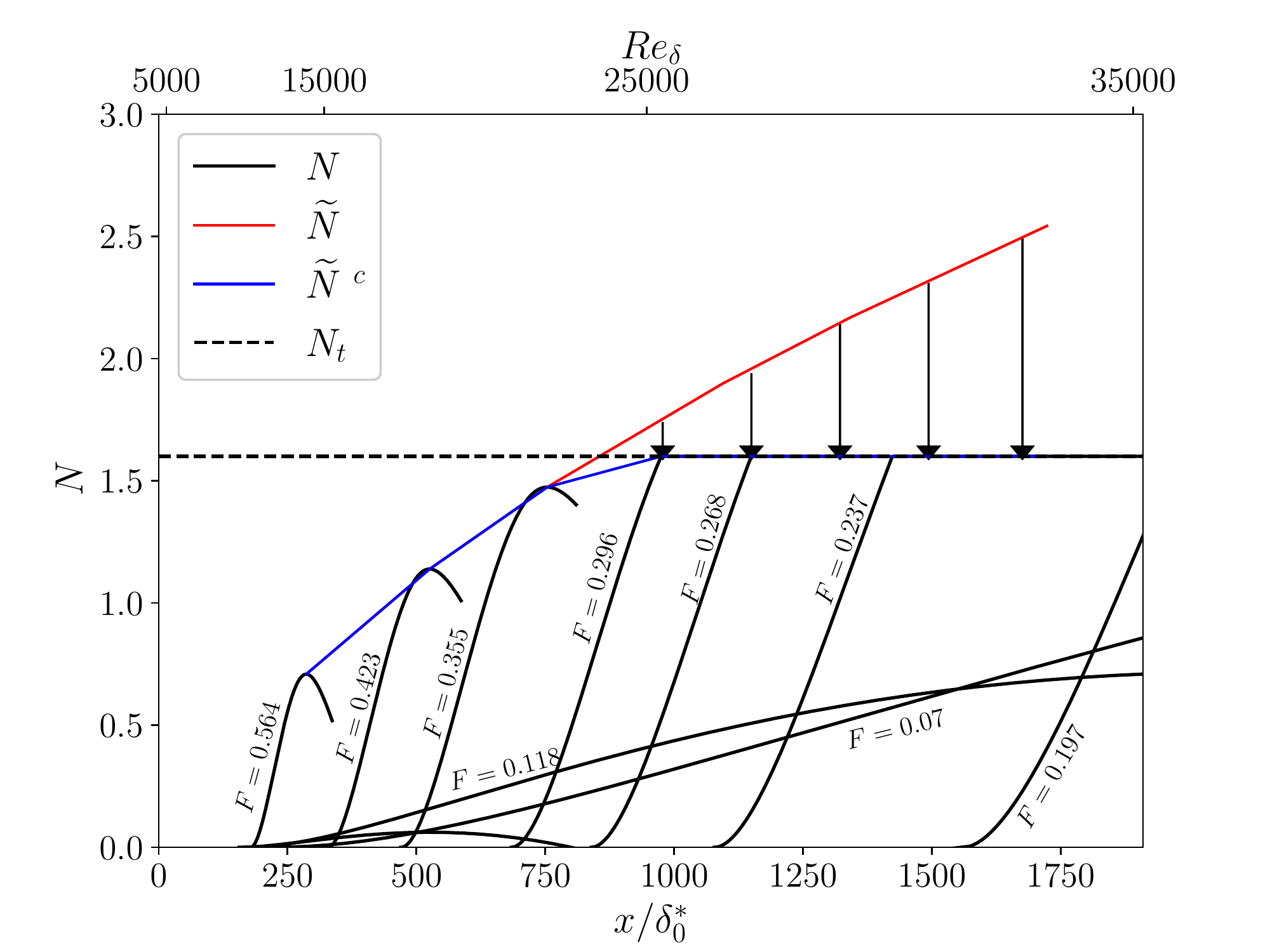}};}}
\tikz{\node (ntest)  [below=0.4cm of 1AR]{};}
\tikz[remember picture,overlay]{\draw[-latex,thick] (1AL) -- (1AR);} 
\tikz[remember picture,overlay]{\draw[-latex,thick] (ntest) -- (2AR);} 
\tikz[remember picture,overlay]{\draw[-latex,thick] (2AR) -- (2AL);}
\caption{a) Stability diagram; red solid lines represent isolines \(\alpha_{i}=0\). b) Calculation of the \(N\)-factors (black solid lines) for transition prediction based on LLST: transition occurs when \(\widetilde{N}>N_{t}\) (notional diagram). c) Performance objective for closed-loop control based on the \(N\)-factor criterion. d) Modification of the N-factor criterion using the \(H_{2}\) norm, in order to reduce conservatism. The quantity \(F=2\pi f\delta_{0}^{*}/U_{\infty}\) represents the dimensionless frequency. }
\label{fig:LLST_allfig}
\end{figure}

%The local stability analysis gives a first idea of the noise-amplifier behaviour of the flow. 
The local spatial stability diagram of spanwise-invariant perturbations is displayed in figure \ref{fig:neutral_curve}, with \(F=2\pi f\delta_{0}^{*}/U_{\infty}\) the dimensionless frequency. It is characterized by two distinct instability regions (i.e. where the spatial growth rate is positive $-\alpha_i>0$): one for the first Mack mode and one for the second Mack mode. For each mode, the instability domain (depicted by the red solid line) for a given frequency is located between branch I (convectively stable/unstable boundary) and branch II (convectively unstable/stable boundary). Each frequency is therefore amplified only on a certain portion of the domain: high frequencies are amplified upstream while low frequencies are found further downstream. Compared to the first mode, the unstable frequencies of the second mode are higher and are associated with higher growth rates. Transition to turbulence is often predicted from LLST using the \(N\)-factor \citep{smith1956transition}
\begin{equation}
N(\omega)=\int_{x_{cr}}^{x}-\alpha_{i}(\omega)dx=\ln\left(\frac{|\phi'|}{|\phi'|_{cr}}\right),
\end{equation}
with \(x_{cr}\) the location of branch I for the considered frequency and \(|\phi'|_{cr}\) the amplitude of the mode at this location. The \(N\)-factors for different frequencies are represented in figure \ref{fig:Nfactor}. Although the instability range of the first Mack mode is larger, the \(N\)-factors of the second mode are greater all along the domain due to their higher growth rates. Transition is often assumed to occur when the quantity \(\widetilde{N}=\underset{\omega}{\mathrm{max}}\ N(\omega)\) (red solid lines in figures \ref{fig:LLST_allfig}(\subref*{fig:Nfactor},\subref*{fig:Nfactor_controlled})) reaches a threshold value \(N_{t}\) (dashed lines in figures \ref{fig:LLST_allfig}(\subref*{fig:Nfactor},\subref*{fig:Nfactor_controlled}), arbitrarily placed for the explanation). This criterion means that the transition process begins when a perturbation has been amplified by a factor of \(e^{N_{t}}\). Thus, in order to delay transition to turbulence, a control action should transform the quantity \(\widetilde{N}\) obtained without control into the quantity \(\widetilde{N}^{c}\) (blue line in figure \ref{fig:Nfactor_controlled}) with control, such that \(\widetilde{N}^{c}<N_{t}\) (see figure \ref{fig:Nfactor_controlled}). The dominant frequency being different at each streamwise location of the domain, a large frequency bandwidth needs to be controlled, which complicates the design of the control law. The \(\widetilde{N}^{c}<N_{t}\) criterion could be directly translated into a \(H_{\infty}\) criterion, because this would mean that the maximum amplification over the entire frequency spectrum must not exceed a threshold over the entire domain, exactly as in the \(N\)-factor method. However, this method may be considered conservative as it is based on the worst perturbation, which is purely harmonic and therefore not quite realistic \citep{mack_1977_N}. \citet{Fedo_Mack_H2} recommended instead the use of a criterion based on both the \(N\)-factors and the entire frequency spectrum of the incoming disturbance $|\phi'|_{cr}$, which amounts to considering an \(H_{2}\) norm rather than an \(H_{\infty}\) norm. We follow this recommendation and choose a performance objective based on an \(H_{2}\) norm. More precisely, our objective will be to maintain the spatially-integrated amplification below a given threshold along the plate, and this integrated amplification will be quantified using an \(H_{2}\) norm (see figure \ref{fig:Nfactor_H2}).

\begin{figure}
    \centering
    \begin{tabular}{cc}
    \adjustbox{valign=b}{\subfloat[\label{fig:resolvent_gain}]{%
          \includegraphics[scale=0.3]{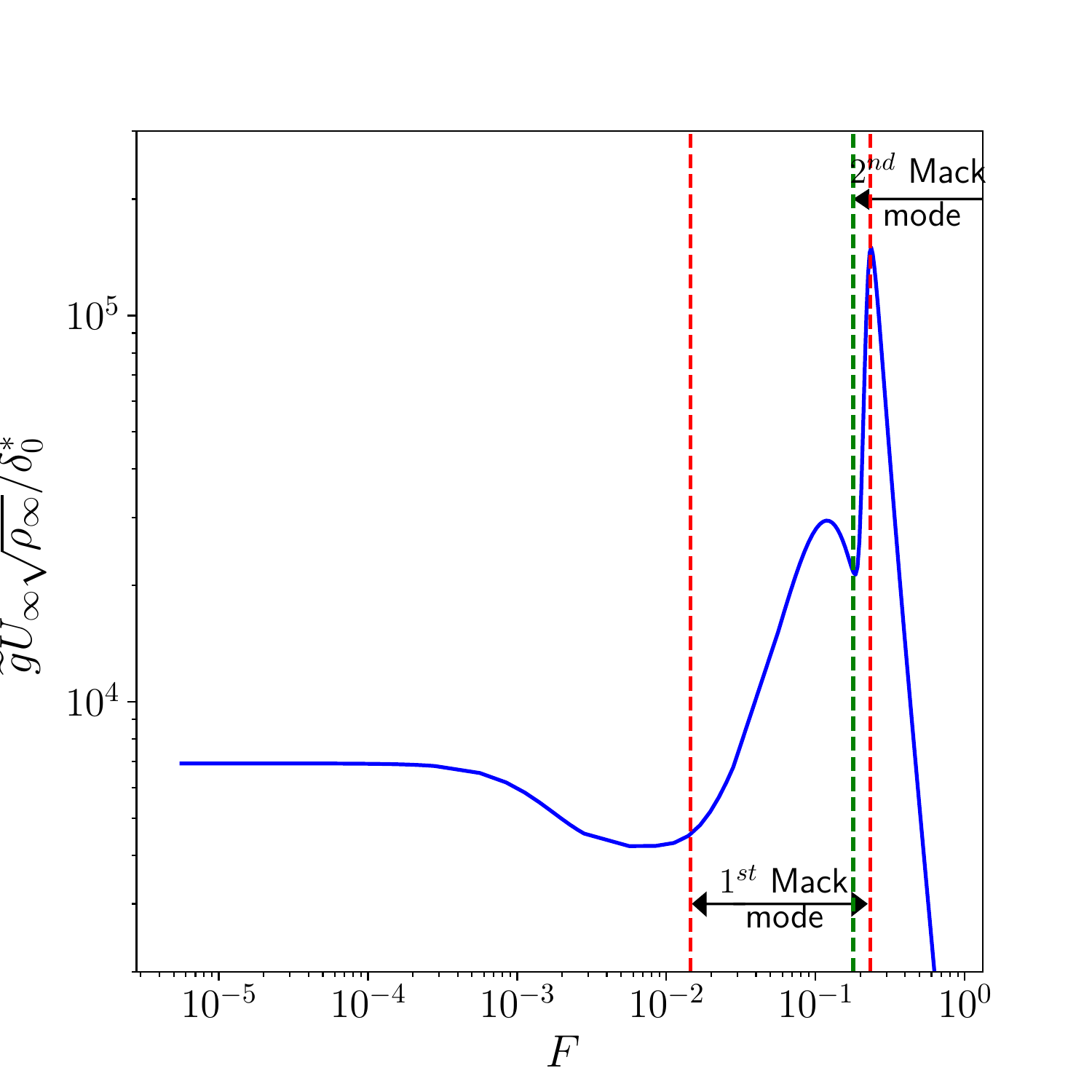}}}
    &      
    \adjustbox{valign=b}{\begin{tabular}{@{}c@{}}
    \subfloat[\label{fig:resolvent_Momentumf}]{%
          \includegraphics[trim=0.2cm 0. 0.cm 0.1cm, clip,height=0.15\linewidth,width=0.6\linewidth]{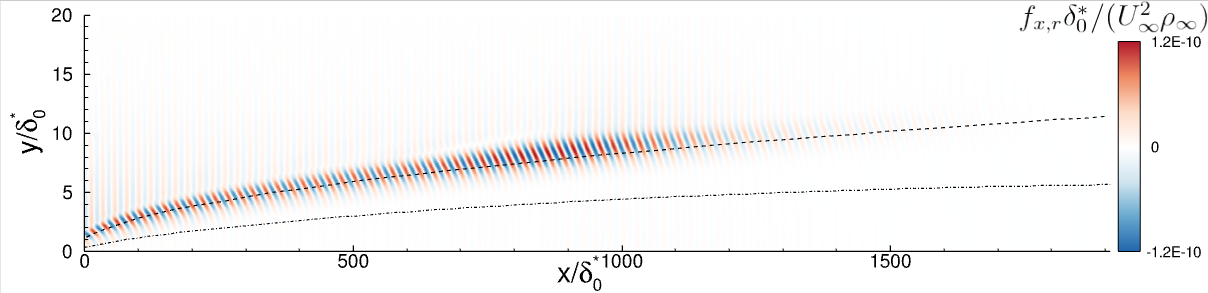}} \\
    \subfloat[\label{fig:resolvent_reponseu}]{%
          \includegraphics[trim=0.2cm 0. 0.cm 0.1cm, clip,height=0.15\linewidth,width=0.6\linewidth]{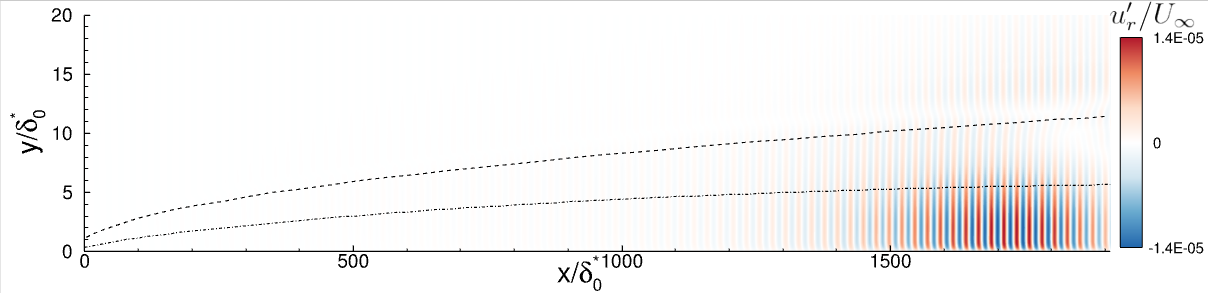}}
    \end{tabular}}
    \end{tabular}
     \subfloat[\label{fig:dchu_df}]{\includegraphics[scale=0.2]{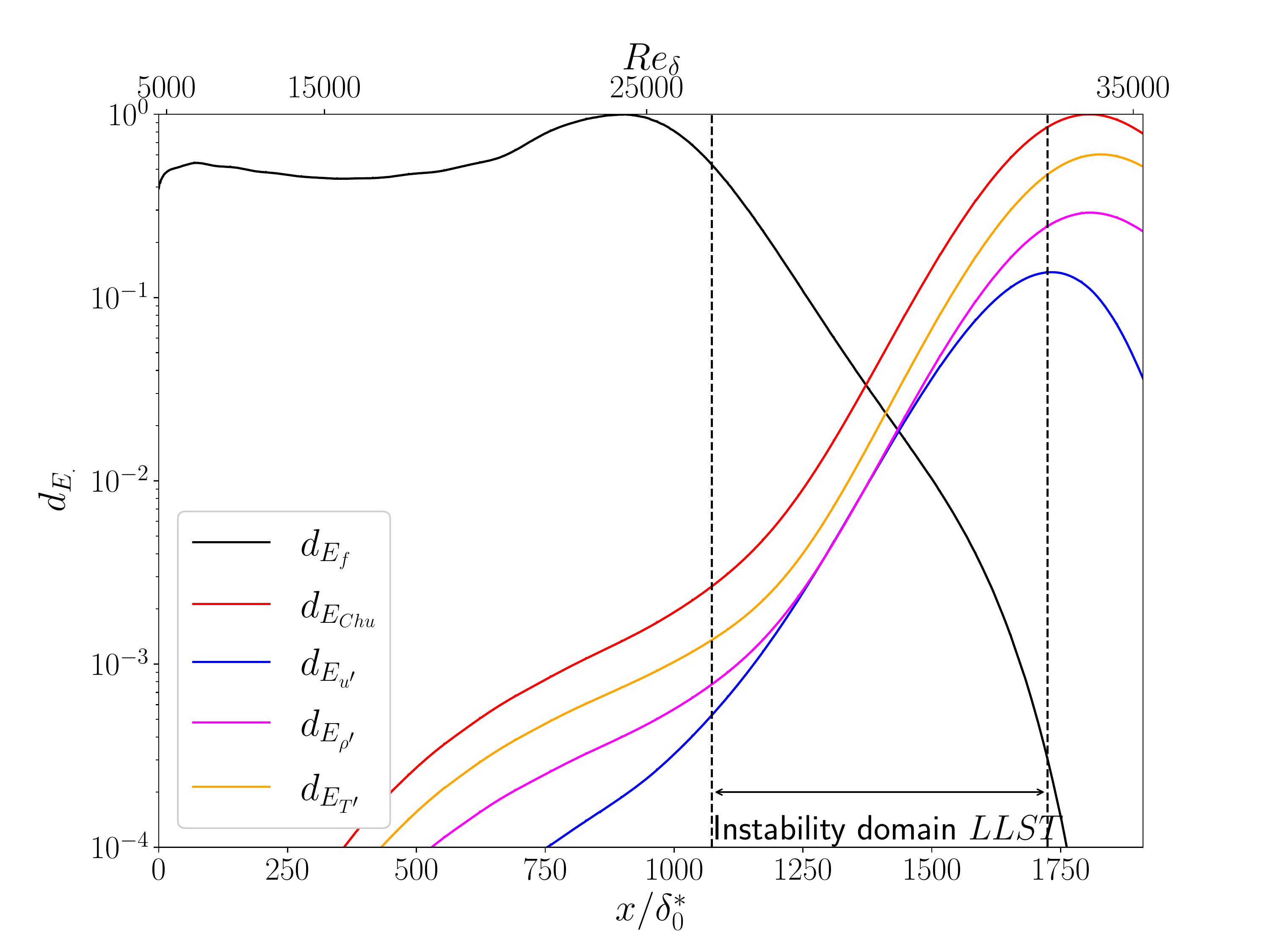}}
     \subfloat[\label{fig:alphai_modal_LST}]{\includegraphics[scale=0.2]{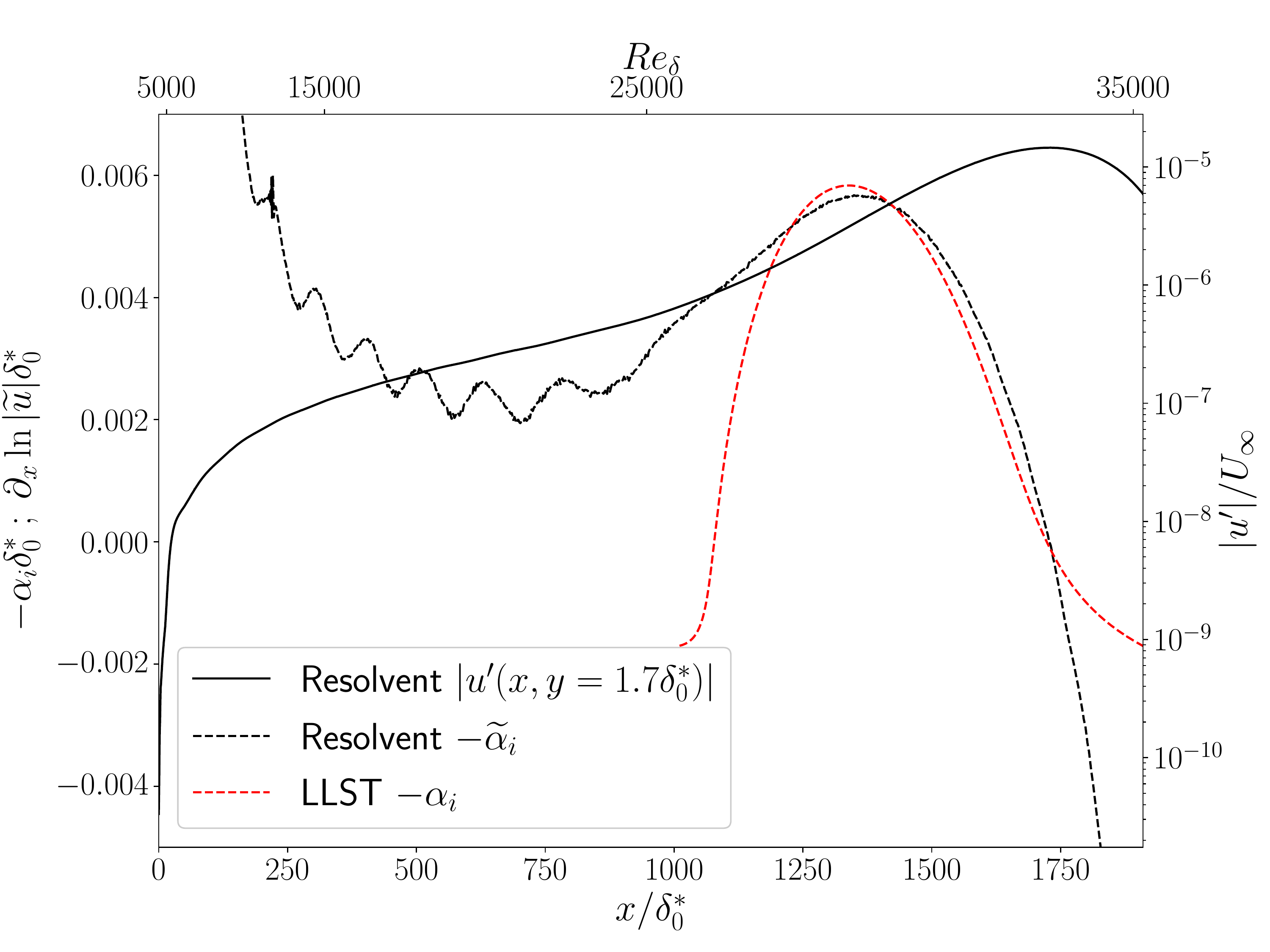}}\\
     \subfloat[\label{fig:profile_forcage}]{\includegraphics[height=0.32\linewidth,width=0.4\linewidth]{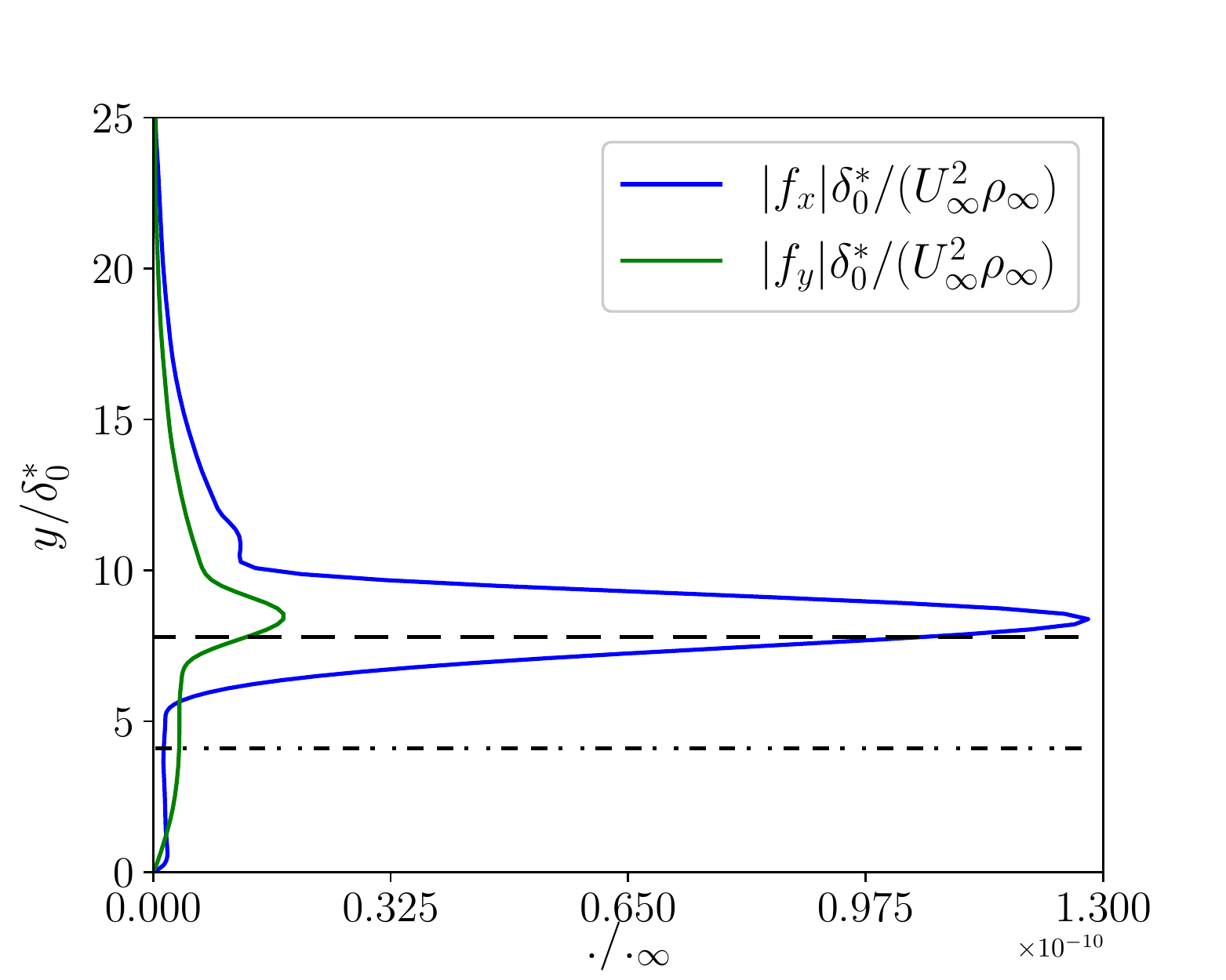}}
     \subfloat[\label{fig:profile_reponse}]{\includegraphics[height=0.32\linewidth,width=0.4\linewidth]{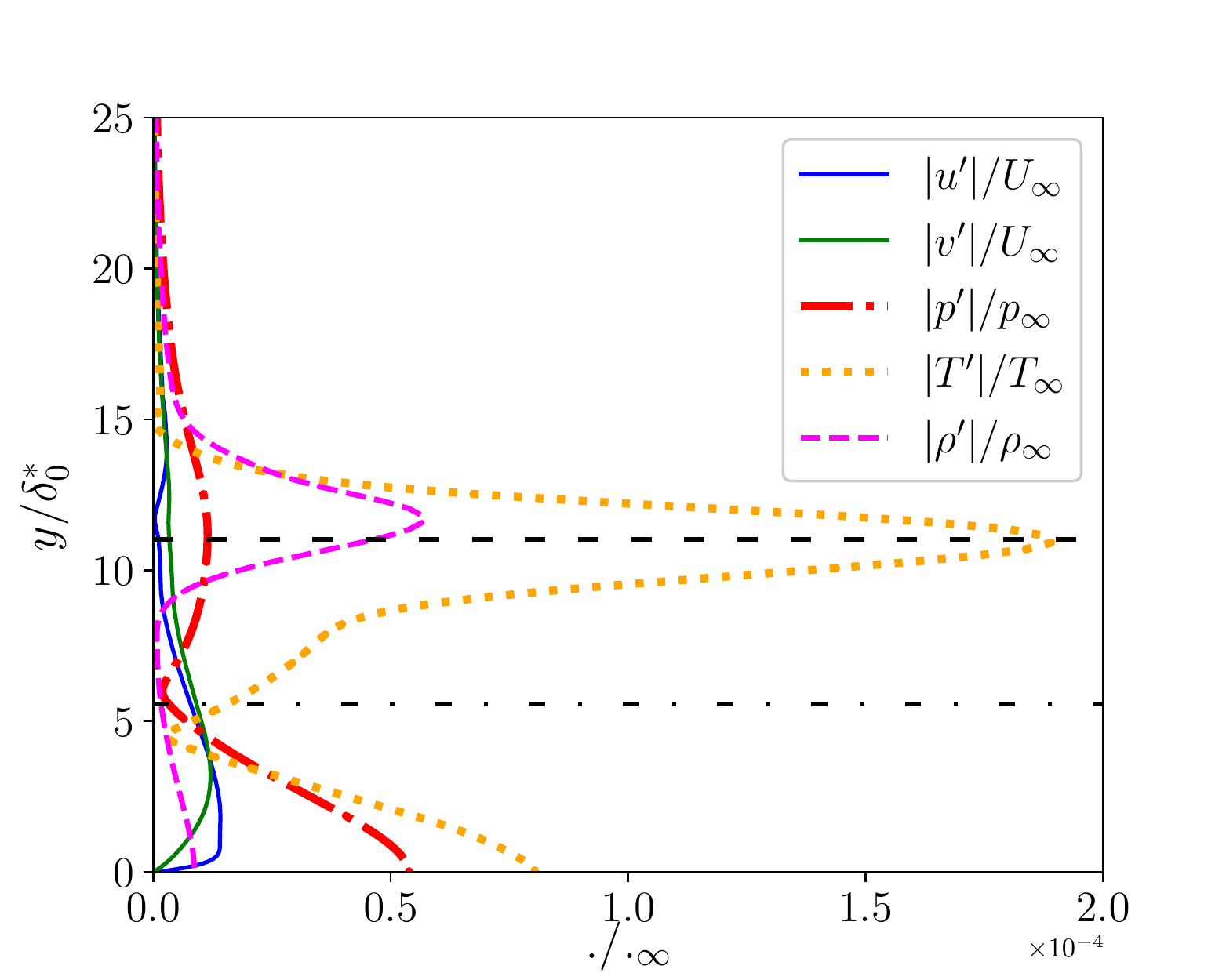}}
 
    \caption{a) Optimal resolvent gain as a function of the dimensionless frequency \(F\). According to LLST, red and green dashed areas represent the unstable frequency range of first and second Mack modes, respectively. The region where both modes are unstable corresponds to an area where the first mode is unstable over a tiny distance. Real part of the streamwise component of the optimal forcing (b) and its associated streamwise velocity response (c) at \(F=0.237\). d) Evolution at \(F=0.237\) of the forcing density and the different contributions to the Chu's energy density normalized by their maximum values. The position of the branch I and II from LLST are symbolized by vertical dashed lines. e) Comparison of $-\alpha_{i}$ and $-\widetilde{\alpha}_{i}$ at \(F=0.237\). Profiles of the optimal forcing components at \(x=867.2\delta_{0}^{*}\) (f) and response at \(x=1766.7\delta_{0}^{*}\) (g) at \(F=0.237\). The black dashed and dashed-dotted lines in (b), (c), (f) and (g) represent the generalised inflection point position and the limit of the region of supersonic instabilities (\(\widehat{M}>1\) below this line), respectively.}
\label{fig:all_image_resolvent}
\end{figure}

The global stability results based on resolvent analysis complement those previously obtained from LLST. The optimal energy gain $ \tilde{g} $ as a function of the forcing frequency $F$ is represented in figure \ref{fig:resolvent_gain}. 
This curve displays two peaks at \(F\approx0.118\) and \(F\approx0.237\), which correspond respectively to the first and second Mack modes identified in LLST. Global resolvent analyses are consistent with those of the local approach, since the optimal energy gain is closely related to \(N\)-factors \citep{Sipp_Marquet_art,Beneddine_art}.
%the \(N\)-factors as the most important \(N\)-factor, which represents the most amplified frequency on the whole domain, will be found considering figure \ref{fig:Nfactor} for a frequency close to \(F=0.237\). %The largest values of gains (and \(N\)-factors) being found for frequencies of the second mode of Mack, we will focus WHAT DOES FOCUS MEAN? ANY WAYS, WE CONTROL ALL FREQUENCIES? on the control of this second mode in our 2D study.

For the frequency \(F=0.237\) leading to the highest gain, the real parts of the streamwise optimal forcing and velocity response are shown in figures \ref{fig:all_image_resolvent}(\subref*{fig:resolvent_Momentumf},\subref*{fig:resolvent_reponseu}). The spatial structure of the forcing is located upstream of the domain while that of the response is located further downstream. This separation of the spatial supports, related to the convective-type non-normality of the Jacobian operator, implies a time-delay between actuation upstream and sensing downsteam, making the design of a robust control law even more complex. Figure \ref{fig:dchu_df} shows that the peak of the forcing density $d_{E_{f}}(x)=\int_{0}^{y=92\delta_{0}^{*}} |\widetilde{\mathbf{f}}|^{2}dy$ (resp. Chu's energy density $d_{E_{Chu}}(x)=\int_{0}^{y=92\delta_{0}^{*}} E_{tot}\,\mathrm{d}y$) is not very far from the position of branch I (resp. II) from LLST \citep{Sipp_Marquet_art}. The energy of the response is dominated at each abscissa by the thermodynamic quantities $E_{T'}$ and $E_{\rho'}$, while quantity $E_{u'}$ has a smaller contribution. A comparison between the spatial amplification rates \(-\alpha_{i}\) from LLST (red dashed line) and \(-\widetilde{\alpha}_{i}=\frac{1}{|\widetilde{u}(x,y=1.7\delta_{0}^{*})|}\partial_x |\widetilde{u}(x,y=1.7\delta_{0}^{*})|\) from resolvent analysis (black dashed line) is depicted in figure \ref{fig:alphai_modal_LST}. The quantity $-\widetilde{\alpha}_{i}$ represents the slope of $\ln{|\widetilde{u}|}$ (black solid line) and can therefore be compared to a growth rate. The growth of the resolvent mode within \(x\in[0;1078\delta_{0}^{*}]\) is due to the optimal forcing that is non-zero in this region (see figure \ref{fig:dchu_df}) and that induces the response. The inclined pattern in the forcing field (see figure \ref{fig:resolvent_Momentumf}) indicates that the response also takes advantage of the Orr mechanism \citep{art_Orr} and more generally of non-modal local interactions.
  %but it results in a velocity response field aligned with the wall-normal direction (see figure \ref{fig:resolvent_reponseu}). A component-wise transfer of energy between both fields occurs leading to algebraic amplitude growth of perturbations.
  After this initial growth region induced by the forcing, both \(-\alpha_{i}\) and \(-\widetilde{\alpha}_{i}\) exhibit similar values in the region between \(x\in[1200\delta_{0}^{*};1730\delta_{0}^{*}]\), which indicates that transient growth is then dominated by the convective instability associated to the second Mack mode.  

To maximize the amplification of the second Mack mode, the forcing field (see figures \ref{fig:all_image_resolvent}(\subref*{fig:resolvent_Momentumf},\subref*{fig:profile_forcage})) must be localised near the generalised inflection point \(y_{g}\) (denoted in figures \ref{fig:all_image_resolvent}(\subref*{fig:resolvent_Momentumf},\subref*{fig:resolvent_reponseu},\subref*{fig:profile_forcage},\subref*{fig:profile_reponse}) with a dashed line), defined as
$\left.\partial_y \left(\bar{\rho} \partial_y \bar{u}\right)\right|_{y_{g}}=0$. A region of supersonic instabilities (below the dashed-dotted line in figures \ref{fig:all_image_resolvent}(\subref*{fig:resolvent_Momentumf},\subref*{fig:resolvent_reponseu},\subref*{fig:profile_forcage},\subref*{fig:profile_reponse})
), defined as \(\widehat{M}=\frac{\left|\bar{u}-\frac{\omega}{\widetilde{\alpha}_{r}}\right|}{\sqrt{\gamma r\bar{T}}}>1\) with \(\widetilde{\alpha}_{r}\) the global resolvent streamwise wavenumber computed as \(\widetilde{\alpha}_{r}=\partial_x \arg(\widetilde{u})\) where \(\arg\) stands for the argument of a complex number \citep[see][]{Beneddine_art}, is detected close to the wall (see figure \ref{fig:resolvent_reponseu}). %This wavenumber \(\alpha_{r}\) resulting from a mixture of different modes in the global analysis is qualified as non-modal and differs from the \(\alpha_{r}\) coming from the modal LLST and corresponding to only one mode.
This confirms that the optimal response mode at $ F=0.237 $ corresponds to a second Mack mode \citep{mack_1984_art}. Note that the critical layer, where $\bar{u}=\frac{\omega}{\widetilde{\alpha}_{r}}$, is not shown here as it is similar to the generalised inflection point; indeed, the phase velocity of an inflectional neutral wave in the LLST is equal to the mean velocity at \(y_{g}\) \citep{mack_1984_art}. 

Finally, we observe in figure \ref{fig:profile_reponse} that the different components of the second Mack mode peak at different locations in the wall-normal direction $y$. Hydrodynamic perturbations (velocity and pressure) peak close to the wall and seem trapped in the region $\widehat{M}>1$ whereas thermodynamic quantities (density and temperature) peak near the generalised inflection point. This observation is in complete agreement with the qualitative results of \citet{Bugeat_art}.

%Indeed, contrary to the first mode which is unstable to inviscid instabilities due to the existence of a generalized inflection point, the inviscid instabilities of the second Mack mode come from the presence of a region where the velocity mean flow, relative to the disturbance phase velocity, is supersonic \citep{mack_1984_art}.
\subsection{Control setup}
\label{sec:specification_h2_hinf}

\begin{figure}
\centering
     \includegraphics[scale=0.6]{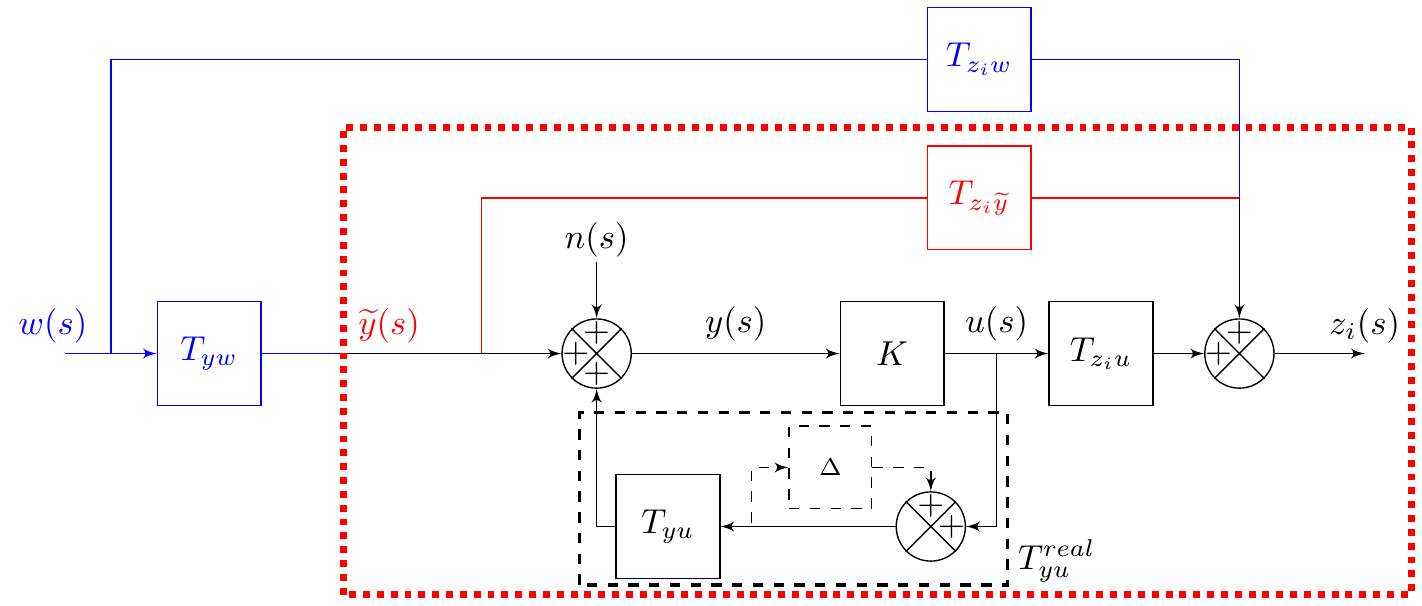}
\caption{Block diagram for noise-amplifier flows for feedforward and feedback configurations in an ideal case (with quantities in blue and black) and in a realistic setup (with quantities in red and black). The quantities in black are common to the ideal and realistic cases. The red dotted square therefore represents the system used with the aim of an experimentally feasible synthesis. In a feedforward setup, \(T_{yu}=\Delta=0\).}
\label{fig:schema_block_both}
\end{figure}

External perturbations are modelled using a random time signal $w$ (see figure \ref{fig:numerical_domain}) that multiplies a time-independent volume force field.
In the case of small amplitude noise considered in this paper, the dynamics is linear and will take advantage of the various instability mechanisms described in the previous section.
If we consider several performance sensors $ z_i$ measuring the flow perturbations along the plate, the transfer functions $ T_{z_iw}=z_i(s)/w(s)$, with \(s \in \mathbb{C}\) the Laplace variable, provide an accurate prediction of the downstream perturbation level without control.
The reactive control setup is depicted in figure \ref{fig:schema_block_both}.
An upstream actuation $ u $ generates small-amplitude perturbations that take again advantage of the instability mechanisms to grow and eventually cancel the fluctuations at the downstream measurements $ z_i $.
The phase of the generated perturbations is therefore important and needs to be tuned with respect to the incoming perturbations that are governed by $ w $.
For this, we introduce an upstream sensor $y$ and   
design a controller $ K$, that actually corresponds to the transfer function $ K=T_{uy} $, and which transforms the noise measurement $ y $ into an actuation signal $ u $.
It is straightforward to show that, in the presence of control, the transfer functions from $ w $ to $ z_i$, denoted with the superscript \(c\), become:
\begin{equation}
    \label{eq:closed_loop}
    T_{z_{i}w}^{c}=T_{z_{i}w}+T_{z_{i}u}K(1-T_{yu}K)^{-1}T_{yw}.
\end{equation}
The design of $K$ therefore requires
additional transfer functions: $ T_{yw}$ characterizes the influence of noise on the upstream measurement $y$, $ T_{z_iu}$ characterizes the influence of the actuator on the downstream performance sensors and, for feedback setups only, $ T_{yu}$ characterizes the influence of the actuator on the upstream sensor. 
In the following, we will assume that \(w\) is a white-noise input and will seek to reduce the expected power of the measurements \(z_{i}\). This expected power, normalized by the intensity of the white-noise input, is measured by the \(H_{2}\) norm of \(T_{z_{i}w}^{c}\). For any stable SISO transfer function $G$, the \(H_{2}\) norm is defined as
\begin{equation}
    ||G||_{2}=\left(\frac{1}{2\pi}\int_{-\infty}^{+\infty}|G|^{2}d\omega\right)^{1/2}.
\end{equation}
%this norm reflects an energy distributed over all frequencies and represents the power of the output controlled signal \(z_{i}\) to unit white noise input \(w\).

%In the absence of a control action, the single-input single-output (\textit{SISO}) transfers \(T_{yw}\) and \(T_{z_{i}w}\) represent the impact of the exogenous input \(w\) on the sensors \(y\) and \(z_{i}\), respectively. The impact of the actuator on the sensors, in the absence of disturbances, is represented by the SISO transfers \(T_{yu}\) and \(T_{z_{i}u}\). 

%Fluidic specifications are as follows: determine the control law \(K\) that allows to reduce as much as possible the amplitude of the disturbances propagating along the plate, while keeping the control law efficient and stable despite modelling errors, new noise sources or inflow condition variations. 

 However, determining the transfers coming from the  noise \(w\) is not possible in realistic cases because the noise environment is unknown (it depends on the characteristics of the wind tunnel or the free-stream turbulence on airplanes). An experimentally feasible control design must therefore not be based on \(T_{z_{i}w}\) and \(T_{yw}\). Following \citet{Herve_art}, the solution proposed here is to introduce an artificial transfer function, $T_{z_{i}\tilde{y}}$, which is intended to predict the downstream measurements $ z_i $ from the upstream measurement $y$ in the absence of a control action. This apparent transfer function ($y$ is not a source) is defined as $T_{z_{i}\tilde{y}}=T_{z_{i}w}T_{yw}^{-1}$ (observability requires that $T_{yw}\neq 0$ when $T_{z_{i}w} \neq 0$) \citep{Sasaki3D_wavecancelling_art,SasakiPSE_art}. In real applications, we can identify this transfer function from uncontrolled $ (y,z_i) $ data. In the following, we will consider $\tilde{y}=T_{yw} w$ as the new exogeneous input of the system. We are therefore led to the modified block diagram framed by the red dotted square in figure \ref{fig:schema_block_both}, where in case of actuation, the upstream measurement reads: $ y=\tilde{y}+T_{yu} u$ (+$n$, which is a measurement noise). In such a case, the controlled transfer function becomes:
  \begin{equation}
    T_{z_{i}\tilde{y}}^ {c}=T_{z_{i}\tilde{y}}+T_{z_{i}u}K(1-T_{yu}K)^{-1}. 
  \end{equation}
  %to expressed \(z_{i}\) as \(z_{i}=T_{z_{i}y}y+r\) where \(T_{z_{i}y}\) is a transfer function design to maximize the extraction of linear information from \(y\) to \(z_{i}\); \(r\) is a residual part that can represent non-linearities, noise on the measurements or linear dynamics not detected by \(y\) but impacting \(z_{i}\). In our 2D linear numerical study where the flow is excited by a single exogenous noise \(w\) and with ideal sensors, we have \(r=0\) and \(T_{z_{i}y}\) is defined such that \(T_{z_{i}y}=\frac{z_{i}}{y}\) as in \citet{SasakiPSE_art}.
The ideal and realistic control schemes shown in figure \ref{fig:schema_block_both} are related through:  
\begin{equation}
    ||\ T_{z_{i}w}^ {c}\ ||_{2}=||\ |T_{yw}|\ T_{z_{i}\tilde{y}}^{c}\ ||_{2}.
    \label{eq:H2_norm_2}
\end{equation}
The term \(|T_{yw}|\) therefore corresponds to a frequency weighting function and can be replaced by \(W_{y}=\sqrt{PSD_{y}(\omega)}\) (where \(PSD_{y}\) is the power spectral density of the estimation sensor \(y\) in the absence of a control action); it represents the fact that the new system input $ \tilde{y} $ is no longer a white noise as $ w $ but a colored noise. Therefore, the four quantities needed for the synthesis are \(T_{z_{i}u}\), \(T_{yu}\), \(T_{z_{i}\tilde{y}}\) and \(W_{y}\). %Noted that \(e^{i\varphi_{yw}}\), representing the phase shift between \(y\) and \(w\), is not required.
They can all be obtained in a realistic setup using input-output data and will be the ones used for identification (see \S \ref{sec:Identif_noise_amplifier}) and controller synthesis (see \S \ref{sec:synthesis_description_algo}).
%The block diagram for an experimentally doable synthesis is shown in figure \ref{fig:synthese_couche_limite}.
For the sake of clarity and to simplify notations, the quantity \(||W_{y}\ T_{z_{i}\tilde{y}}^{c}||_{2}\) will be replaced in the rest of the paper by \(||T_{z_{i}w}^ {c}||_{2}\).

Maintaining closed-loop performance in spite of modelling errors or inflow conditions variations around the nominal case requires first and foremost the stability robustness of the control law. From a control design point of view, this implies considering uncertainties $ \Delta $ representing a model error on $T_{yu}$ that can lead to the instability of the feedback loop.
For example, for the block $ \Delta $ represented in figure \ref{fig:schema_block_both}, if no upstream noise is considered, we have $ y=\frac{T_{yu}}{1-\Delta} u $, so that  
$\Delta$ represents an inverse multiplicative uncertainty on $T_{yu}$ such that $\Delta=\frac{T_{yu}^{real}-T_{yu}}{T_{yu}^{real}}$. This type of uncertainty has the advantage of representing a relative error, which facilitates its interpretation. 
%between the measurement $ y $ and the noise \(n\) injected in this same measurement (see figures \ref{fig:schema_block_both}(\subref*{fig:block_couche_limite},\subref*{fig:synthese_couche_limite})), such that the closed-loop transfer-function becomes:
%Hence $ \Delta $ appears as an additive perturbation on $ T_{yu}K $.
Since \(-T_{yu}^{real}K\) does not exhibit any unstable pole ($ T^{real}_{yu}$ is stable because the boundary layer flow is globally stable while $ K $ is stable by design), the closed loop system is stable if and only if the Nyquist plot of \(-T_{yu}^{real}K\) does not encircle the critical point (-1, 0), which is equivalent to $|1-T_{yu}K|>|\Delta|$ \citep{Skogestad_art}. Therefore, the stability of the closed loop can be guaranteed by working on the sensitivity function
\begin{equation}
    \label{eq:sensitivity_function}
    S=(1-T_{yu}K)^{-1}.
\end{equation}  
Defining the \(H_{\infty}\) norm of a stable SISO transfer function $G$ as
\begin{equation}
    ||G(s)||_{\infty}=\underset{\omega \in \mathbb{R}}{\sup}\ |G(i\omega)|,
\end{equation}
we request to maintain the \(H_{\infty}\) norm of the sensitivity function \(S\) below a threshold, which allows to keep adequate stability margins. By directly measuring the minimal distance between the Nyquist plot and the critical point (-1, 0) after which the closed loop becomes unstable for a negative feedback loop, the modulus margin \(||S||_{\infty}^{-1}\) appears to be the most generic measure for quantifying the available stability margin \citep{Skogestad_art}.

Finally, maintaining optimal performance despite uncertainties on certain frequency range of the measurement $ y$ means minimizing the \(H_{\infty}\) norm of the transfer function
\begin{equation}
    \frac{u}{n}=KS.
\end{equation}
Desensitizing the control output $u$ on certain frequency ranges allows to be robust to noise $n$ on the estimation sensor \(y\). %As a matter of fact, a measurement noise \(n\) is strongly amplified in case of high gain \(KS\) on its frequency bandwidth, resulting in a control signal \(u\) of high amplitude on the this bandwidth.
Even if these frequencies are attenuated far downstream of the actuator (if they are convectively stable, resulting in low $|T_{z_i\tilde{y}}|$), strong injection of energy may occur in the direct vicinity of the actuator, which may in turn provoke transition to turbulence in a 3D setup.  

In summary, the fluidic specifications for noise-amplifier flows may be reformulated from a control point of view as an optimization problem based on \(H_{2}\) and \(H_{\infty}\) norms, in order to guarantee both performance and robustness. The constrained minimization problem for our specific study will be formulated in \S \ref{sec:synthesis_description_algo}.

\subsection{Selecting actuator and sensors}
\label{sec:Actuator_description}

For a given external perturbation, the choice of appropriate actuator and sensors is essential to ensure effective flow control. The input perturbation, representing an external disturbance (acoustic noise, roughness, freestream turbulence, etc.) is modelled by a volume forcing $ w(t) \mathbf{B}_{w}(x,y) $ in the right-hand-side of the momentum equations \eqref{eq:Navier_Stokes}(b),
where the noise $ w(t)$ is chosen white (with a variance sufficiently small for the perturbation to remain in the linear regime) and
\(\mathbf{B}_{w}(x,y)\) is divergence-free and compact in space \citep{Bagheri_art,Semeraro_art,Belson_art}:
\begin{equation}
    \mathbf{B}_{w}=\mathbf{h}(\frac{10.66}{\delta_{0}^{*^{2}}},4.1\delta_{0}^{*}, \delta_{0}^{*}, 1.5\delta_{0}^{*}, 0.15\delta_{0}^{*}),
\end{equation}
with
\begin{equation}
\mathbf{h}(A_{h},x_{0},y_{0},\sigma_{x},\sigma_{y})=A_{h}
\begin{pmatrix}
   (y-y_{0})\sigma_{x}/\sigma_{y}\\
   -(x-x_{0})\sigma_{y}/\sigma_{x}\\
\end{pmatrix}  
\mathrm{exp}^{-\left(\frac{x-x_{0}}{\sigma_{x}}\right)^{2}-\left(\frac{y-y_{0}}{\sigma_{y}}\right)^{2}}.
\label{eq:divergence_freeeq}
\end{equation}
It is centred around the generalised inflection point in the wall-normal direction in order to maximize the receptivity process by exciting the optimal mechanisms of the second Mack mode, which is the most amplified, as shown by the resolvent analysis results in section \ref{sec:Result_local_global}. The position of \(\mathbf{B}_{w}\) in the streamwise direction is upstream of branch I (locally stable regions) for all frequencies according to the LLST.

For the sensors, in order to have strong observability of the disturbances, we choose \(y\) and \(z(x)\) to be wall-pressure fluctuation sensors. This choice is supported by the fact that second Mack modes exhibit strong pressure fluctuations close to the wall, as shown by the optimal response profiles in figure \ref{fig:profile_reponse}. Also, that kind of sensors is commonly used in supersonic experimental studies \citep{Lugrin_expe}. 

\begin{figure}
    \centering
     \subfloat[\label{fig:Tzwx_evol}]{\includegraphics[scale=0.3]{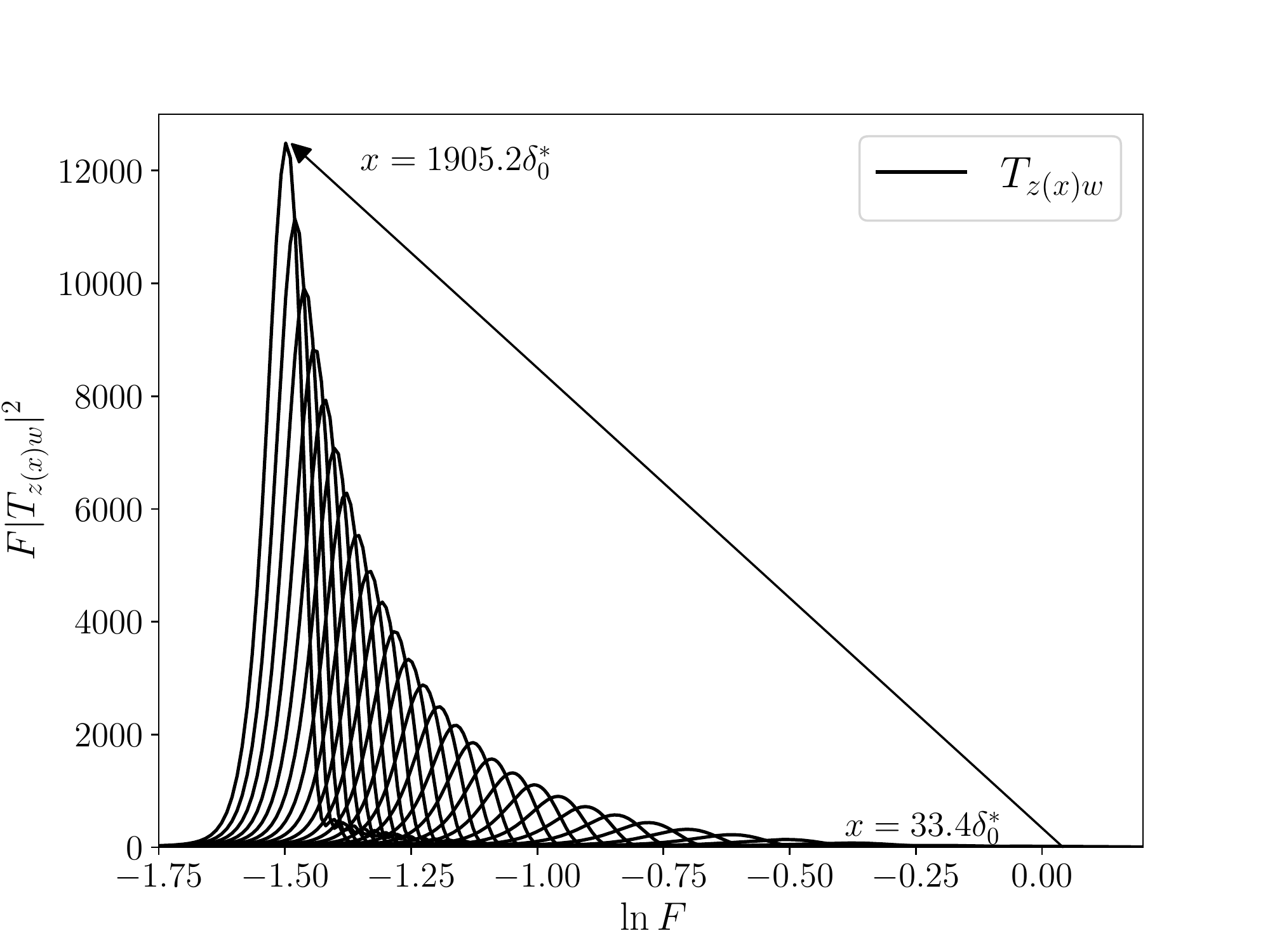}}
      \subfloat[\label{fig:Tzwx_FactorN}]{\includegraphics[scale=0.3]{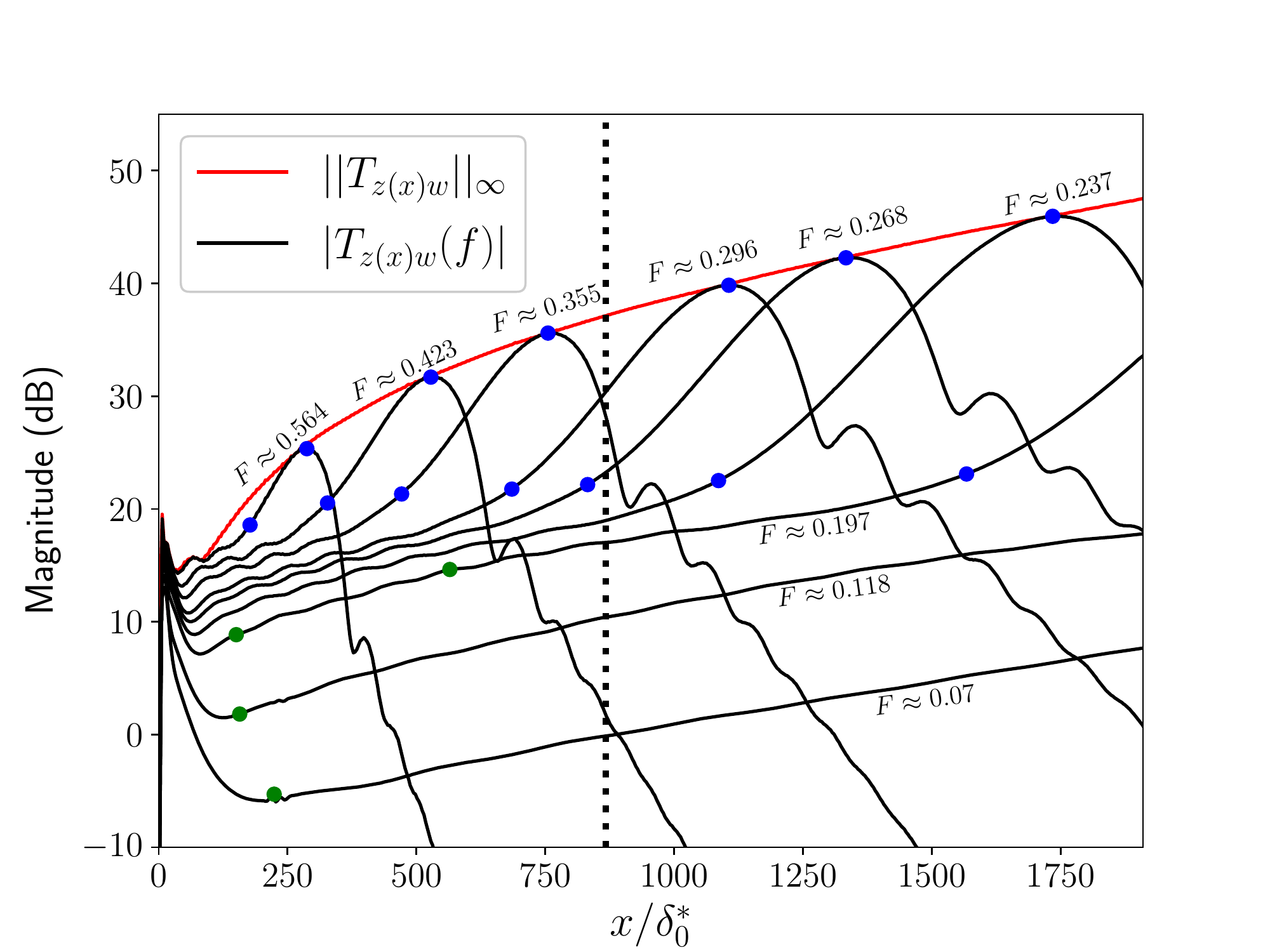}} \\
      \subfloat[\label{fig:Tzwx_h2}]{\includegraphics[scale=0.3]{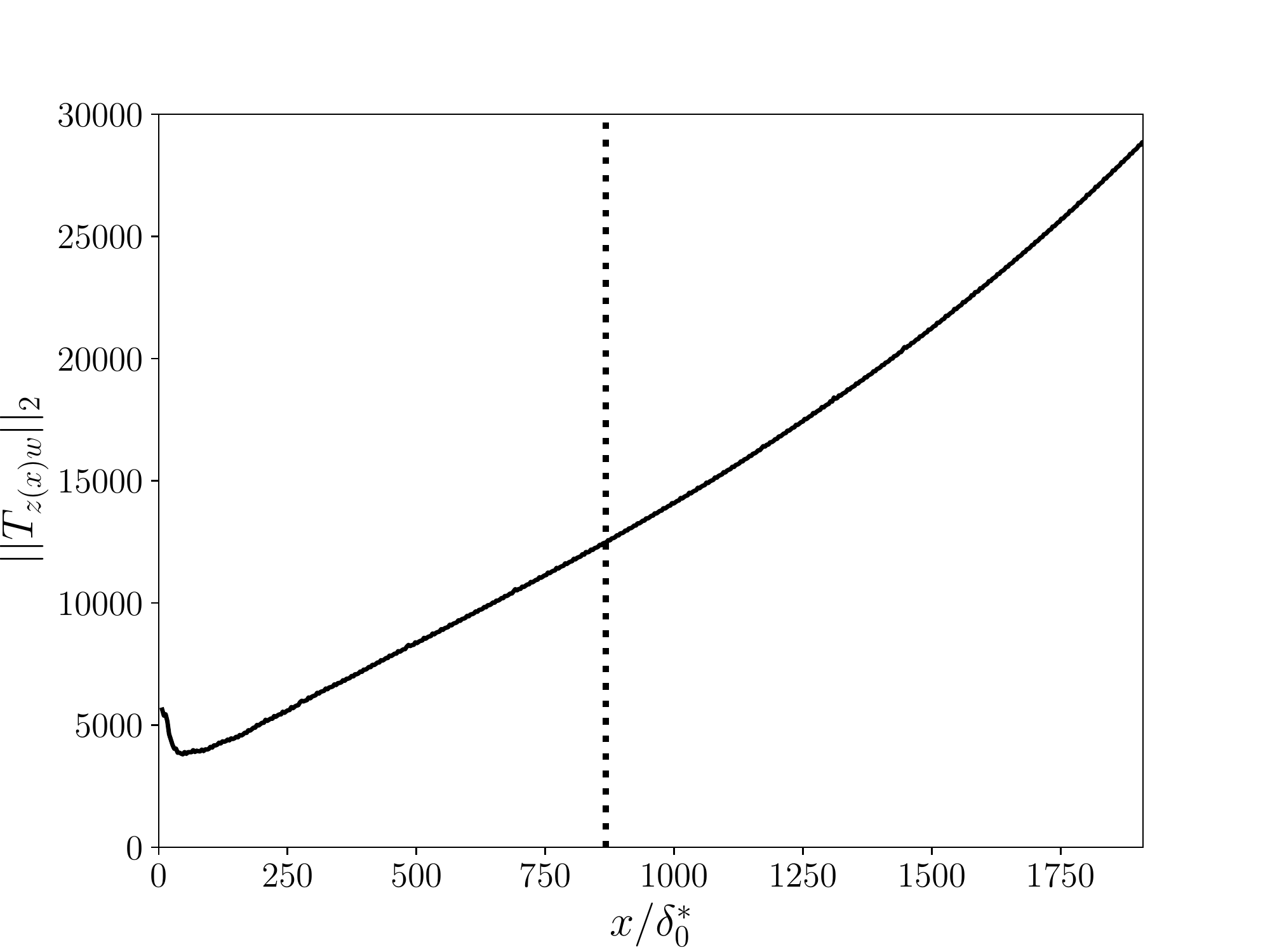}}%Tzwx_h2-eps-converted-to.pdf
      \subfloat[\label{fig:non_modal_TzuTzw}]{\includegraphics[scale=0.3]{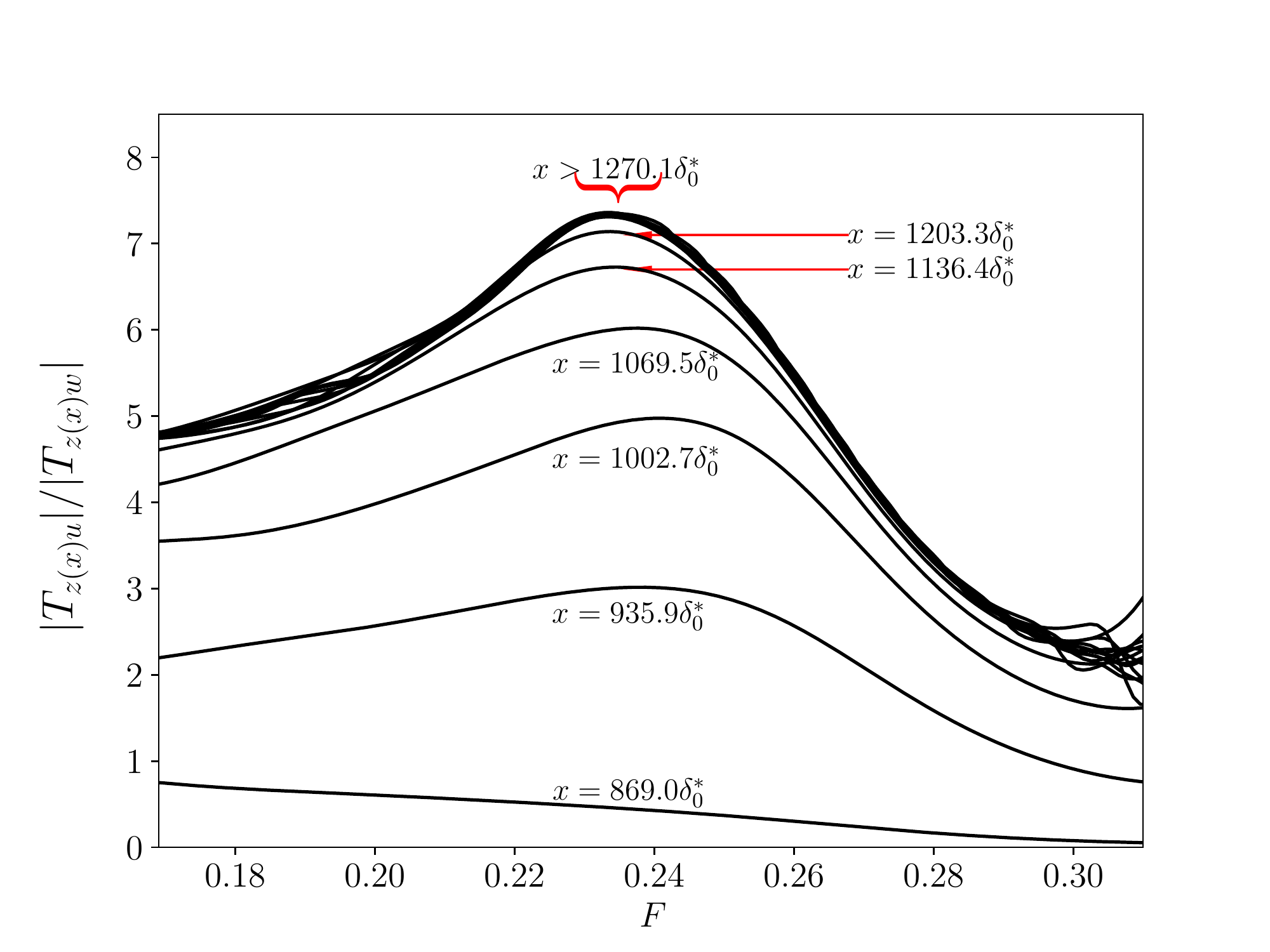}}
    \caption{a) Evolution of \(F|T_{z(x)w}|^{2}\). b) Variation of the frequency magnitude as a function of the plate abscissa. For each frequency, the green (resp. blue) dots represent branch I and branch II of the first (resp. second) Mack modes according to LLST. c) Evolution of the \(H_{2}\) norm. The vertical dashed line in (b) and (c) shows the streamwise location of the actuator \(\mathbf{B}_{u}\), denoted \(x_{u}\). d) Evolution of the ratio \(|T_{z(x)u}|/|T_{z(x)w}|\) as a function of frequency $ F $ for several plate abscissa.} 
\label{fig:Tzw_sensors_explain}
\end{figure}

In figure \ref{fig:Tzwx_evol}, we represent the quantity $F|T_{z(x)w}|^{2}$ as a function of $\ln{F}$, where $F$ is the frequency, such that the integral represents the $H_{2}$ norm of $T_{z(x)w}$. The module of $|T_{z(x)w}|$ is obtained by Fourier transform of the signals from an impulse response. At each abscissa $x$ of the plate, the energy contribution to the sensor \(z(x)\) is only due to a certain frequency bandwidth. Indeed, after reaching a peak, the magnitude associated with a frequency rapidly decreases, as can be seen in figure \ref{fig:Tzwx_FactorN}. Therefore, for control, we will need to use several performance sensors \(z_{i}\) to obtain a
suitable frequential representation at different streamwise positions and capture the entire amplified bandwidth. As the spectrum of $F|T_{z(x)w}|^{2}$ is narrow (especially downstream of the domain), reducing $||T_{z_{i}w}^ {c}||_{2}$ should also lead to a significant reduction in $||T_{z_{i}w}^ {c}||_{\infty}$.
Sufficiently far downstream from \(\mathbf{B}_{w}\), the most amplified frequency at each abscissa of the domain (red line in figure \ref{fig:Tzwx_FactorN}) is similar to the one that could be found with the \(N\)-factors (see figure \ref{fig:Nfactor}). As the magnitude of the perturbations increases for all frequencies in spatially stable regions upstream of branch I (see first dot symbols in figure \ref{fig:Tzwx_FactorN}), the perturbations seem to be subject first to a growth due to the non-modal Orr mechanism, before being dominated by the "modal" growth of the unstable Mack mode. The highest value of \(|T_{z(x)w}|\) is found at the end of the domain of interest, at a frequency of \(F=0.223\), close to the frequency leading to the highest gain in the global resolvent analysis (\(F=0.237\)). Therefore, the optimal response mechanisms already observed in \S \ref{sec:Result_local_global} are well triggered by the chosen disturbance \(\mathbf{B}_{w}\), which is therefore representative of a more general transition scenario due to the second Mack mode.

The control goal is to create a destructive interference by generating a second wave of appropriate amplitude and phase, which will oppose the one generated by the upstream noise $w(t)$ \citep{Herve_art,Sasaki3D_wavecancelling_art}. Thus, in order to maximize the impact of the control action, the perturbations generated by the actuator must match those induced by the upstream noise. The incoming disturbance being mainly due to second Mack mode instabilities, an efficient actuator can be obtained with a volume forcing around the generalised inflection point in the wall-normal direction, in order to maximize the receptivity process. We therefore consider $ \mathbf{B}_{u} u(t)$ in the right-hand-side of equations \eqref{eq:Navier_Stokes}(b) to model the actuator, with the same divergence free spatial support as for the disturbance \(\mathbf{B}_{w}\):  
\begin{equation}
    \mathbf{B}_{u}=\mathbf{h}(\frac{10.66}{\delta_{0}^{*^{2}}},867.2\delta_{0}^{*}, 7.79\delta_{0}^{*}, 1.5\delta_{0}^{*}, 0.5\delta_{0}^{*}).
\end{equation}
The actuator is placed sufficiently far downstream of $\mathbf{B}_w$ ($x_u=867.2\delta_{0}^{*}$) for two reasons. The first one is to allow disturbances to strengthen sufficiently (see figure \ref{fig:Tzwx_h2}) to be easily detected by the estimation sensor $y$ (which is close to the actuator), which in an experimental configuration would mean placing the actuator a little upstream of the beginning of the transition process. The second reason is to limit the bandwidth of the frequencies to be controlled (see figure \ref{fig:Tzwx_FactorN}) in order to keep the complexity of the control problem reasonable. Hence, for the chosen streamwise position of the actuator, the frequency range to be controlled is around \(F \in [0.225,0.324]\); a more upstream actuator should have controlled a wider bandwidth. The streamwise position of the actuator remains sufficiently upstream so that incoming perturbations are controlled over a sufficiently long domain (\(\sim 0.34\ \mathrm{m}\)) representative of an experimental configuration (the plate of the experimental tests of \citet{Kendall_art} measured $0.35\ \mathrm{m}$). 

A comparison of \(|T_{z(x)w}|\) and \(|T_{z(x)u}|\) is shown in figure \ref{fig:non_modal_TzuTzw}. It can be noted that in the vicinity of the actuator, the ratio \(|T_{z(x)w}|/|T_{z(x)u}|\) evolves with the \(x\) abscissa. As this phenomenon no longer appears for abscissas further away from the actuator and the ratio becomes constant, it could be attributed to a non-modal transient behaviour. Indeed, we have:
\begin{equation}
    \frac{|T_{z(x)u}|}{|T_{z(x)w}|}	\propto \frac{e^{\int^{x}_{x_{u}} -(\widetilde{\alpha}_{i})_{u} dx}}{e^{\int^{x}_{x_{u}} -(\widetilde{\alpha}_{i})_{w} dx}},
\end{equation}
where  \(-(\widetilde{\alpha_{i}})_{u}\) and \(-(\widetilde{\alpha_{i}})_{w}\) represent the slope of $\ln{|T_{z(x)u}|}$ and $\ln{|T_{z(x)w}|}$, respectively. Therefore, a constant ratio implies having the same slope  from a certain distance \(x\). This distance \(x\) represents the non-modal distance due to the receptivity of multiple modes to the volume forcing of the actuator on the flow. 

The impact of the position of the estimation sensor \(y\) has been already extensively studied in the noise-amplifier flow control literature \citep{Barbagallo_art,Belson_art,Juillet_art,Freire_art}, hence the detailed analysis for the case of the supersonic boundary layer is left to appendix \ref{sec:estimation_sensor_perfo}. It is just reminded that, for a feedback design, the estimation sensor \(y\) has to be close enough to the actuator to avoid sending outdated information and limit the effective delay impacting the maximum achievable performance. For a feedforward design where the impact of the actuator on the estimation sensor \(y\) is assumed to be negligible in the synthesis step (\(T_{yu}=0\)), the estimation sensor  has to be located sufficiently upstream of the actuator for the hypothesis to be valid. 

Regarding the number of performance sensors \(z_{i}\) used in the identification/synthesis step, it was found by numerical simulations that six probes are required to achieve nearly uniform performance along the domain because of the need to capture the entire amplified bandwidth and the non-modal effects due to the actuator (see appendix \ref{sec:number_perfo_sensor}). The streamwise positions of the input perturbation, the actuator and the sensors used for the identification and synthesis steps are summarized in table \ref{tab:reca_pos_all}.

\begin{table}
    \centering
 \begin{tabular}{lcccc}
      
      &
      \(\mathbf{B}_{w}\)
      &
       \(\mathbf{B}_{u}\)
       &
       \(y\)
       &
       \(z_{i}\)\\
      %\hline
      \midrule
      %\cline{1-5}
      Streamwise position
      &
      \(x_{w}=4.1\delta_{0}^{*}\)
      &
      \(x_{u}=867.2\delta_{0}^{*}\)
      &
      \makecell{\(x_{\textit{ff}}=801.2\delta_{0}^{*}\) \\ \(x_{\textit{fb}}=885.7\delta_{0}^{*}\)}
      
      &\makecell{\(x_{1}=933.2\delta_{0}^{*}\) \\ \(x_{2}=1029.4\delta_{0}^{*}\) \\ \(x_{3}=1125.6\delta_{0}^{*}\) \\ \(x_{4}=1317.9\delta_{0}^{*}\) \\ \(x_{5}=1510.2\delta_{0}^{*}\) \\ \(x_{6}=1766.7\delta_{0}^{*}\)}\\
 \end{tabular}
    \caption{Streamwise positions of the input perturbation, the actuator and the sensors used for the identification and synthesis steps. The position of the estimation sensor for feedforward and feedback configurations are denoted \(x_{\textit{ff}}\) and \(x_{\textit{fb}}\), respectively.}
    \label{tab:reca_pos_all}
\end{table}

\section{Identification and synthesis methods}
\label{sec:indentif_synthesis}
\subsection{Identification of a state-space model}
\label{sec:Identif_noise_amplifier}
Most synthesis methods require the use of state-space ROMs corresponding to the transfers involved in the controller synthesis.
%Four quantities are required for the control design \ref{sec:specification_h2_hinf}:  \(T_{z_{i}u}\), \(T_{yu}\), \(T_{z_{i}y}\) and \(W_{y}\).
For the model reduction step, some of the input/output delays linked to the convective nature of the flow may be discarded due to the fact that the \(H_{2}\) norm is not modified by dead-time delays. In a feedback configuration $ (u,y,z_i) $, the dead-time delays verify $ \tau_{z_iu}=\tau_{z_i\tilde{y}}+\tau_{yu}$, so that 
\begin{equation}
\label{eq:fb_h2_delay}
\begin{split}
||T_{z_{i}w}^ {c}||_{2}&=||\ e^{-\tau_{z_{i}\tilde{y}} s} W_{y}(T'_{z_{i}\tilde{y}}+e^{-\tau_{yu} s}T'_{z_{i}u}KS)\ ||_{2}\\
   &=||\ W_{y}(T'_{z_{i}\tilde{y}}+e^{-\tau_{yu} s}T'_{z_{i}u}KS)\ ||_{2},
\end{split}
\end{equation}
where \(T'(s)\) designates the ``dead-time-free" transfer function associated to $ T(s)$.
The same idea can be applied also to a feedforward design $ (y,u,z_i)$ with the result below:
\begin{equation}
\label{eq:ff_h2_delay}
\begin{split}
  ||T_{z_{i}w}^ {c}||_{2}=||\ W_{y}(e^{-\tau_{uy}s}T'_{z_{i}\tilde{y}}+T'_{z_{i}u}K)\ ||_{2}.
\end{split}
\end{equation}
%In feedback (resp. feedforward) configuration, the residual delay \(\tau_{yu}\) (respectively \(\tau_{uy}\)) represents the time taken by the wavepacket to travel from the actuator \(u\) to the estimation sensor \(y\) (respectively from the estimation sensor \(y\) to the actuator \(u\)).
Thus, the only remaining delay is the one between the actuator and the estimation sensor, $\tau_{yu}$ or $ \tau_{uy}$, which is reasonably small (compared to the delays involving $z_i$.)
%Consequently, we aim at obtaining reduced-order models for the free dead-time transfer-functions $ T'$.
Removing unnecessary delays (for example $ \tau_{z_i\tilde{y}} $ in the feedback case) leads to a significant reduction in the size of the ROMs when the dead time scale is important compared to the time scale of the physical phenomenon to be captured (the period of the second Mack mode). This reduction in the order of the ROMs is beneficial both for the identification and the synthesis step:  the higher the order, the more difficult the identification %Moreover, syntheses which solve Riccati equations leading to controller of the same order as the plant, having a low-order model will avoid the need for a post-processing step to reduce the number of states of the controller to make it usable.
and the larger the cost of the controller synthesis.

\begin{figure}
    \centering
     \subfloat[\label{fig:ERA_1}]{\includegraphics[scale=0.42]{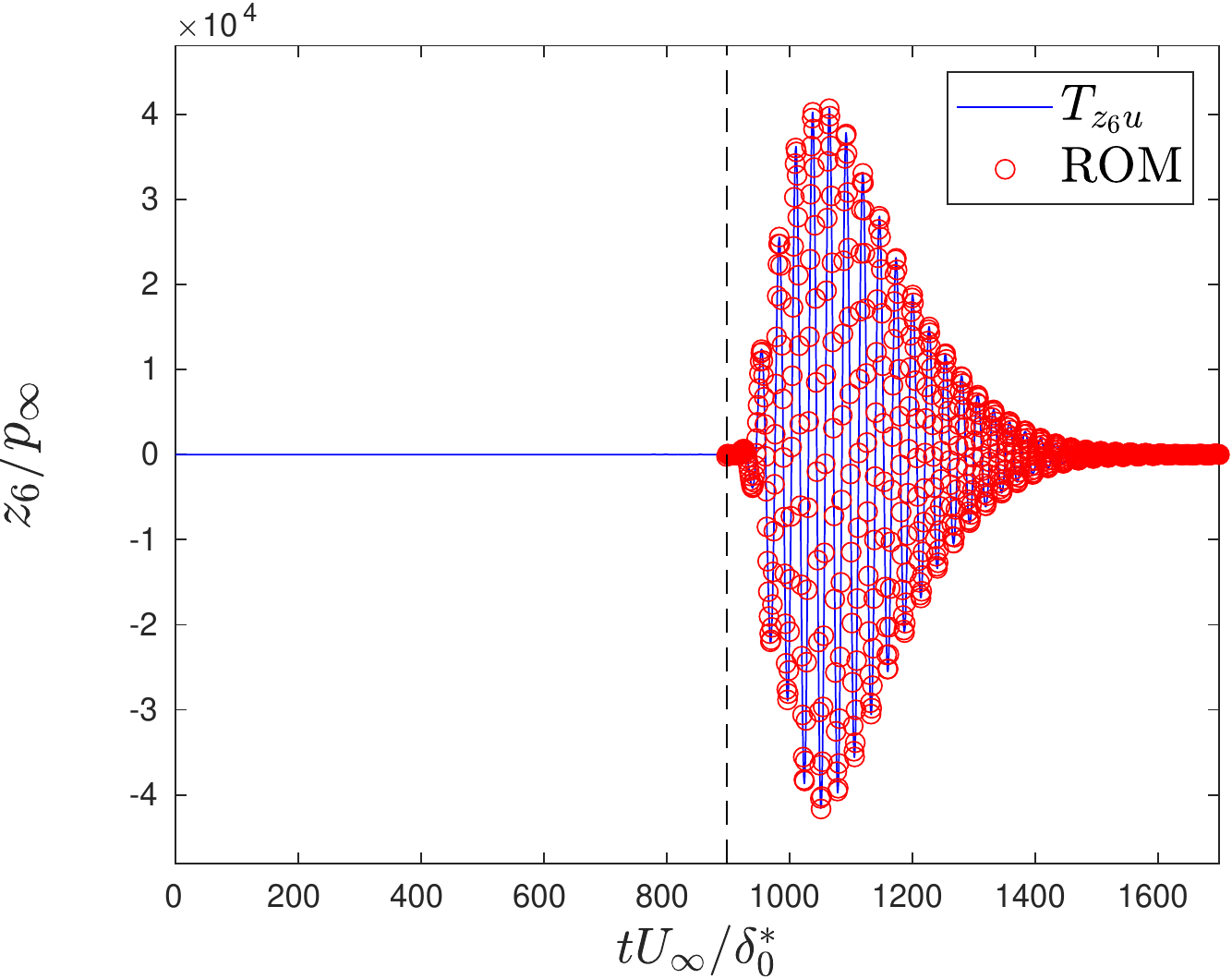}}
     \hspace{2mm}
      \subfloat[\label{fig:ERA_2}]{\includegraphics[scale=0.42]{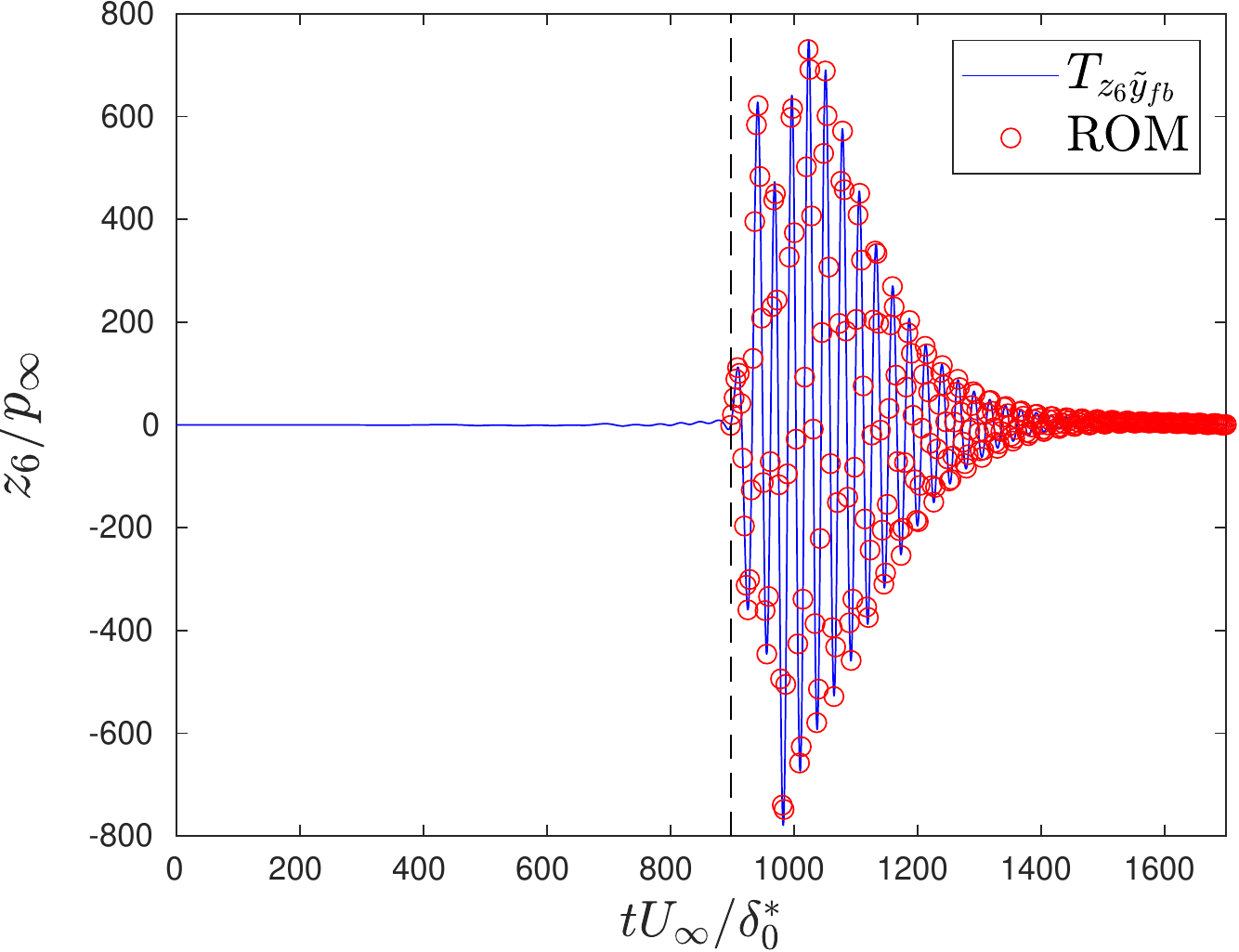}} \\
      \subfloat[\label{fig:ERA_3}]{\includegraphics[scale=0.42]{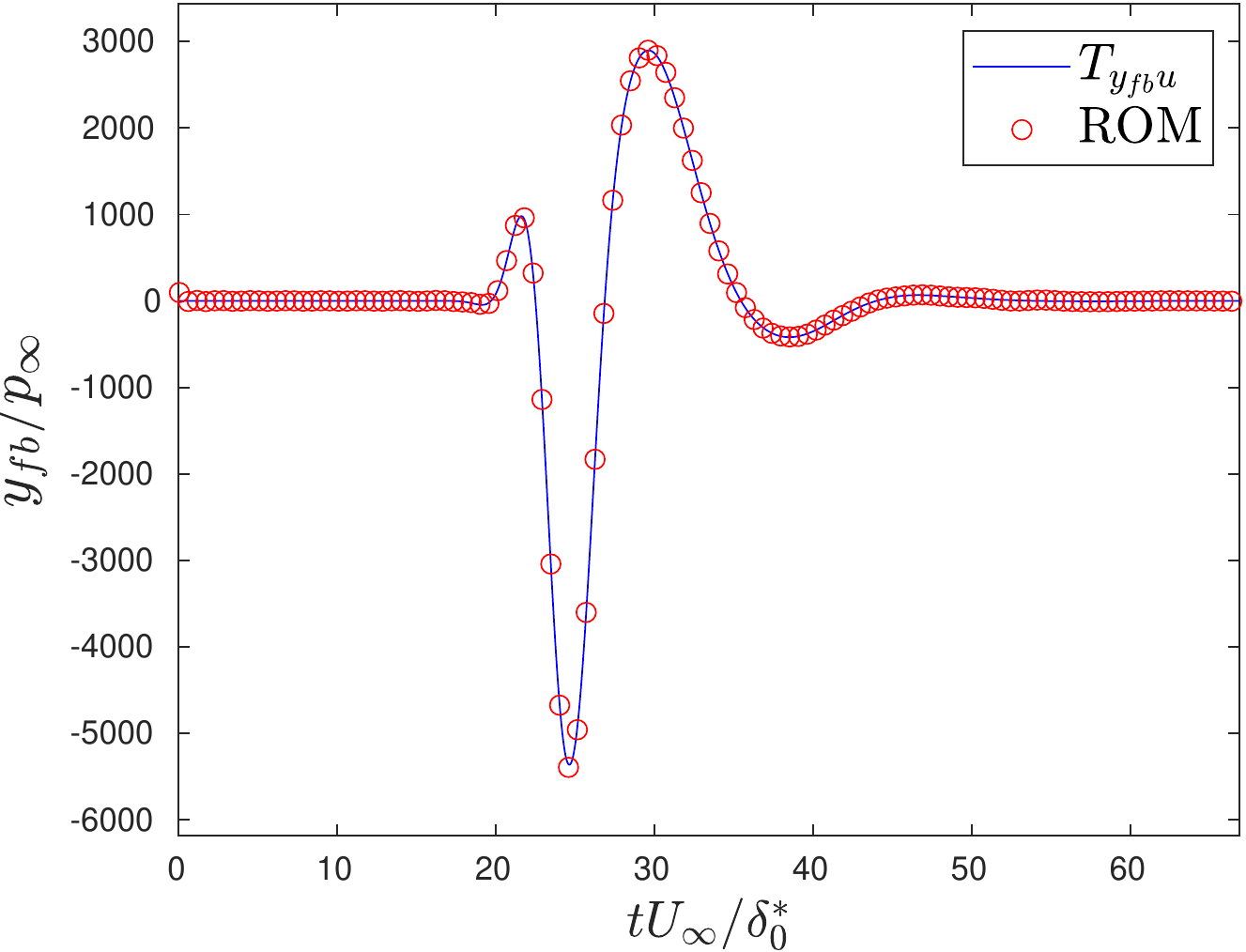}}
      \hspace{2mm}
      \subfloat[\label{fig:tfest_1}]{\includegraphics[scale=0.42]{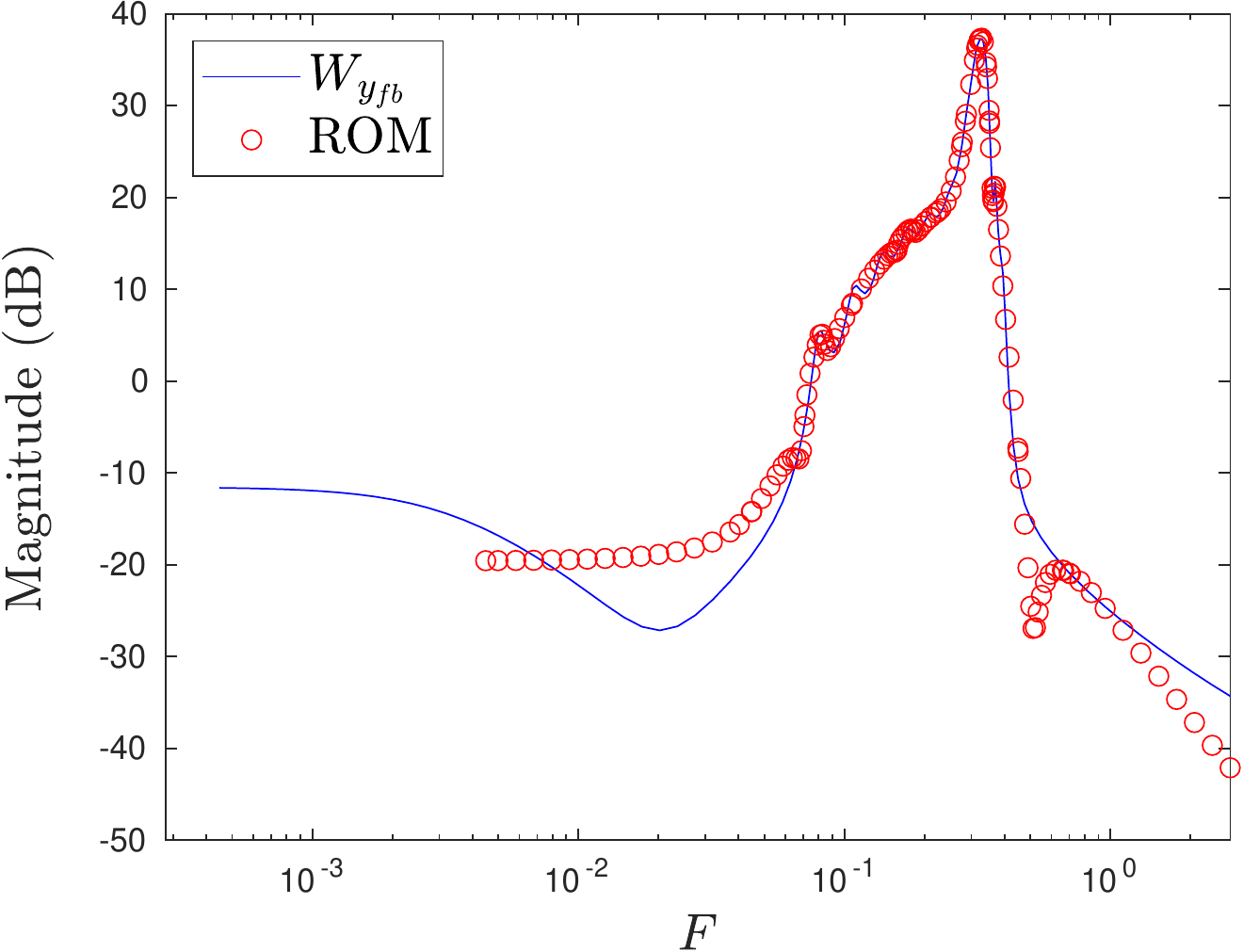}}
    \caption{(a,b,c) Comparison between impulse responses (blue lines) and the ROMs (red circles) for the performance sensor \(z_{6}\) and for the feedback estimation sensor \(y_{\textit{fb}}\). Note that for the ROMs of \(T_{z_{6}u}\) and \(T_{z_{6}\tilde{y}_{\textit{fb}}}\), the time axis of the impulse responses are shifted by \(\frac{t_{shift}U_{\infty}}{\delta_{0}^{*}}\simeq 897\) (black dashed lines) which corresponds to the suppression of unnecessary dead times. d) Comparison between the quantity \(W_{y_{\textit{fb}}}\) from the linear simulation (blue line) and the ROM (red circles).}
    \label{fig:Identification_ROM}
\end{figure}

The quantities required for the synthesis are obtained by impulse responses of $w$ and $u$. %Although these pulses are not experimentally feasible and we have normally to use white noise inputs to obtain these quantities in a realistic setup, the pulses allow in our purely numerical study to construct all the quantities needed for the synthesis with only two simulations of a maximum duration equivalent to one convective time; this considerably reduces the calculation time compared to using white noise inputs.
The state-space ROMs associated to the transfer functions \(T_{z_{i}u}\), \(T_{yu}\) and \(T_{z_{i}\tilde{y}}\) are obtained by the subspace identification method ERA, which requires impulse responses for each of the inputs and involves performing a singular value decomposition to compress the state \citep{juang_pappa_art}. This method has been used several times for the control of 2D \citep{Belson_art} or 3D \citep{Morra_art,Sasaki3D_wavecancelling_art} incompressible boundary layers. The ERA algorithm is applied after removing (just by shifting the time axis)
either $ \tau_{z_i\tilde{y}}$ (in the feedback case) or $ \tau_{z_iu}$ (in the feedforward case) within the impulses from $y$ and $u$ to $ z_i $.
 %Regarding the residual delay linked to the actuator/measurement sensor distance, it is directly identified with ERA.
 The impulse responses from $y$ to $ z_i$ are obtained by inverse Fourier transform of $ T_{z_iw}T_{yw}^{-1}$, each individual transfer function being obtained by Fourier transform of an impulse from $w$. The sampling time for ERA is \(5\times dt\); the discrete time models obtained are then converted to continuous time models by first-order hold method \citep{book_foh}. As shown in figures \ref{fig:Identification_ROM}(\subref*{fig:ERA_1},\subref*{fig:ERA_2},\subref*{fig:ERA_3}) for the performance sensor \(z_{6}\) and for the feedback estimation sensor \(y_{\textit{fb}}\), the constructed ROMs capture most of the dynamics. 

The identification of the quantity \(W_{y}\) is obtained by a vector-fitting method (Matlab function \textit{tfest}) designed to  fit frequency response measurements \citep{tfest}. For this quantity, there is no uniqueness of the identified model as the phase can vary from one model to another without impacting the results of the synthesis (see (\ref{eq:H2_norm_2})); the ROM just needs to be stable and causal. Hence, we simply choose to define $W_y$ as $W_y=\sqrt{PSD_{y}(\omega)}$ where $y$ is the response from an impulse in $w$. A good agreement is achieved between \(W_{y_{\textit{fb}}}\) and the ROM in the case of the feedback estimation sensor \(y_{\textit{fb}}\) (see figure \ref{fig:tfest_1}).     

For the current application and with the six performance sensors \(z_{i}\), the sum of the orders of each ROM is \(130\) for the case of the feedback configuration and \(115\) for the feedforward one. By comparison, identifying the single transfer function \(T_{z_{6}u}\) (corresponding to the farthest performance sensor downstream) without suppressing the dead time leads to a ROM of order \(220\), which is already greater than the sum of the orders of each ROM without their unnecessary dead-times.%; this highlights the importance of removing unnecessary dynamics.

In the control result section \S \ref{sec:Result_ff_vs_fb}, because the models are of excellent quality (see figure \ref{fig:Identification_ROM}), the distinction between ROMs and real transfer functions is not deemed necessary and the depicted results are those on the complete system after implementation of the controllers in \textit{elsA}. 

\subsection{Multi-objective structured \(H_{2}\)/\(H_{\infty}\) synthesis}
\label{sec:synthesis_description_algo}

In this study, control laws are designed following a structured mixed \(H_{2}\)/\(H_{\infty}\) synthesis implemented in the Matlab function \textit{systune} \citep{apkarian_systune_art}.
\begin{figure}
\centering
\includegraphics{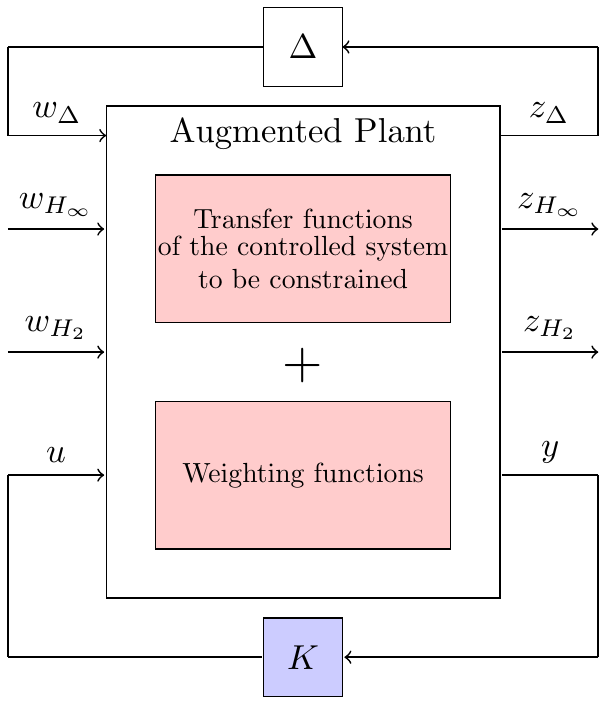}
\caption{Multiple requirements \(H_{2}\)/\(H_{\infty}\) synthesis.}
\label{fig:H2_Hinf_Picture}
\end{figure}
The general framework of this modern synthesis is illustrated in figure \ref{fig:H2_Hinf_Picture}. In this figure, \(w_{H_{\infty}}\) (resp. \(w_{H_{2}}\)) and \(z_{H_{\infty}}\) (resp. \(z_{H_{2}}\)) represent the set of inputs and outputs whose associated transfers are subject to \(H_{\infty}\) (resp. \(H_{2}\)) norms. This synthesis thus allows to minimize different \(H_{2}/H_{\infty}\) norms under closed-loop stability constraints despite model uncertainties $\Delta$. The structure of the controller \(K\) is defined by the user independently from the order of the state-space model to be controlled, which makes it a particularly powerful and flexible synthesis method. The set of transfer functions subject to an \(H_{2}/H_{\infty}\) norm minimization or constraints constitutes the \textit{augmented plant}; these transfer functions are composed of the transfers of the controlled system allowing to respect the specifications, along with weighting functions \citep{Skogestad_art}. Weighting functions act as frequency domain constraints in order to shape adequately the transfer functions to achieve specific design goals. Furthermore, weighting functions allow to normalize the different requirements to be able to balance them during the constrained minimization problem.

In our specific study, the structure of the controller \(K\) is imposed beforehand in the following way: (1) the controller \(K\) is searched in a state-space representation form; (2) the controller \(K\) must be stable; (3) we limit the controller order to \(5\) as high-order controllers are less easily implemented in practice \citep{Reduction_perfo_Goddard}; (4) we impose a tridiagonal state matrix which has significantly fewer parameters to determine than the full matrix given that any real square matrix is similar to a real tridiagonal form \citep{tridiagonal}; (5) we impose a strictly proper controller involving a natural roll-off of the high frequencies of \(-20\ \mathrm{dB}\) per decade in order to neglect dynamics in high frequencies and to be robust to high frequency noise on the estimation sensor \(y\) naturally present in every experimental setup. For the controller structure imposed above, the algorithm then solves the following constrained minimization problem:
\begin{equation}
\label{eq:min_problem}
\begin{split}
     \text{minimize} &\underset{i=1,...,6}{\mathrm{max}}(||T_{z_{i}w}^{c}||_{2})\\
     &\text{subject to}\ ||W_{S}S||_{\infty}<1 \text{ and } ||W_{KS}KS||_{\infty}<1.
\end{split}
\end{equation}
This constrained minimization problem is the transcription of the fluidic specifications established throughout \S \ref{sec:amplifier_behaviour}.

Firstly, the minimization of \(H_{2}\) norms of \(T_{z_{i}w}^{c}\) directly allows the reduction of the expected power for the six performance sensors \(z_{i}\) used in the synthesis when they are excited by white-noise perturbations \(w\) and sensed by the estimation sensor $ y $. A multi-objective synthesis approach is necessary for our problem by minimizing the expected power of sensors at different abscissa of the flat plate instead of minimizing an overall energy. Indeed, the disturbance energy growing as it is convected downstream, an overall energy would then essentially account for the fluctuating energy downstream of the domain, leaving aside the structures further upstream in the case of a very large computational domain.  Transition to turbulence appearing locally above a certain perturbation energy threshold (see \S \ref{sec:Result_local_global}), we advocate the need for minimizing the largest \(H_{2}\) norm of the controlled system over the set of performance sensors $ z_i$ used to assess the local character of transition to turbulence. 

Secondly, the \(H_{\infty}\) constraint on \(W_{S}S\) maintains adequate stability margins. To prevent the closed loop from being unstable in a feedback design, a frequent choice is to ensure that \(||S||_{\infty}<2\) \citep{Skogestad_art,Belson_art}. Thus, the weighting function \(W_{S}\) has a constant frequency template such as \(W_{S}(s)=0.5\) because the \(H_{\infty}\) constraint on \(W_{S}S\) is equivalent to \(|S|<1/|W_{S}|\) \(\forall\ \omega \in \mathbb{R}\). This means that the system will be guaranteed stable up to 50\% of relative model errors $\Delta$ on \(T_{yu}\) (see \S \ref{sec:specification_h2_hinf}). In the case of a feedforward design, \(S(s)=1\) (because \(T_{yu}=0\)) and this \(H_{\infty}\) constraint is always respected, which explains the unconditional stability of the feedforward configuration.

Finally, the  \(H_{\infty}\) constraint on \(W_{KS}KS\) is here to desensitize the controller to new noise sources on a certain bandwidth. Our controller being already robust to high frequency uncertainties due to the strictly proper structure imposed, \(W_{KS}\) is just designed to limit low frequency actuator activity in case, for example, of low frequency noise on the estimation sensor \(y\).

By minimizing the maximum value between several transfer functions and using \(H_{\infty}\) norm constraints, a non-smooth optimization is performed; as non-smooth optimization is computationally intensive (compared to LQG), it is all the more important to obtain ROMs with the least possible states (see \S \ref{sec:Identif_noise_amplifier}), giving in our case computations of several tens of minutes. %because a hundred random controller initializations are required to deal with the non-convex nature of the optimization problem, giving in our case computations of several tens of minutes.

\section{Feedforward versus Feedback control}
\label{sec:Result_ff_vs_fb}

\subsection{Performance on the nominal case}
\label{sec:performance_analysis}
The results of both feedforward (denoted 'Ff') and feedback (denoted 'Fb') controllers resulting from the constraint minimization problem (\ref{eq:min_problem}) are evaluated by implementing the controllers in the DNS solver \textit{elsA}. 

\begin{figure}
    \centering
     \subfloat[\label{fig:S_fb}]{\includegraphics[scale=0.45]{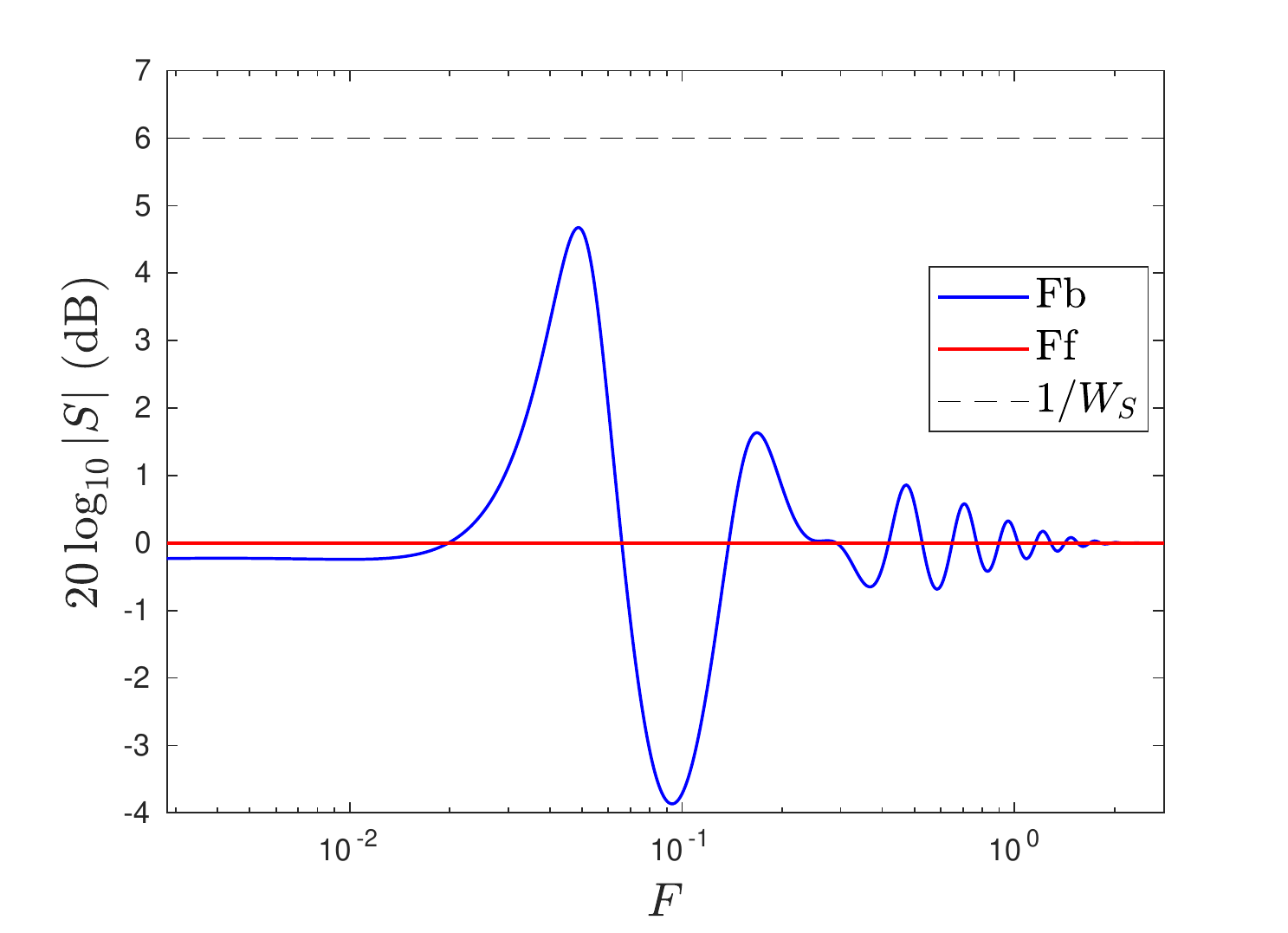}}
      \subfloat[\label{fig:KS_fb_ff}]{\includegraphics[scale=0.45]{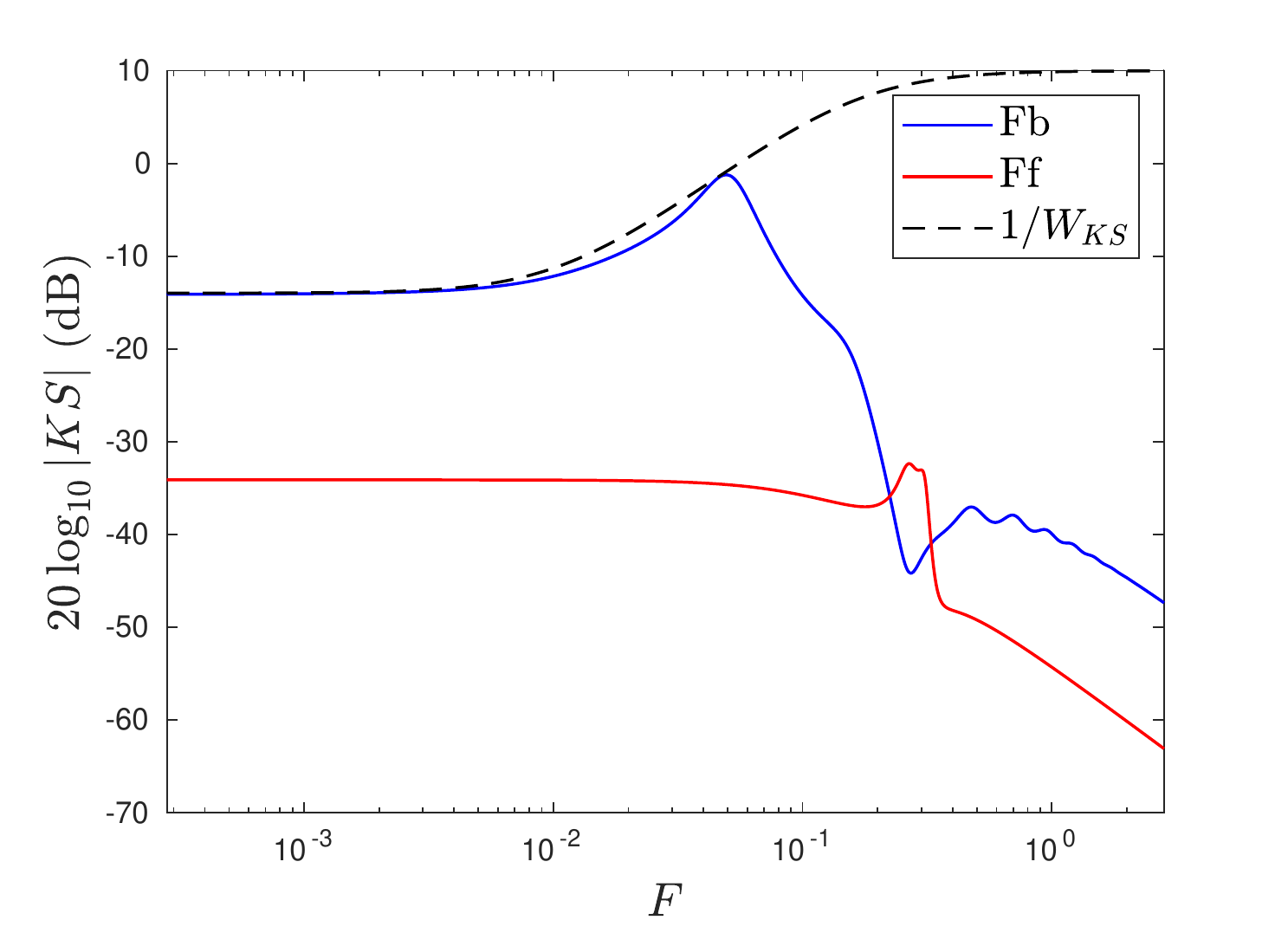}} \\
      \subfloat[\label{fig:h2_final}]{\includegraphics[scale=0.33]{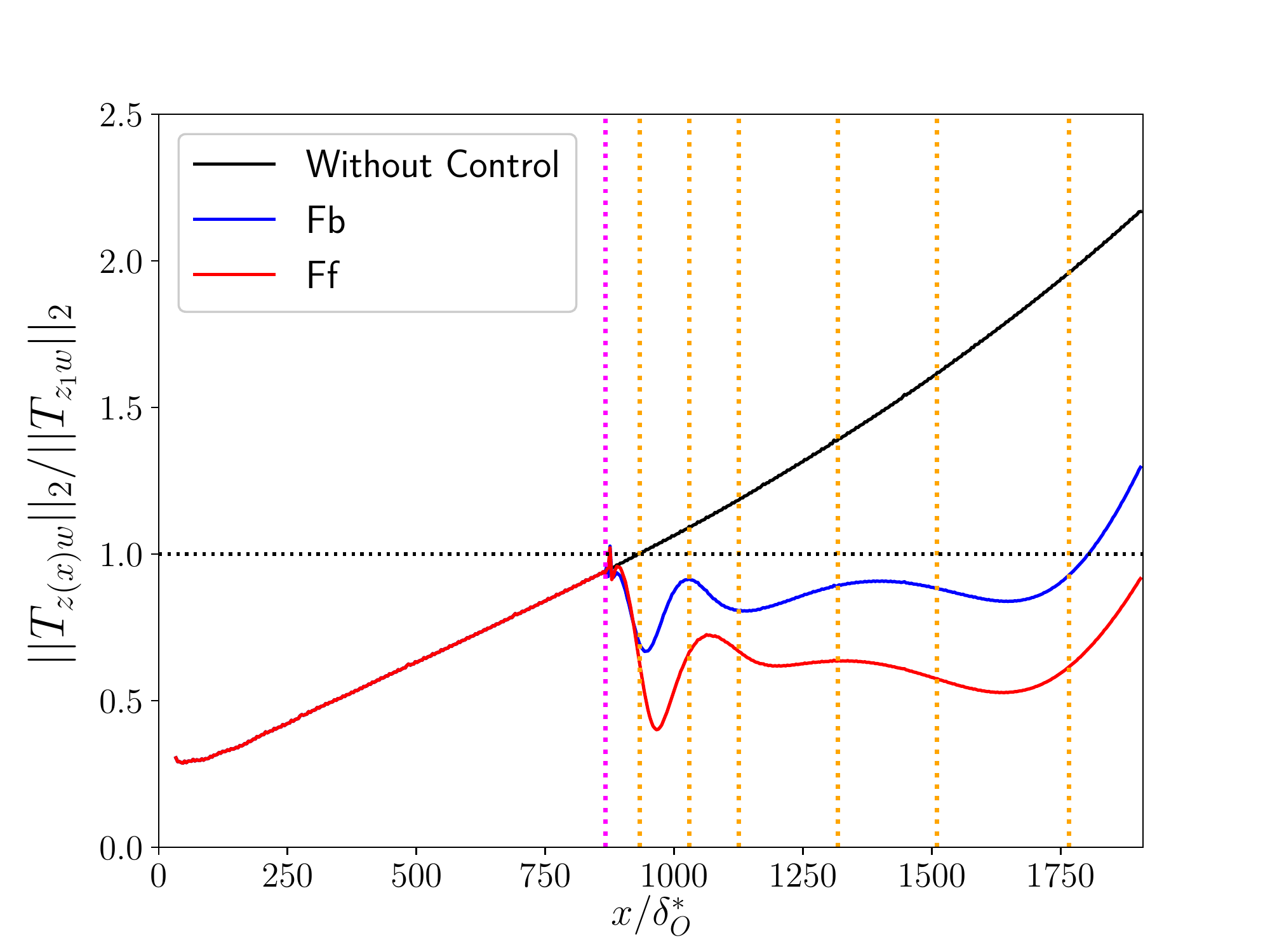}}
      %\subfloat[\label{fig:pressure_fb_convective}]{\includegraphics[scale=0.33]{IMAGE/convective_pressure_fb-tiff-converted-to.png}}
    \caption{Blue, red and dashed lines represent the feedback case, the feedforward case and the constraints (weighting functions) imposed for the control design, respectively. a) Magnitude of \(S\). b) Magnitude of \(KS\). c) Evolution of the local \(H_{2}\) norm of the transfer \(T_{z(x)w}\) as obtained from the DNS simulation (those obtained with the ROMs are actually identical at $ x=x_i$ since the ROMs are very accurate). The vertical magenta and orange dotted lines represent, respectively, the position of the actuator (with the sensors \(y_{\textit{fb}}\) and \(y_{\textit{ff}}\) nearby) and the six performance sensors \(z_{i}\) used for synthesis. The values are normalized by \(||T_{z_{1}w}||_{2}\). The horizontal black dotted line depicts the energy threshold \(||T_{z_{1}w}||_{2}\) respected until \(x_{6}\) following the minimization of the cost functional \(\underset{i=1,...,6}{\mathrm{max}}(||T_{z_{i}w}^{c}||_{2})\).}
    \label{fig:Performance_final_ff_fb}
\end{figure}
Figure \ref{fig:S_fb} shows the sensitivity function \(S\) for the feedback design which respects the \(H_{\infty}\) constraint  on  the  sensitivity  function (\textit{i.e.} \(|S|<1/|W_{S}|=6\ \mathrm{dB}\))  imposed  in  the  minimization problem (\ref{eq:min_problem}) (represented by the black dashed line). As previously explained, for the feedforward design, \(|S|=1\) (red line) and the constraint is automatically satisfied. Figure \ref{fig:KS_fb_ff} represents \(|KS|\) for both the feedforward and feedback cases. The weighting function \(W_{KS}\), which allows to limit actuator activity in case of low frequency disturbances, is also shown and we verify that \(|KS|<1/|W_{KS}|\) \(\forall\ \omega \in \mathbb{R}\). For the feedback design, \(|KS|\) is close to \(1/|W_{KS}|\) at low frequencies, meaning that there is a trade-off between minimizing \(H_{2}\) norms and desensitizing the controller in the low-frequency range. We notice the natural roll-off of the controllers of \(-20\ \mathrm{dB}\) per decade at high frequencies related to the strictly proper structure imposed in the synthesis.

The control action results in a significant reduction in the local \(H_{2}\) norm of the transfers \(T_{z(x)w}\) at each abscissa of the plate (see figure \ref{fig:h2_final}) for both the feedforward and feedback configurations. As expected from the literature \citep{Belson_art,Juillet_art,Semeraro_fb_art,Tol_fbff_art}, the feedforward design minimizes even more the local \(H_{2}\) norm than the feedback one. Nevertheless, for both configurations, the minimization of the cost functional \(\underset{i=1,...,6}{\mathrm{max}}(||T_{z_{i}w}^{c}||_{2})\) allowed the local $H_2$ norm of \(T_{z(x)w}\) not to exceed, before \(x=x_{6}\), a threshold given by the $ H_2$ norm at $x=x_1$. Thus both configurations successfully achieve the control strategy set forth in figure \ref{fig:LLST_allfig}. The use of an \(H_{2}\) performance criterion alongside the \(H_{\infty}\) criterion on stability margin allows to address both performance in terms of disturbance rejection and stability robustness in the design of the feedback loop. %  during the synthesis of a feedback controller.  enabled to have an explicit performance objective of disturbance rejection, which was sometimes missing during previous studies with a feedback design that focused mainly on stability robustness. their synthesis method not being multi-criteria \citep{Belson_art,Vemuri_fb}. 

In addition to the reduction of the local \(H_{2}\) norm along the plate, the local \(H_{\infty}\) norm \(||T_{z(x)w}||_{\infty}\) has also decreased for both the feedforward and feedback designs; this variation is directly  related to the \(N\)-factor envelope \(\widetilde{N}\) by:
\begin{equation}
\underset{x_{1}<x<x_{6}}{max}\ln{||T_{z(x)w}||_{\infty}}-\underset{x_{1}<x<x_{6}}{max}\ln{||T^{c}_{z(x)w}||_{\infty}}=\underset{x_{1}<x<x_{6}}{max}\widetilde{N}_{x}-\underset{x_{1}<x<x_{6}}{max}\widetilde{N}^{c}_{x}.
\end{equation} 
More precisely, feedforward and feedback designs respectively ``save" \(1.13\) and \(0.89\) points of \(N\)-factor. One might ask which is the most effective setup for delaying transition, between minimizing \(\underset{i}{\mathrm{max}}(||T_{z_{i}w}^{c}||_{2})\) or minimizing \(\underset{i}{\mathrm{max}}(||T_{z_{i}w}^{c}||_{\infty})\), but answering the question is beyond the scope of this study.  
\begin{figure}
    \centering
     \subfloat[\label{fig:Hinf_Tzw_fb}]{\includegraphics[scale=0.35]{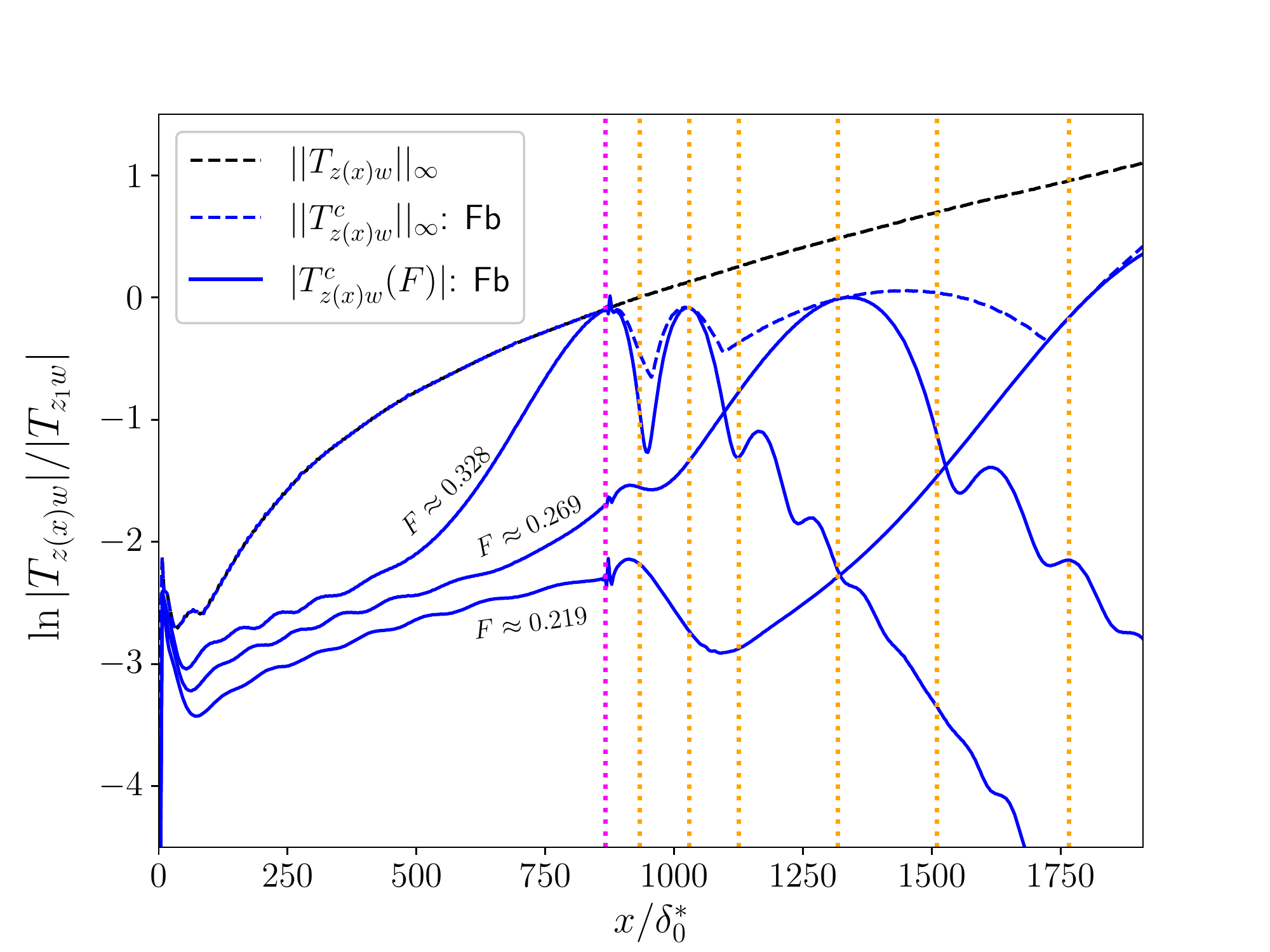}}
      \subfloat[\label{fig:Hinf_Tzw_ff}]{\includegraphics[scale=0.35]{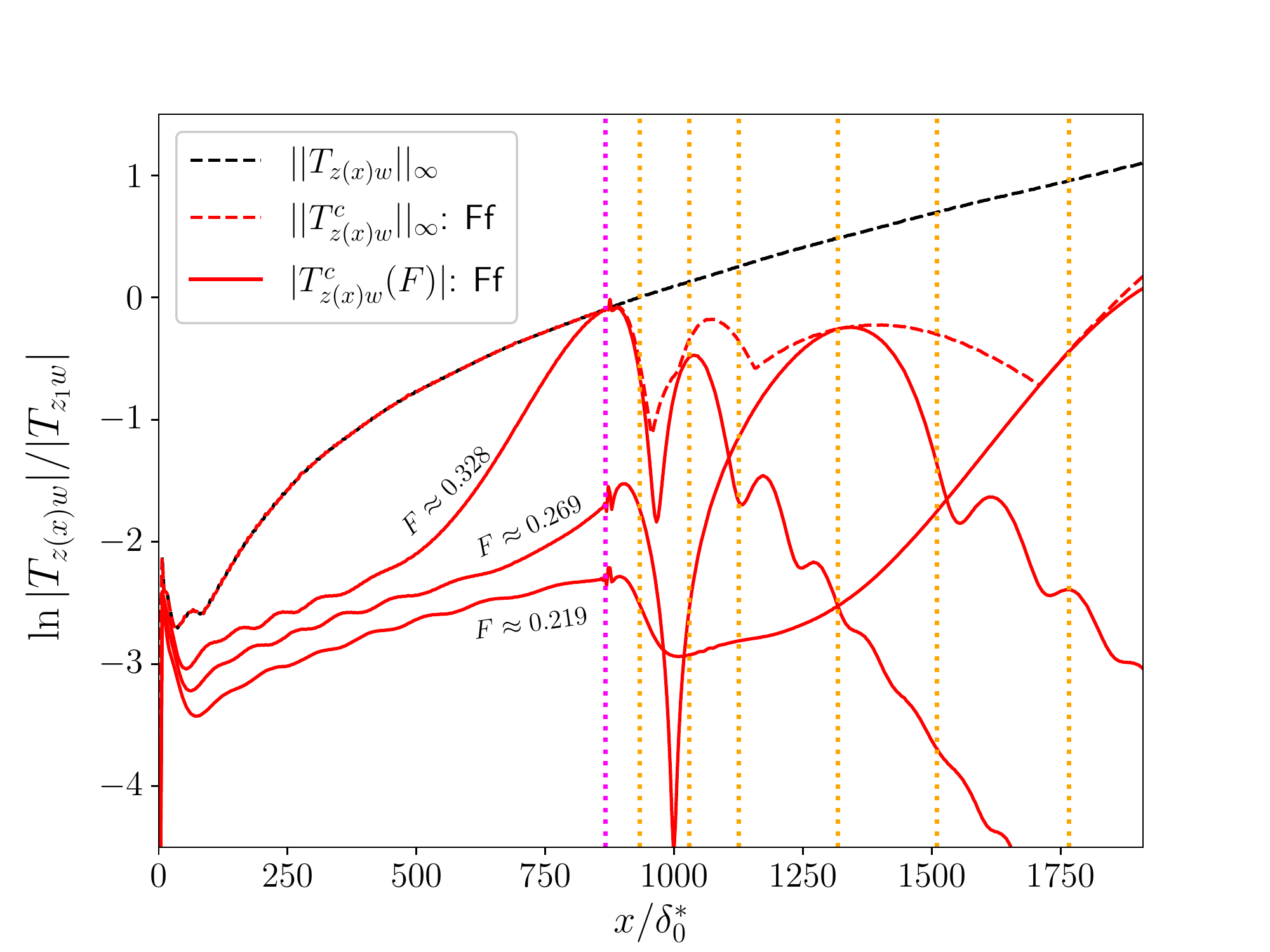}} 
    \caption{Evolution of the local \(H_{\infty}\) norm of the transfer \(T_{z(x)w}\) as a function of the plate abscissa for the uncontrolled (black lines), feedback (a) and feedforward (b) cases. The evolution of $|T_{z(x)w}^{c}(F)|$ for some frequencies is also shown for the controlled cases. For vertical lines, it is the same legend as in figure \ref{fig:h2_final}.}
    \label{fig:Hinf_fb_ff_Tzw}
\end{figure}

\begin{figure}
    \centering
      %\subfloat[\label{fig:profile_urms}]{\includegraphics[scale=0.2]{IMAGE/profile_urms-eps-converted-to.pdf}}
     % \subfloat[\label{fig:v_max}]{\includegraphics[scale=0.33]{IMAGE/Max_urms-eps-converted-to.pdf}}
       \subfloat[\label{fig:Trms_NC}]{\includegraphics[scale=0.22]{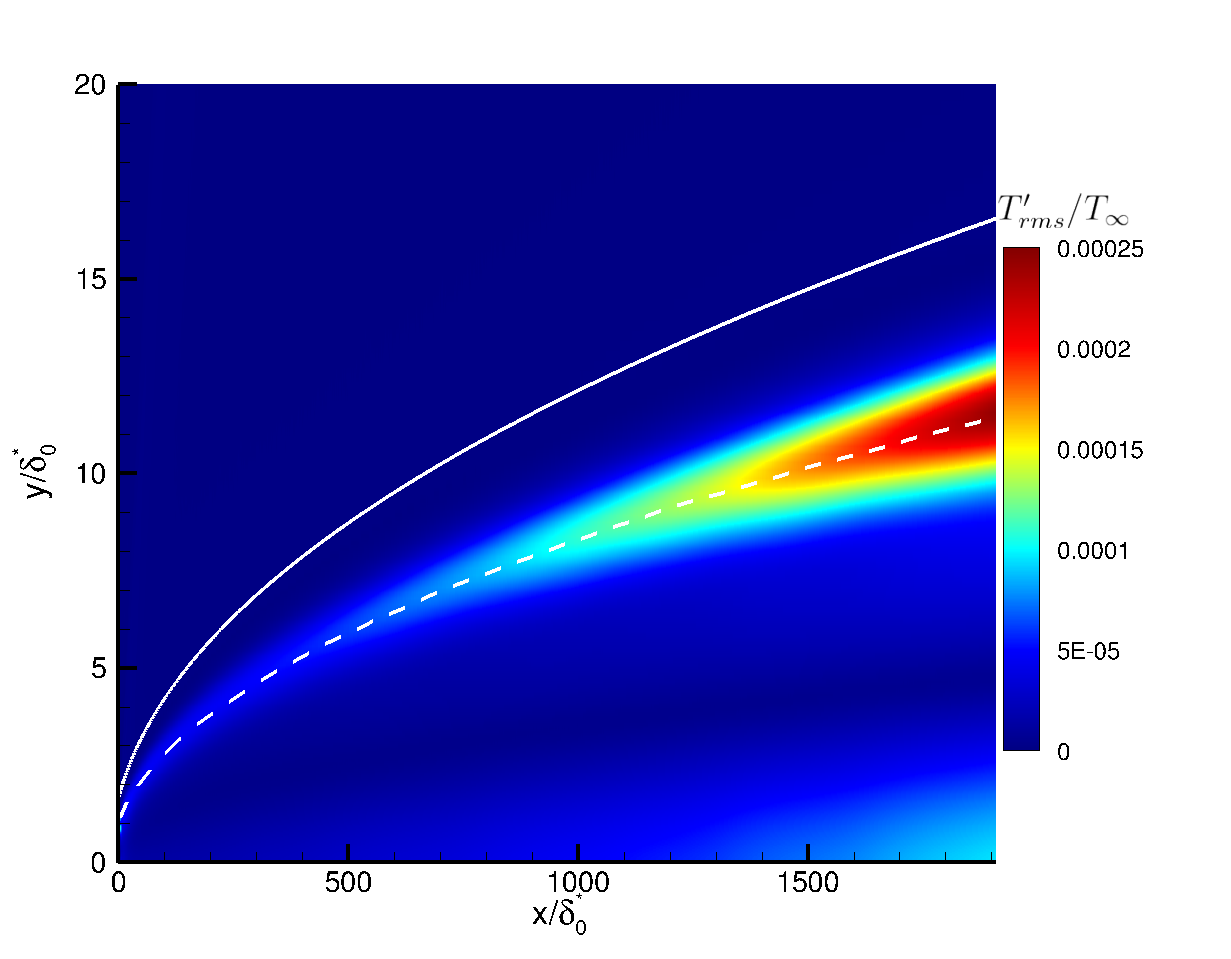}} 
        \subfloat[\label{fig:Trms_Fb}]{\includegraphics[scale=0.22]{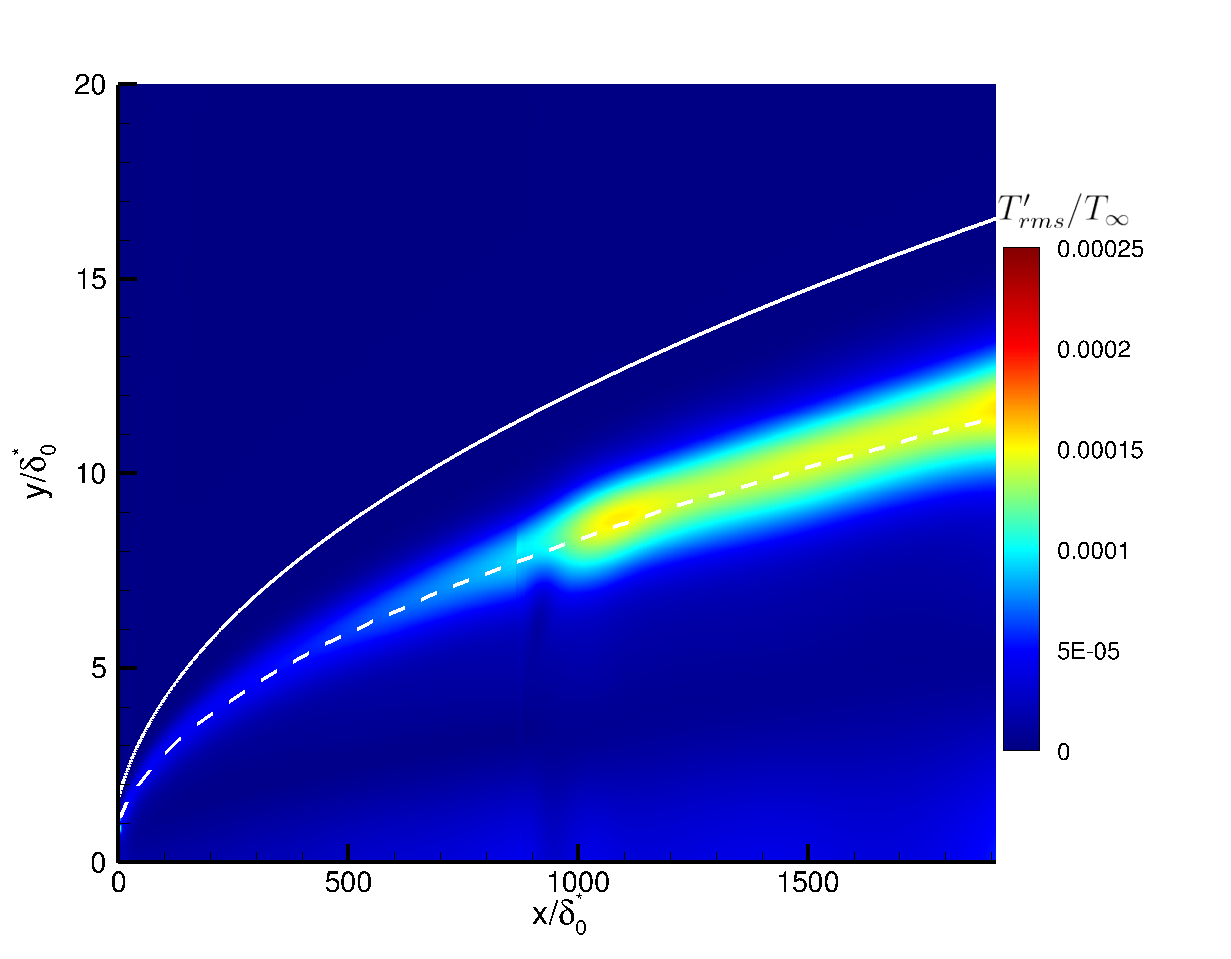}} \\
        \subfloat[\label{fig:urms_NC}]{\includegraphics[scale=0.22]{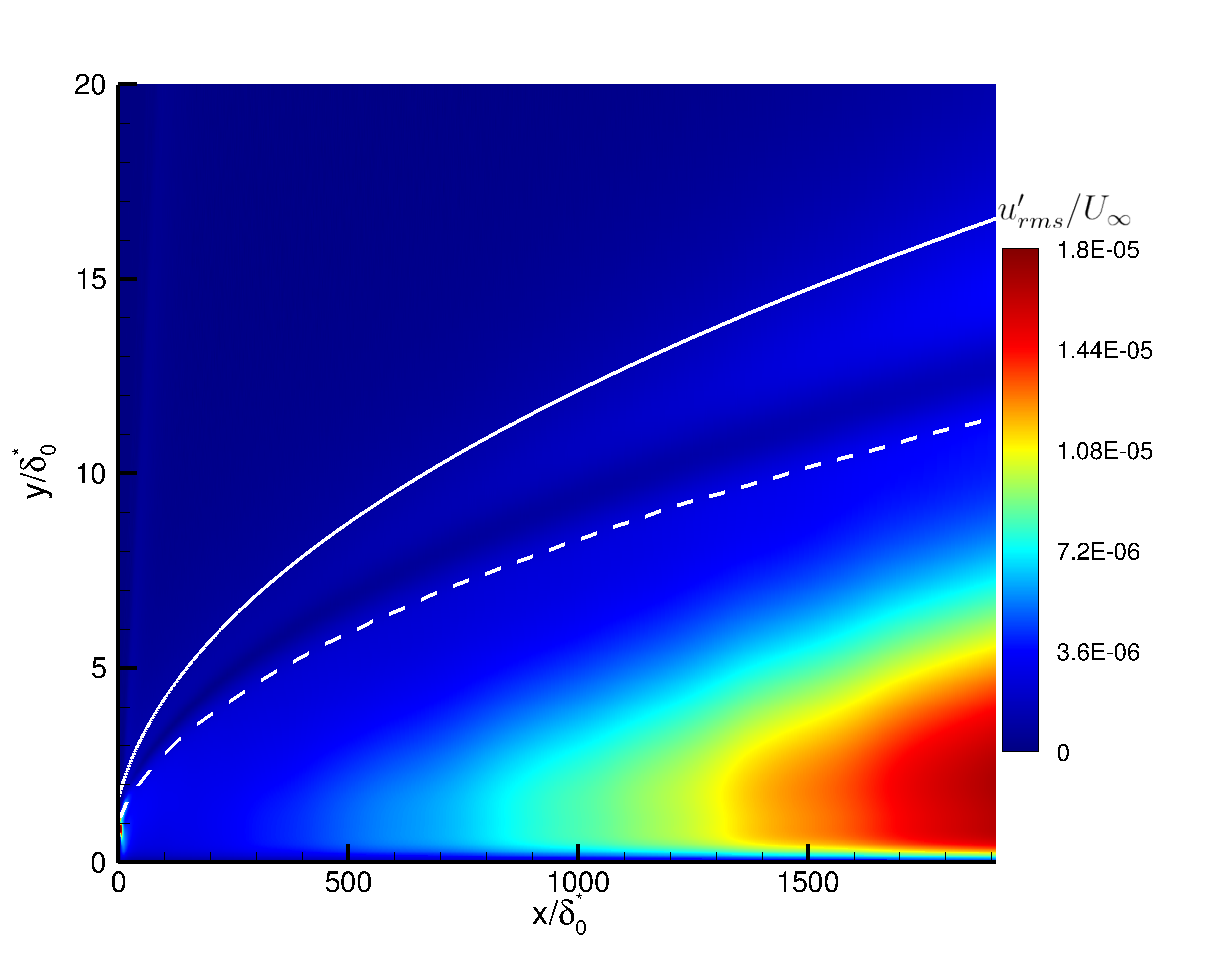}}
      \subfloat[\label{fig:urms_Fb}]{\includegraphics[scale=0.22]{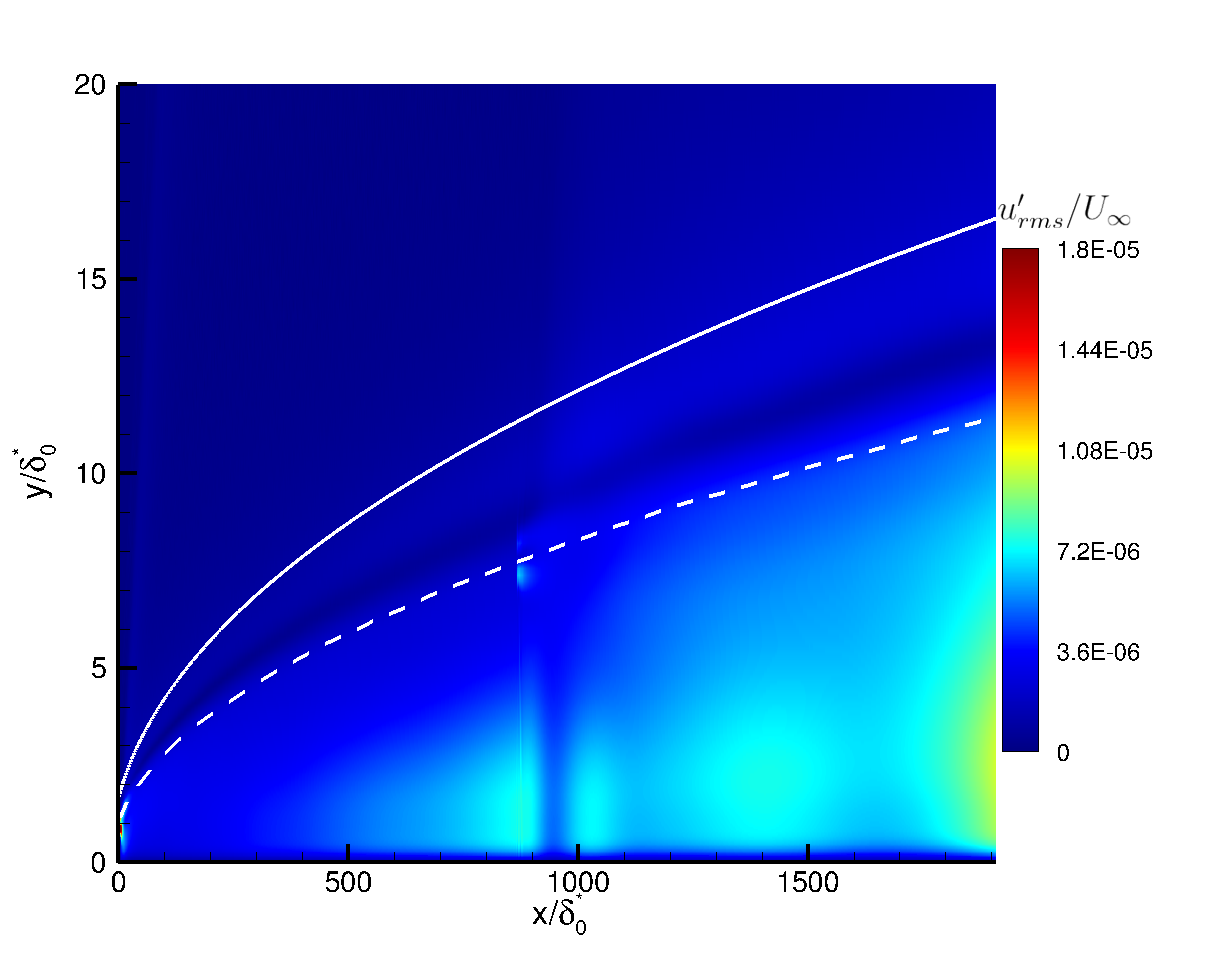}} \\
    \caption{Contours of \(T'_{rms}\) (a,b) and \(u'_{rms}\) (c,d)  for the uncontrolled (a,c) and feedback (b,d) cases. The white solid lines and dashed lines represent the boundary layer thickness $\delta$ and the generalised inflection point position $y_{g}$, respectively.} 
    \label{fig:White_Noise_urms}
\end{figure}

In addition to these results on wall-pressure fluctuation sensors coming from impulse responses, we consider the global root-mean-square (r.m.s.) temperature field (denoted \(T'_{rms}\)) and streamwise velocity field (denoted \(u'_{rms}\)). For \(T'_{rms}\), whose high values are located around the generalized inflection point (white dashed line) in the uncontrolled case (see figure \ref{fig:Trms_NC}), the control action reduces the amplitude of the perturbations  (see figure \ref{fig:Trms_Fb} for the feedback case). For the field \(u'_{rms}\) in the uncontrolled case (see figure \ref{fig:urms_NC}), high level regions are localized close to the wall. These levels are drastically decreased when the control action is present (see figure \ref{fig:urms_Fb} for the feedback case). Regarding the feedforward design (not shown here), it further reduces the amplitude of disturbances (as in figure \ref{fig:h2_final}). By drastically reducing the amplitude of velocity disturbances in both feedforward and feedback configurations, while the controllers were built from wall pressure fluctuation performance sensors, one may hope to strongly delay transition to turbulence due to the second Mack mode in a 3D setup. Stability and performance robustness are further addressed next.% However, this result does not provide any information regarding robustness to stability and performance.

\subsection{Stability robustness}
\label{sec:stab_robustness}
In the case of the feedback design, the configuration can be unstable and it is necessary to quantify the evolution of the stability margins following inflow condition variations or uncertainties. The closed loop system is stable if and only if the Nyquist plot of the loop gain \(-T_{yu}^{real}K\) (which is stable) does not encircle the critical point (-1, 0). As already discussed in \S \ref{sec:specification_h2_hinf}, the Nyquist plot of \(-T_{yu}K\) therefore allows to quantify the available stability margins related to the distance to the critical point by visualizing the maximum amount of error \(|\Delta|\) admissible before instability sets in. The gain and phase margins (denoted \(GM\) and \(PM\)) respectively represent the minimum amount of gain and phase variations required to lose stability. In our case, the gain and phase margins respectively stand for an estimation error in the instability's growth rate and convection speed which can lead to an instability of the feedback loop \citep{Sipp_Important_art}. Inlet velocity variation is considered here to be the most problematic variation (compared to other primitive variable variations) as it involves multiple changes: (i) variation in time delays due to change in convection velocity; (ii) modification of the Reynolds number \(Re_{x}\) implying that for a given abscissa on the domain, the dominant frequencies are higher (respectively lower) after an increase (respectively decrease) of \(Re_{x}\); (iii) variation of the Mach number \(M_{\infty}\) implying a modification of the neutral curves and by extension a modification of the growth rates. 
\begin{figure}
    \centering
     \subfloat[\label{fig:nyq_module_Tyu}]{\includegraphics[scale=0.45]{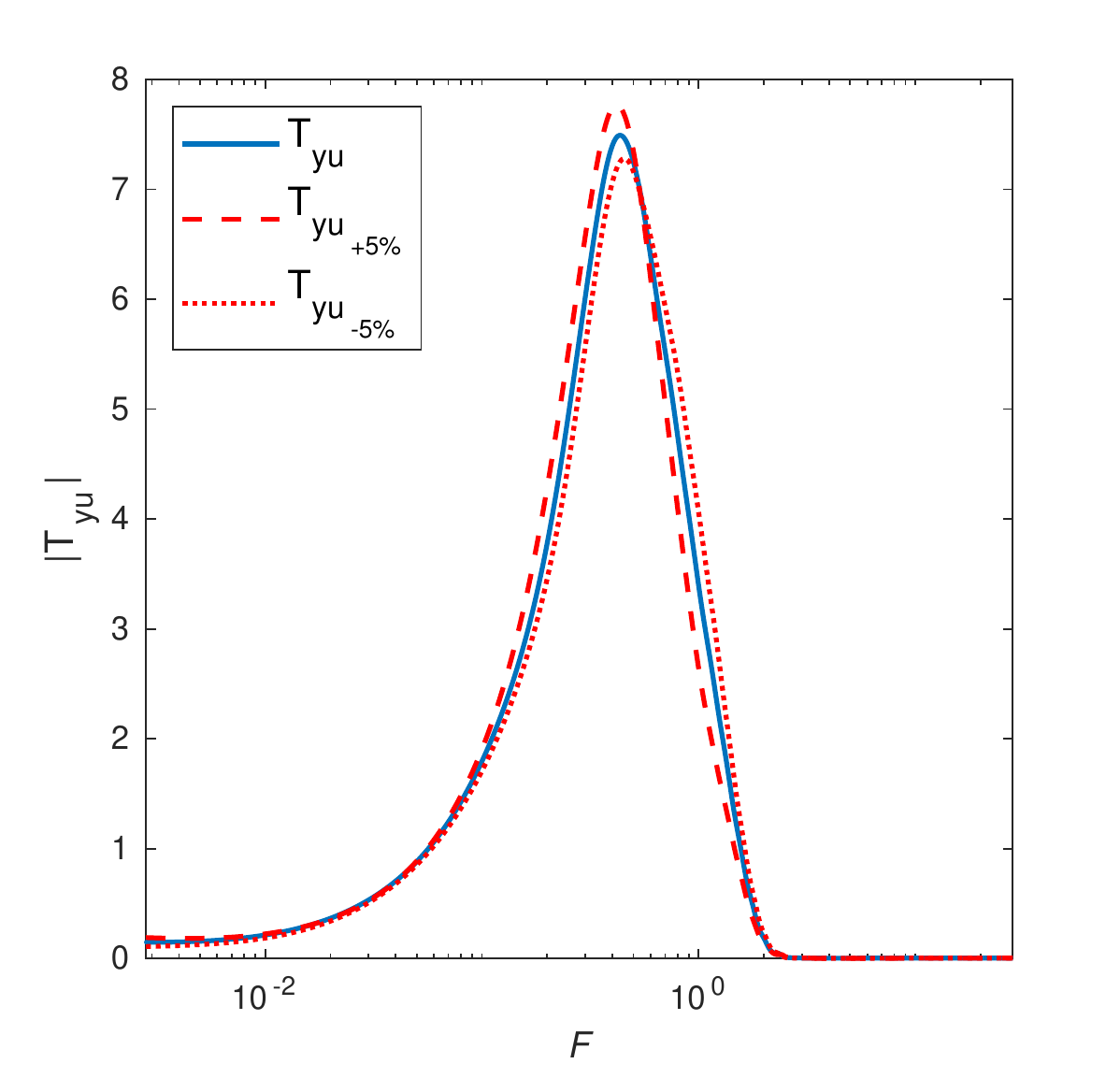}}
      \subfloat[\label{fig:nyq_Phase_Tyu}]{\includegraphics[scale=0.45]{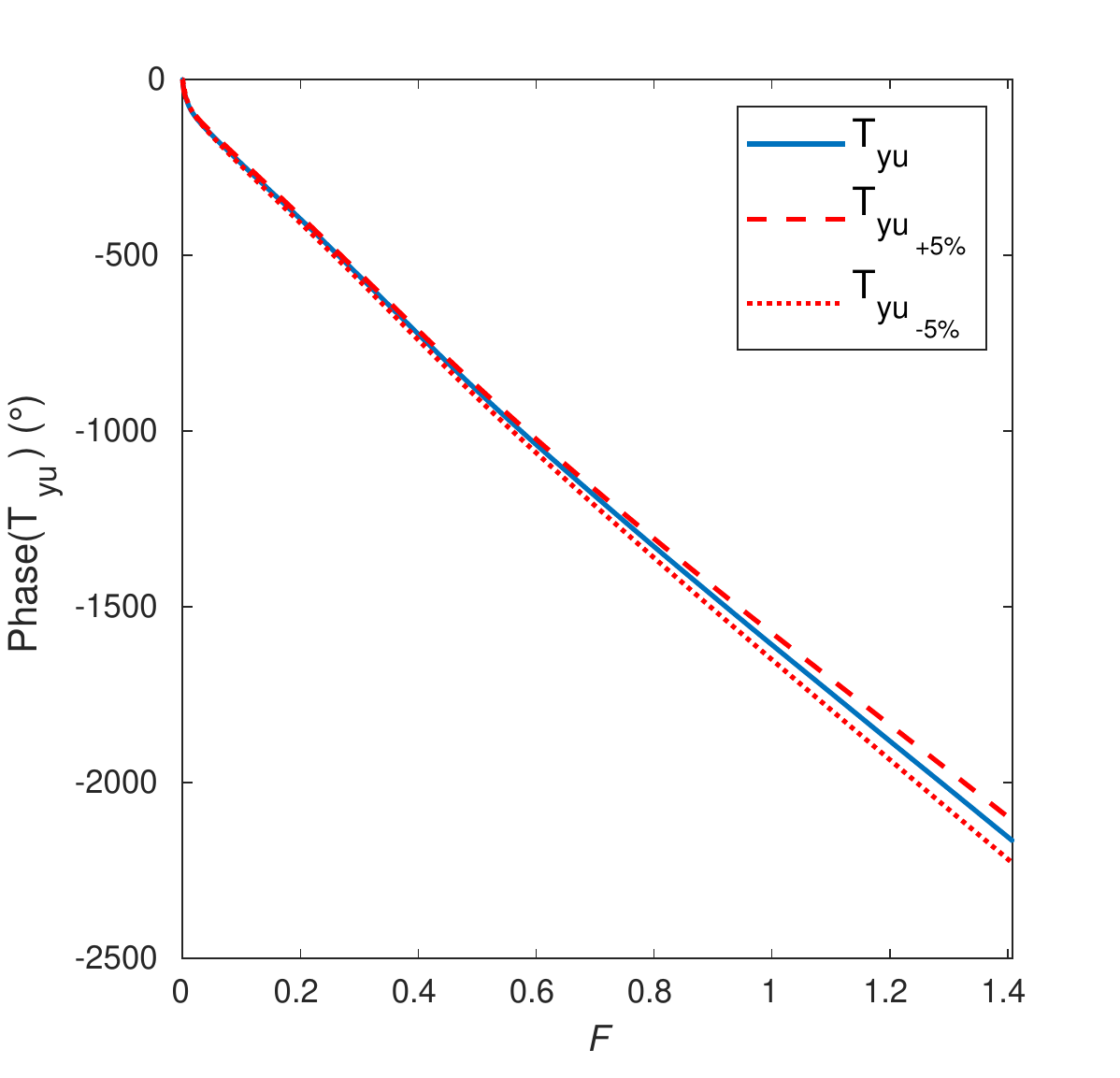}} \\
      \subfloat[\label{fig:nyquist_all}]{\includegraphics[scale=0.45]{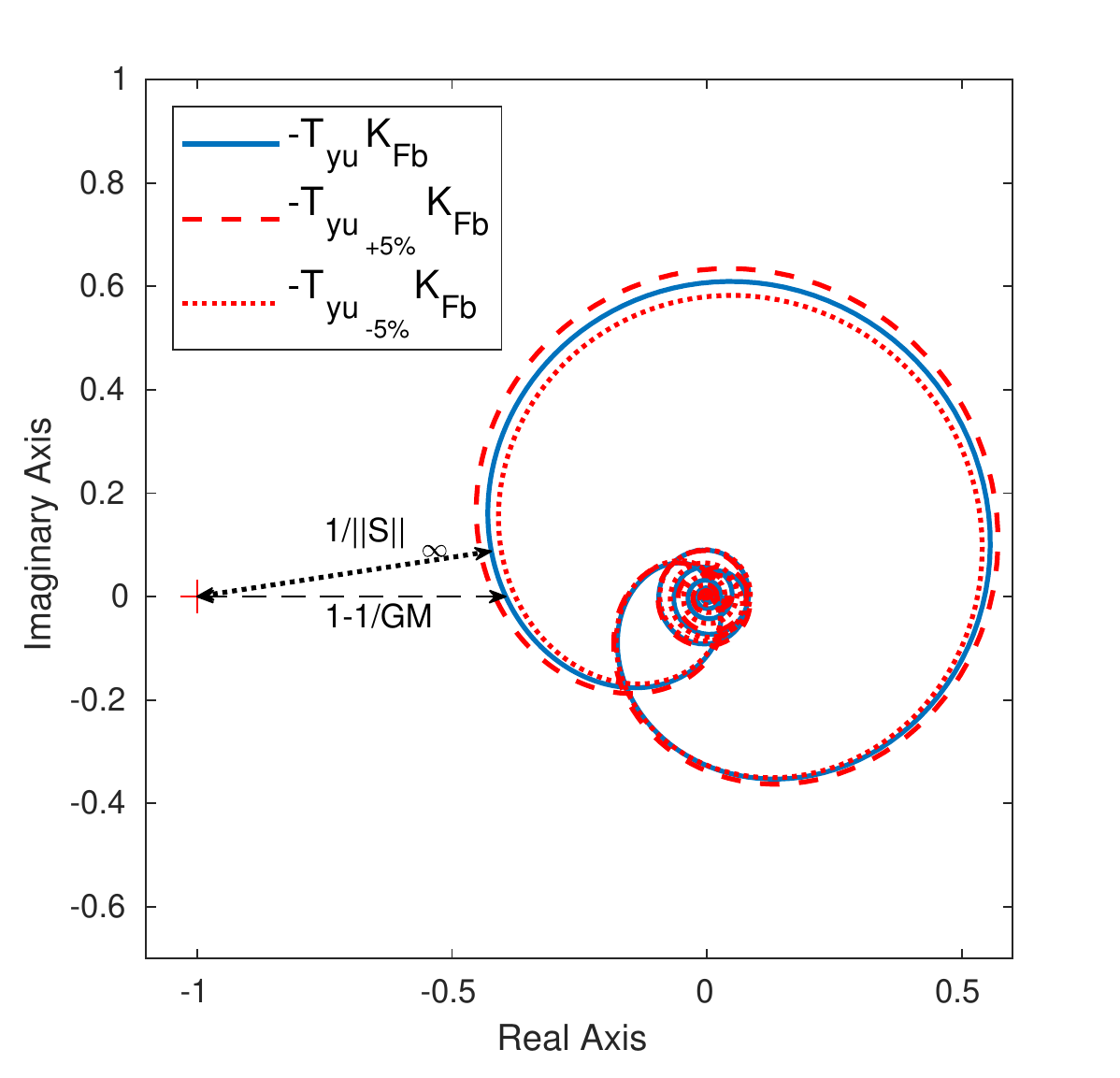}}
      \subfloat[\label{fig:nyquist_zoom}]{\includegraphics[scale=0.45]{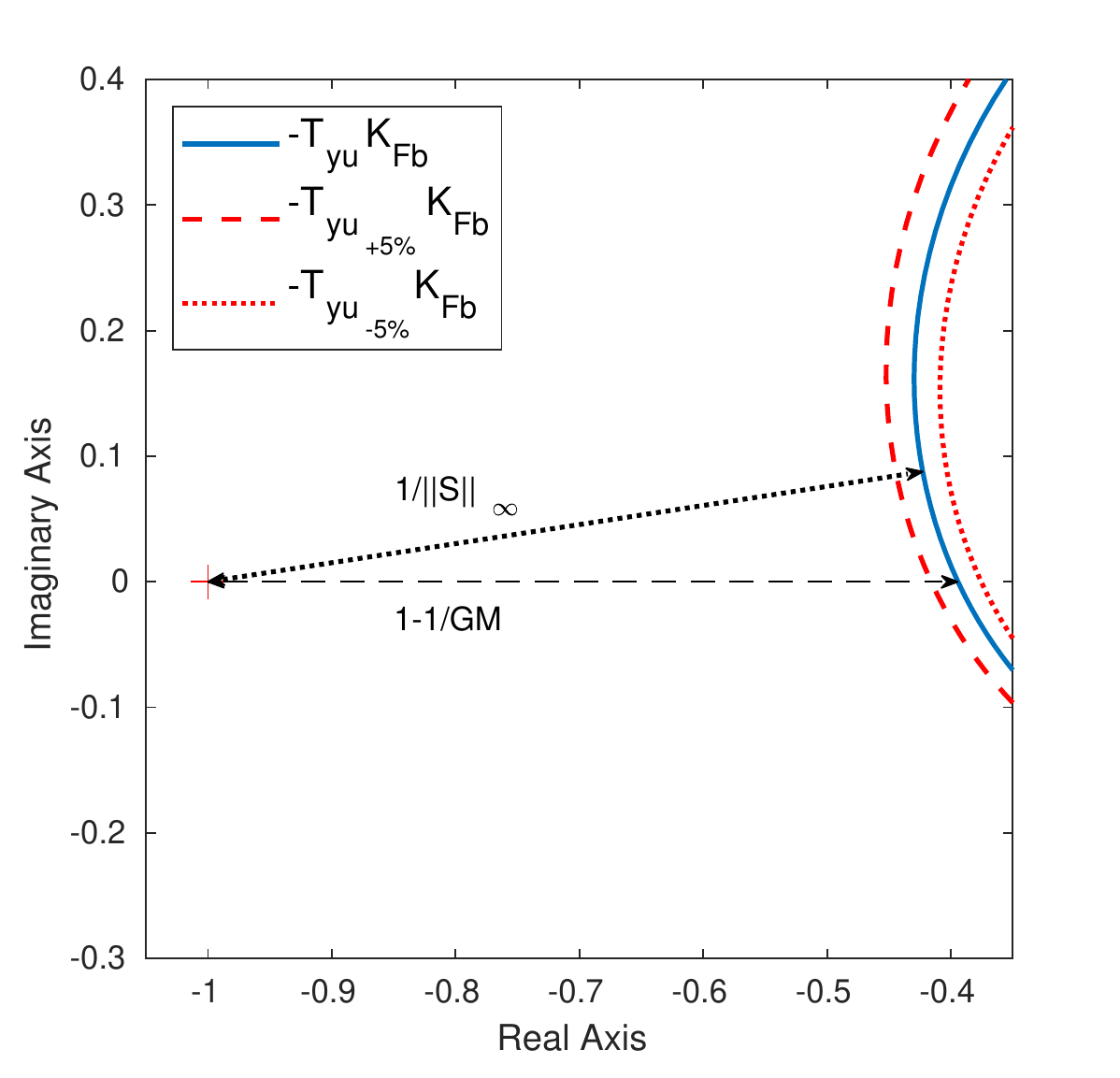}}
    \caption{Evolution of the module (a) and phase (b) of \(T_{yu}\) after a variation of \(\pm 5\%\) of the inlet velocity. Global view (c) and zoom near the critical point (-1,0) (d) of the Nyquist plot of the loop gain \(-T_{yu}K_{Fb}\) (solid blue line) and \(-T_{yu_{\pm 5\%}}K_{Fb}\) (dashed and dotted red lines). The black dotted line represents the modulus margin \(||S||_{\infty}^{-1}\) (the minimal distance to instability). The black dashed line represents the gain difference before instability and is linked to the gain margin $GM$.}
    \label{fig:Stability_Robustness}
\end{figure}
For a variation of the upstream velocity at the entry of the domain \(U_{\infty}\) of \(\pm 5\%\), which induces $ M_\infty \in [4.275,4.725]$, the new transfer functions \(T_{yu_{\pm 5\%}}\) are compared with the reference one \(T_{yu}\) in figures \ref{fig:nyq_module_Tyu} and \ref{fig:nyq_Phase_Tyu}. The greatest variations for the module appear to be around \(F=0.423\); we notice that a \(5\%\) increase of the upstream velocity implies a greater maximum value for the module at a slightly lower frequency whereas a \(5\%\) decrease in velocity implies a smaller maximum value for the module at a slightly higher frequency (see figure \ref{fig:nyq_module_Tyu}). The variation of \(\pm 5\%\) of the inlet velocity leads to the modification of the delays, represented by the slope of the phase versus frequency plot (see figure \ref{fig:nyq_Phase_Tyu}): for the \(5\%\) increase of the upstream velocity, the absolute value of the slope is less and the delay is therefore shorter (with a relative variation for the delay of \(3.4\%\) compared to the reference case), whereas the opposite is obtained in the case \(-5\%\) (with a relative variation of \(3.9\%\) for the delay). Figures \ref{fig:nyquist_all} and \ref{fig:nyquist_zoom} show the Nyquist plot of the loop gains \(-T_{yu}K_{Fb}\) and \(-T_{yu_{\pm 5\%}}K_{Fb}\). The variations of the upstream velocity slightly alter the stability margins compared to those obtained in the reference case: the phase margin stays infinite, while the gain margin $GM$ (black dashed lines) and the modulus margin \(||S||_{\infty}^{-1}\) (black dotted lines) fluctuate respectively by a maximum of \(3.6\%\) and \(5.1\%\), while remaining far from the critical point. Given the small impact of the inflow velocity variations of \(\pm 5\%\) on all margins, the feedback design may be stable for even greater velocity variation. Therefore, unlike previous feedback studies using LQG \citep{Barbagallo_art,Tol_fbff_art}, the robustness to stability for a feedback design obtained with a robust synthesis method is not a problem. Next, we examine performance robustness, which is a different issue.

\subsection{Performance robustness}
\label{sec:perfo_robustness}

Robustness to performance is evaluated by checking that the control laws remain efficient in terms of expected power reduction of the different performance sensors $z_i$ despite new noise sources or differences between on-design and off-design operating conditions.
\newline
\begin{itemize}
\item \textit{Noisy sensors}
\newline
\phantom{text}Noisy estimation sensors are modelled by adding white Gaussian noise on both \(y_{\textit{fb}}\) and \(y_{\textit{ff}}\) (see figure \ref{fig:noise_yfb}). Both estimation sensors are corrupted by the same amount of noise (\(50\%\) of the r.m.s. value without control action of \(y_{\textit{fb}}\)), which models an intrinsic defect of the sensor, such as electronic noise, that does not depend on its position along the domain. Nevertheless, the streamwise position of \(y_{\textit{ff}}\) being quite close to that of \(y_{\textit{fb}}\), the ideal signal-to-noise ratio remains very similar for both configurations and only varies by a few percents.  
\begin{figure}
    \centering
     %% cc pour dire deux colonnes CENTRE ll pour deux colonnes alignes a gauche er rr pour droite
     \begin{tabular}{cc}
      %\hspace*{-1cm}
      \subfloat[\label{fig:noise_yfb}]{\includegraphics[scale=0.33]{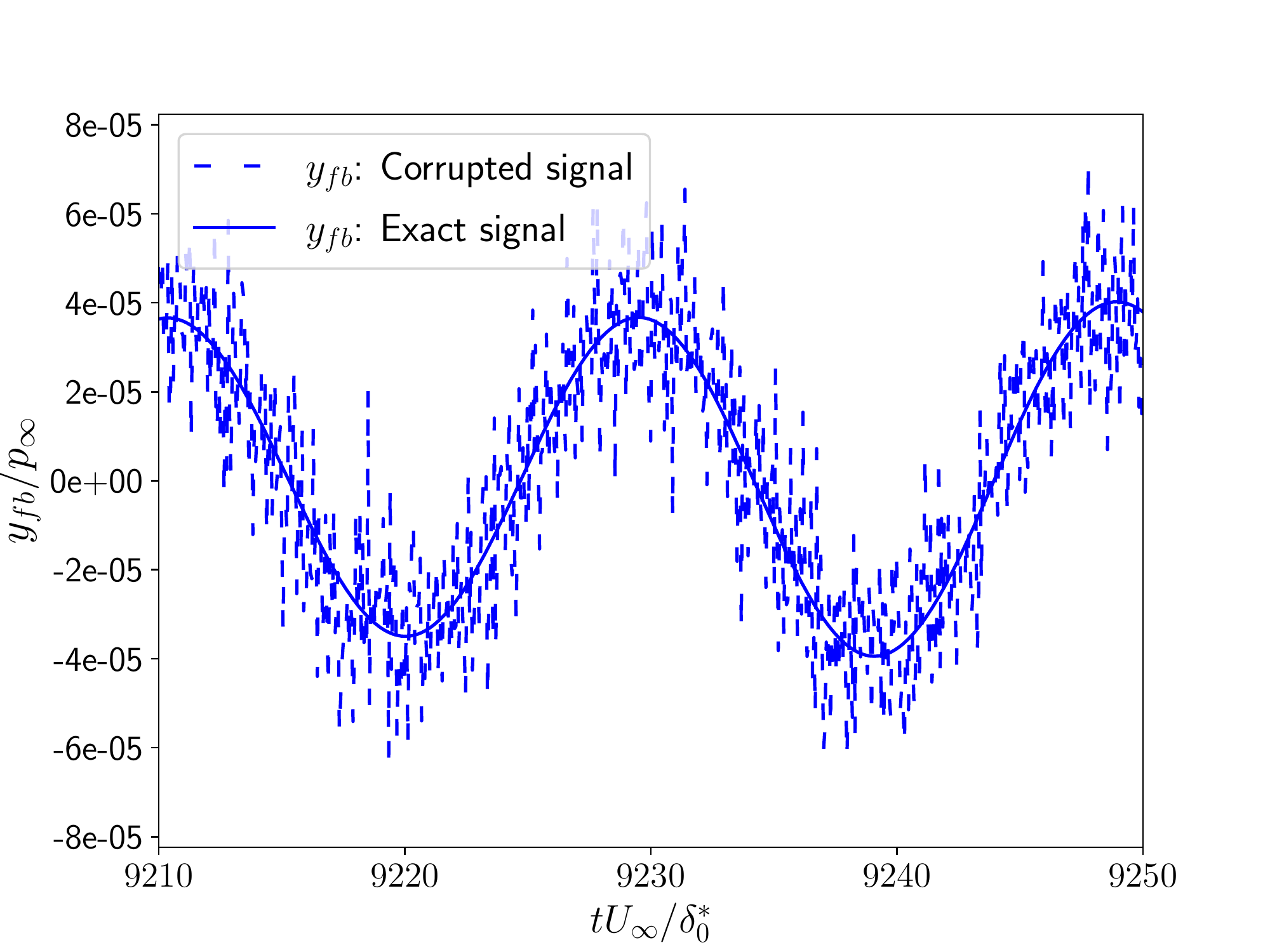}}
      &
      %\hspace*{-1cm}
     \subfloat[\label{fig:PSDy_noisyy}]{\includegraphics[scale=0.33]{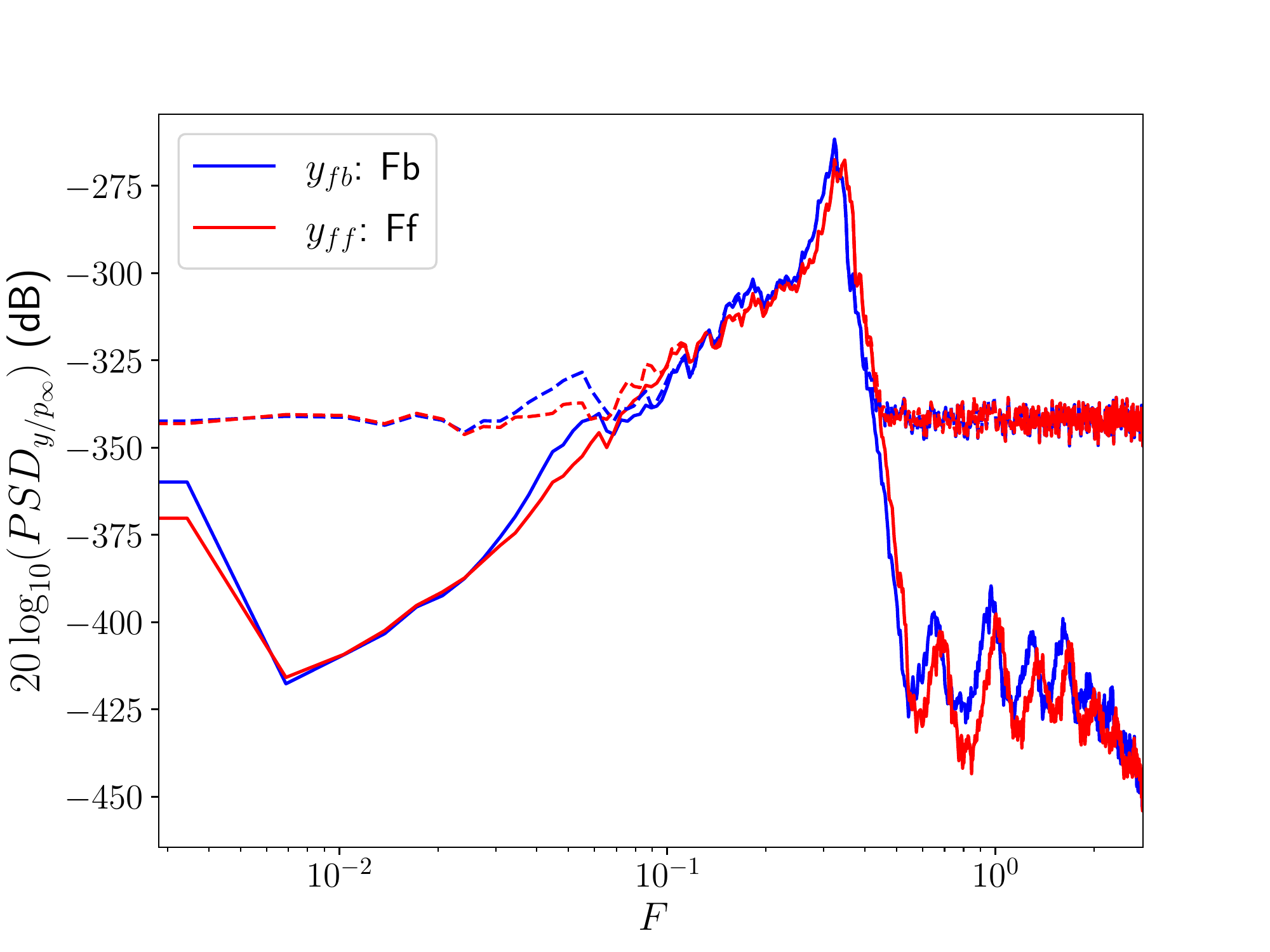}} \\
      \subfloat[\label{fig:PSDu_noisyy}]{\includegraphics[scale=0.33]{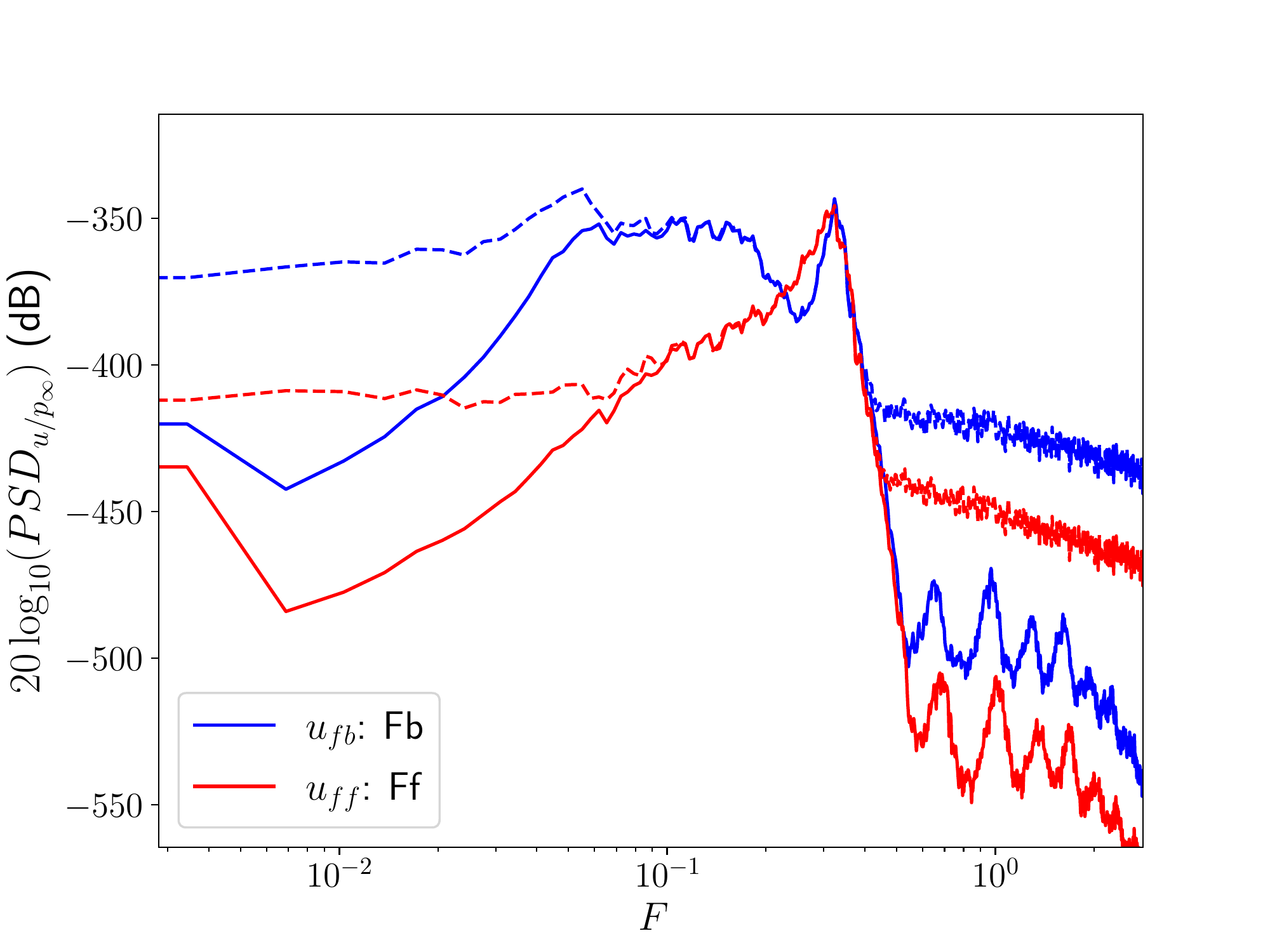}} 
      &
    \subfloat[\label{fig:urms_noisyy}]{\includegraphics[scale=0.33]{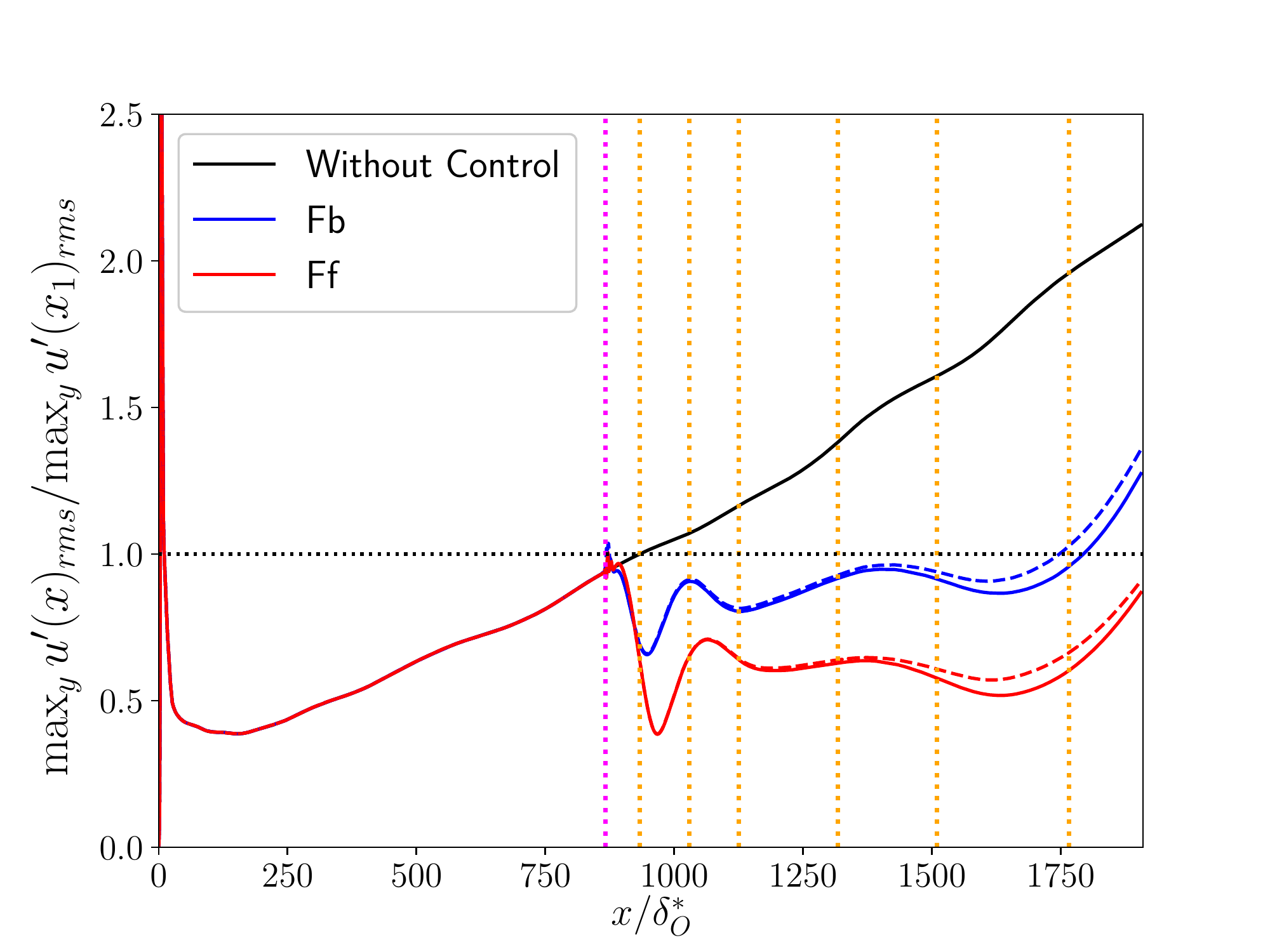}}
     \end{tabular}
    \caption{In all subplots, feedback and feedforward designs are in blue and red lines, respectively. Controlled systems with ideal and noisy estimation sensors are represented by solid and dashed lines, respectively. a) Short sequence of \(y_{\textit{fb}}\) corrupted by \(50\%\) of the r.m.s. value without control action of \(y_{\textit{fb}}\). Comparison of the evolution of \(PSD_{y}\) (b), \(PSD_{u}\) (c) and \(\mathrm{max}_{y}\ u'_{rms}\) (d) for the controlled systems with ideal and noisy estimation sensors. The legend in (d) is the same as in figure \ref{fig:h2_final}.}
    \label{fig:urms_noise_sensor}
\end{figure}
The PSD of the corrupted estimation sensors remain unchanged in the frequency band of the second Mack mode but exhibit much larger values in low and high frequencies (see figure \ref{fig:PSDy_noisyy}). This is because the PSD of white noise being constant, the ideal signal-to-noise ratio is particularly low for frequencies where the ideal signal energy is low. The signal \(y\) is given to the controller \(K\), which generates the actuator signal \(u\); the control signal PSD for corrupted signals \(y\) becomes stronger on the previously mentioned low and high frequency bands, compared to the PSD of $u$ for ideal signals \(y\) (see figure \ref{fig:PSDu_noisyy}). Nevertheless, thanks to the strictly proper structure and the filter \(W_{KS}\) imposed in the synthesis step, $ |KS| $ have been constrained in these frequency bands. Thus, the actuator activity remains limited in these regions despite the important added noise and, if we look at the evolution of the maximum along the wall-normal direction of \(u'_{rms}\) (denoted \(\mathrm{max}_{y}\ u'_{rms}\)), we keep a performance close to the ideal case (see figure \ref{fig:urms_noisyy}). Both feedback and feedforward configurations stay below the velocity energy threshold until $x_{6}$ and these two designs are robust to noise on the estimation sensors. If even noisier sensors were used, it would suffice to decrease the amplitude of the weighting function \(1/W_{KS}\) to recover performance robustness (especially in low frequencies for the feedback configuration). For the case illustrated in figure \ref{fig:urms_noise_sensor}, a higher \(1/W_{KS}\) (involving a less constrained controller) could lead to an excessive injection of energy in the vicinity of the actuator (see appendix \ref{sec:Impact_WKS_noisy_sensors}).
\newline
\item \textit{Off-design operating conditions}
\newline
\phantom{text}Performance robustness to off-design operating conditions is assessed by considering the evolution of the local \(H_{2}\) norm of \(T_{z(x)w}\) after a variation of free-stream density \(\rho_{\infty}\) and velocity \(U_{\infty}\) of \(\pm 5\%\), for both feedback and feedforward cases. The density variation may correspond in practice to a change in altitude whereas the velocity variation may correspond to a change in cruise speed. When \(\rho_{\infty}\) is modified, the temperature and velocity inlet values are kept constant, which means that \(M_{\infty}\) and hydrodynamic delays related to the convective behaviour are maintained (see green dashed line figure \ref{fig:impulse_den_velo}) while only \(Re_{x}\) is modified (which implies a change in the dominant frequencies at a given abscissa as seen in figure \ref{fig:PSD_den_velo}). A modification of \(U_{\infty}\) on the other hand has a much more dramatic effect since it implies variations of time delays (see purple dashed line figure \ref{fig:impulse_den_velo}) which will ultimately impact the only important residual delay which is the one between the actuator and the estimation sensors. It will also impact the values of \(Re_{x}\) and \(M_{\infty}\), which modify the  base flow profiles. Changing the base flow impacts the stability characteristics of the boundary layer, and, in turn, the dominant frequencies along the plate, as seen figure \ref{fig:PSD_den_velo}.%; it is important to vary \(U_{\infty}\) as it is the only quantity that can modify the hydrodynamic delays related to the convection, contrary to a variation of \(M_{\infty}\) which would not necessarily have impacted these delays. 
\newline
\phantom{text}With density variations of \(\pm 5\%\) (see figure \ref{fig:Robustness_density}), despite degraded off-design performance, both feedback and feedforward controllers manage to reduce the local \(H_{2}\) norm compared to the case without control over a fairly large distance on the flat plate. However, while the feedforward design minimized the local \(H_{2}\) norm more than the feedback one for the nominal case (solid lines), it seems that this is no longer necessarily the case in off-design situations (dotted and dashed lines). The variation in performance between the nominal and off-design cases in the feedback configuration appears less pronounced than in the feedforward setup, which is allowed by the sensitivity function \(S\). Although this transfer function, because of the delay due to the actuator/estimation sensor distance, limits the achievable performance on the nominal case for a feedback setup (see appendix \ref{sec:estimation_sensor_perfo}), it allows to desensitize the system to modelling errors or to variations in system characteristics over a certain bandwidth. Even if both designs exceed the $H_2$ norm threshold at some point, they have some robustness to performance with respect to density variations by staying below the uncontrolled system $H_2$ norm all along the domain. 
\begin{figure}
    \centering
    \subfloat[\label{fig:impulse_den_velo}]{\includegraphics[scale=0.33]{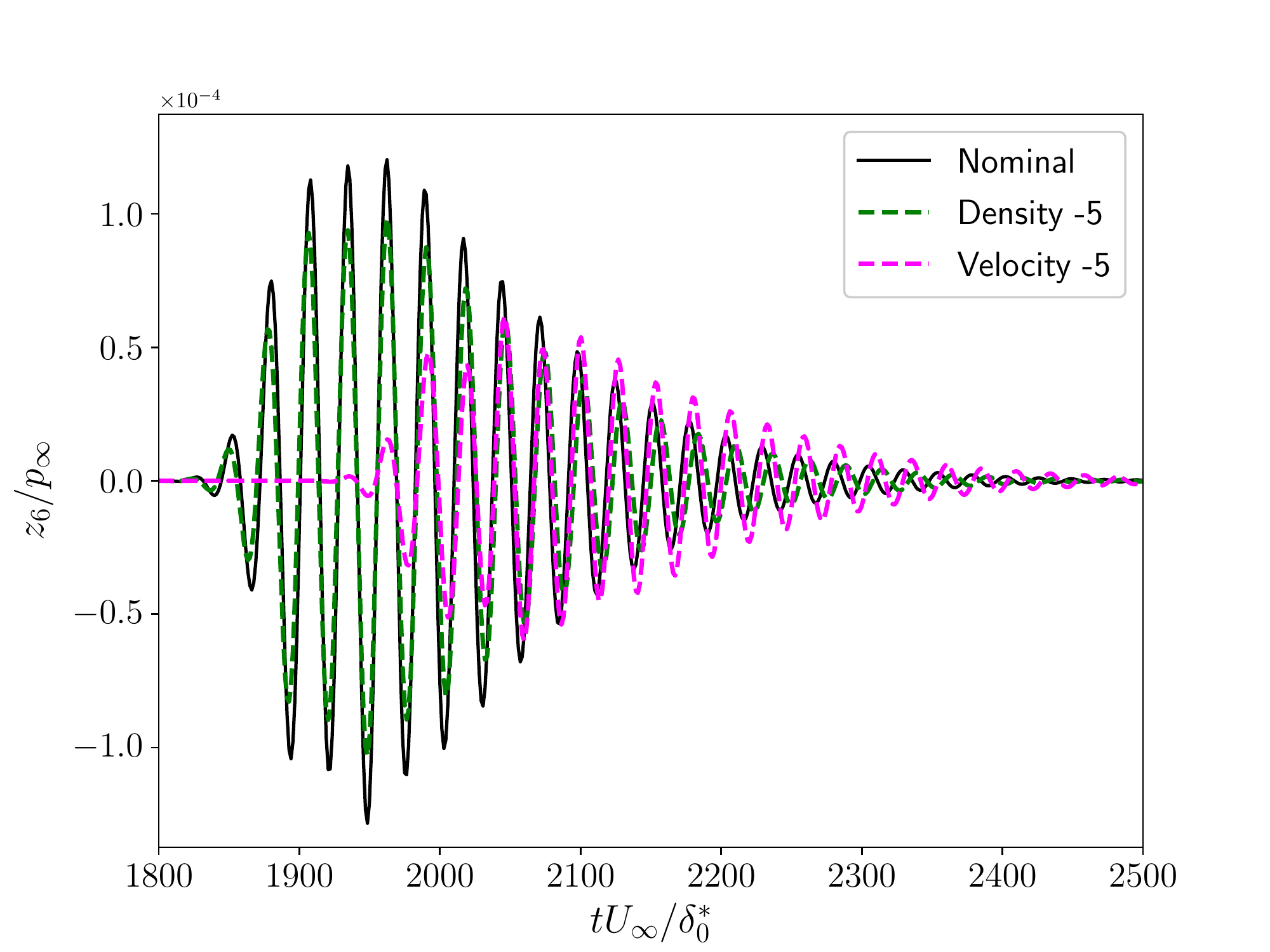}}
      \subfloat[\label{fig:PSD_den_velo}]{\includegraphics[scale=0.33]{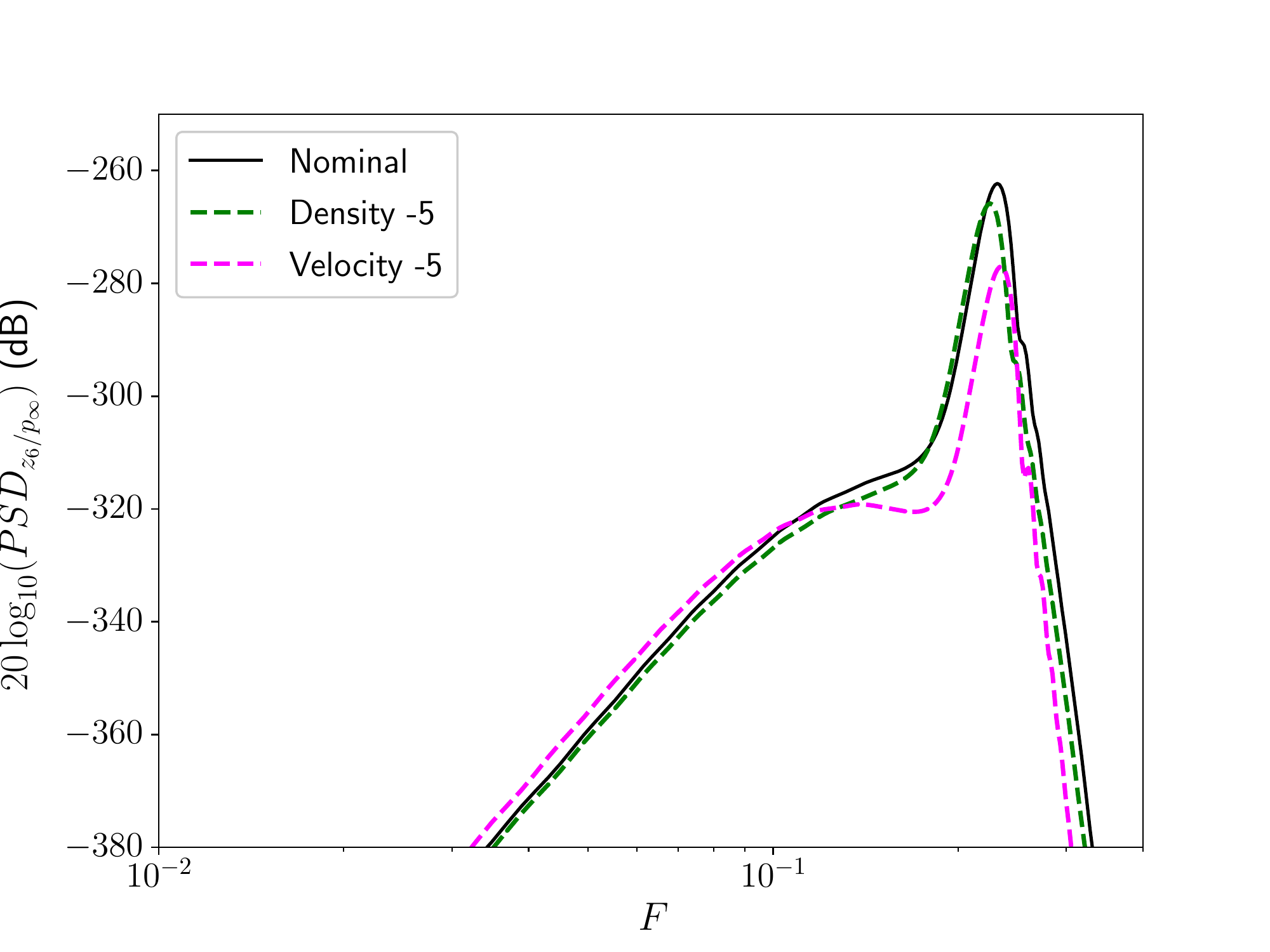}}\\
     \subfloat[\label{fig:Robustness_density}]{\includegraphics[scale=0.33]{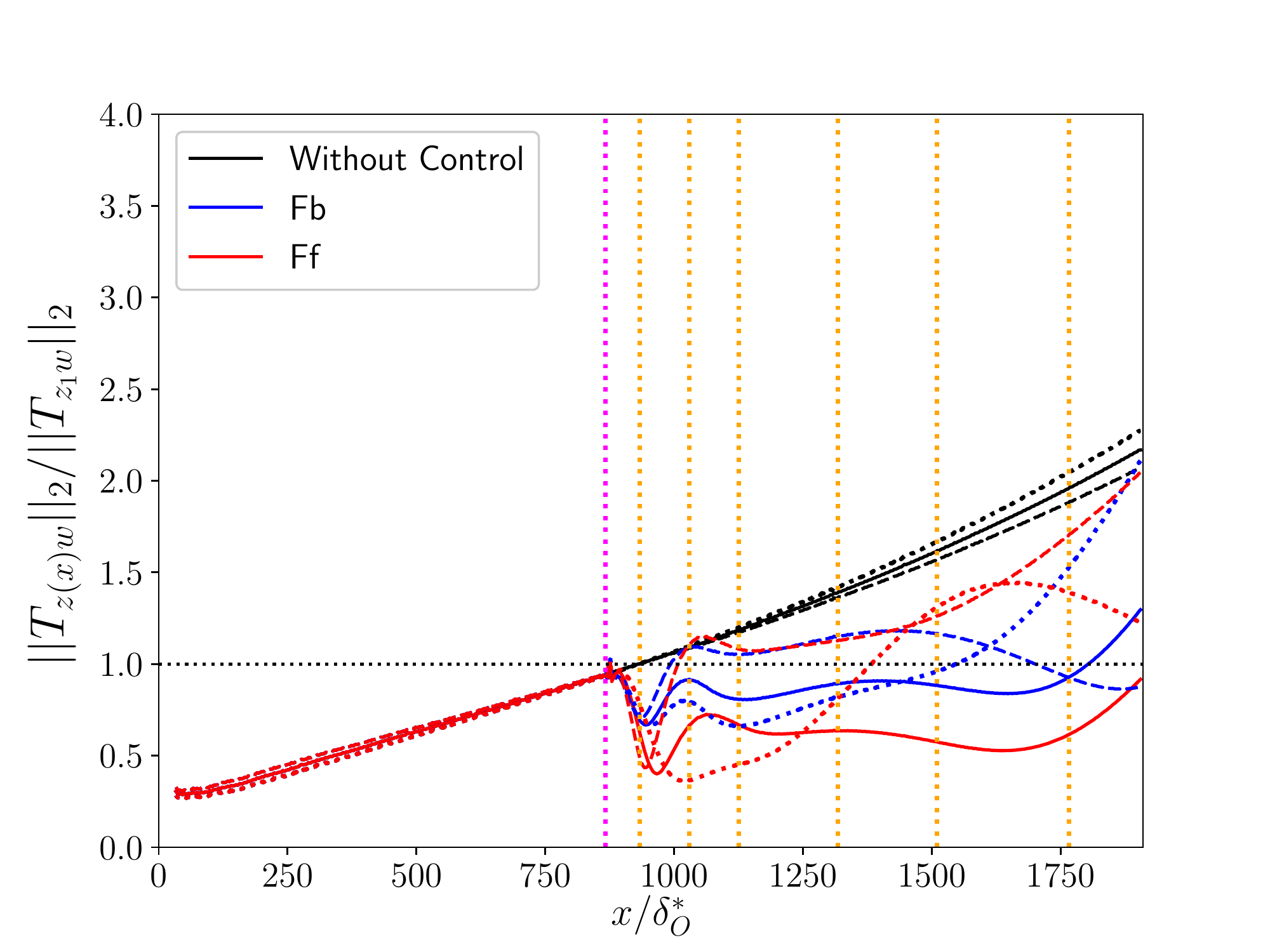}}
      \subfloat[\label{fig:Robustness_velocity}]{\includegraphics[scale=0.33]{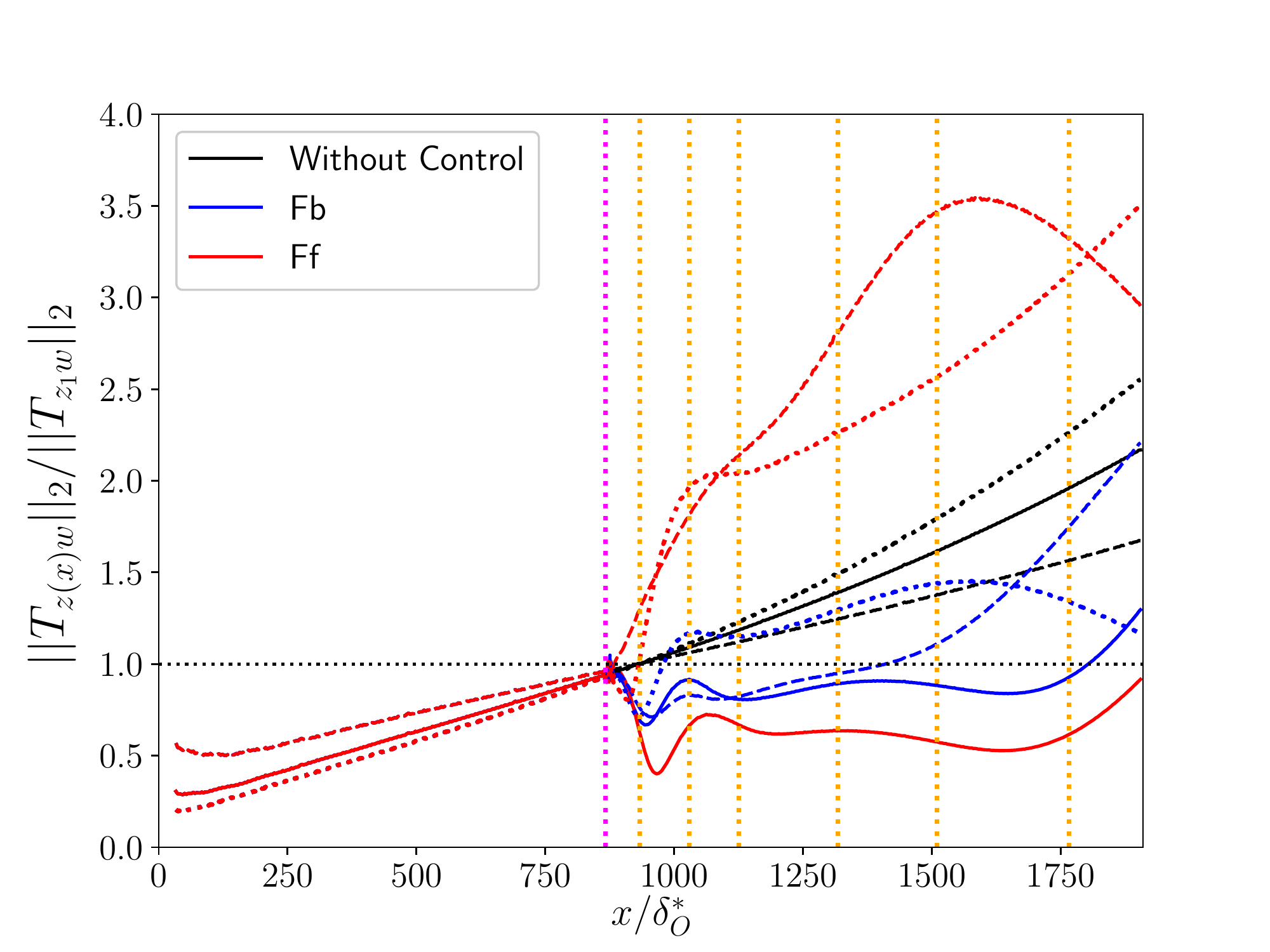}}
    \caption{Comparison of uncontrolled pressure wavepackets generated by an impulse of $w$ (a) and their PSD (b) at \(x_{6}\) after a variation of \(\rho_{\infty}\) and \(U_{\infty}\) of \(- 5\%\). Evolution of the local \(H_{2}\) norm of \(T_{z(x)w}\) after a variation of \(\rho_{\infty}\) (c) and \(U_{\infty}\) (d) of \(\pm 5\%\) (dotted and dashed lines). The nominal cases are in solid lines. Same legend as in figure \ref{fig:h2_final}.}
    \label{fig:Robustness_performance}
\end{figure}
\newline
\phantom{text}The real strength and superiority of the feedback design over the feedforward one lies in its ability to maintain correct performance during velocity variations (see figure \ref{fig:Robustness_velocity}). While the feedback setup manages to maintain some performance in off-design conditions by staying below the local \(H_{2}\) norm of the uncontrolled system over a fairly long distance along the plate, the feedforward design fails to maintain the performance requirement by amplifying the local \(H_{2}\) norm. This increase in the feedforward setup may then lead to a faster transition to turbulence, which is the opposite of the desired objective. The velocity variations, regardless of changes in Reynolds and Mach numbers, are indeed particularly problematic as they modify the residual delay \(\tau_{yu}\) which directly impacts (\ref{eq:fb_h2_delay}) and (\ref{eq:ff_h2_delay}) and therefore may cause the controllers to activate out of phase. Thus, in the case of noise-amplifier flows, we underline the importance to assess the robustness to performance with respect to velocity variations, as in \citet{Fabbiane_JFMrapids_art}. One could be tempted to robustify the feedforward setup by using an adaptive controller structure \citep{Fabbiane_review_art}, but this type of approach is only robust at long times (subject to convergence of the method) and the problem of robustness following abrupt velocity variations would remain. Therefore, as soon as variations or uncertainties on the inflow velocity are present, the best trade-off between performance and robustness is a feedback configuration.  
\newline
\end{itemize}

\section{Conclusions}
\label{sec:conclusions}
A robust reactive control method has been developed in order to control the linear growth of the second Mack mode in a 2D boundary layer over a flat plate at Mach 4.5. The control tools (for identification and synthesis) being mathematically well-established in a linear framework, they are perfectly suited for this precise scenario of transition to turbulence where we seek to control the linear growth of small perturbations. 

The choice of the type and position of the actuator and sensors are based on the study of the noise-amplifier behaviour of our flow, in order to trigger the optimal growth mechanisms and ensure efficient flow control. During the identification step, some unnecessary dead times related to the convective nature of the flow are removed, allowing a significant reduction in the size of the ROMs, which is beneficial both for the identification and synthesis steps. Moreover, we strive to identify only quantities that could be obtained in an experimental setup. 

After identifying these useful transfers through data-driven methods, the synthesis of the controllers is achieved with a structured mixed \(H_{2}\)/\(H_{\infty}\) synthesis. This robust synthesis method allows to limit the order of the controller, to impose its structure upfront and to constrain simultaneously several transfer functions to obtain at the same time performance and robustness. Instead of simply minimizing a global energy, the constraint minimization problem is posed in such a way that a shaping of the spatial evolution of different local energy measures is realized, which seems a more suitable approach to delay transition to turbulence. Multiple performance sensors in the streamwise direction are therefore needed in this study to cover the entire spectrum of amplified frequencies along the domain and to capture the non-modal transient growth effect generated by the actuator.

After implementing the control laws in the \textit{elsA} solver, we find that feedforward and feedback designs both manage not to exceed a certain energy threshold on the nominal case. Moreover, the stability robustness for the feedback design is not a problem thanks to the robust synthesis and the constraints imposed. Regarding performance robustness, both feedforward and feedback designs manage to reduce the amplitude of disturbances compared to the uncontrolled case despite noisy estimation sensors or inflow density variations. Nevertheless, for noise-amplifier flows, we stress the importance to assess robustness to performance by changing the inflow velocity. Indeed, this type of variation may cause the controller to activate out of phase. It appears that the feedforward setup is completely unable to follow inflow condition variations while the feedback setup keeps reasonable performance over a large velocity variation of \(\pm 5\%\). Therefore, the best trade-off between performance and robustness requires a feedback configuration (in the case of a linear time-invariant controller). This result looks contradictory with conventional wisdom which favors a feedforward setup for noise-amplifier flows. The widespread use of a feedforward structure is likely rooted in the massive use of LQG synthesis for the control of noise-amplifiers. Indeed, LQG comes with no guaranteed stability margin, which hinders its practical application to a feedback setup. \citet{Belson_art} were among the first to recognize the superiority of a feedback design for performance robustness in noise-amplifier flows. The authors used loop-shaping on a simple PI controller, but much richer feedback laws may be designed in a systematic way using the modern robust synthesis tools of the present paper. Such tools are already commonly used for the control of oscillator flows, where feedback is mandatory to stabilize the unstable base flow \citep{Flinois_art,Leclercq_art,Shaqarin_art}. We expect the methodology of the present paper, based on data-driven identification and robust synthesis on a feedback setup to be relevant to other convectively unstable flows. 

We are currently extending this study to a three-dimensional case with the goal of delaying transition to turbulence. This implies placing multiple estimation/performance sensors and actuators in the transverse direction and controlling oblique waves of the first Mack mode as well as non-linearities.

%\section*{Acknowledgements}
%This work is partially funded by the French Agency for Innovation and Defence (AID). Their support is gratefully acknowledged. We are also grateful to Xavier Chanteux who made his python LLST code available to us. This study expands the work presented at the 55$^\text{th}$ 3AF International Conference on Applied Aerodynamics 12 --- 14 April 2021, Poitiers -- France.

\bigskip
\noindent
\textbf{Acknowledgements}. This work is partially funded by the French Agency for Innovation and Defence (AID). Their support is gratefully acknowledged. We are also grateful to Xavier Chanteux who made his python LLST code available to us. This study expands the work presented at the 55$^\text{th}$ 3AF International Conference on Applied Aerodynamics 12 --- 14 April 2021, Poitiers -- France.

\bigskip
\noindent
\textbf{Declaration of interests}. The authors report no conflict of interest.

\appendix

\section{Position of the estimation sensor}
\label{sec:estimation_sensor_perfo}
To obtain the quantitative position of the sensor \(y\) in our supersonic boundary layer study for both feedforward and feedback configurations, a quick analysis is carried out; it consists in looking at the impact of the actuator/measurement sensor distance on the maximum achievable performance in terms of \(H_{2}\) norm reduction on the performance sensor \(z_{6}\) regardless of the desensitization to low frequency disturbances. We only look at the performance sensor \(z_{6}\) because it is the one furthest downstream from the domain; the further downstream we are, the more we have to reduce the local \(H_{2}\) norm in order not to exceed a given threshold (see the principle diagram in figure \ref{fig:Nfactor_H2}). This performance sensor therefore plays a central role and the position of the estimation sensor \(y\) must allow a consequent reduction of the energy of the sensor \(z_{6}\). Since this analysis is only done off-line on the ROMs and the resulting controller is not implemented on the real complete system, the \(||W_{KS}KS||_{\infty}\) constraint which was only useful in case of new noise sources (as noisy estimation sensor) is disabled. 

\begin{figure}
\centering
\includegraphics[scale=0.5]{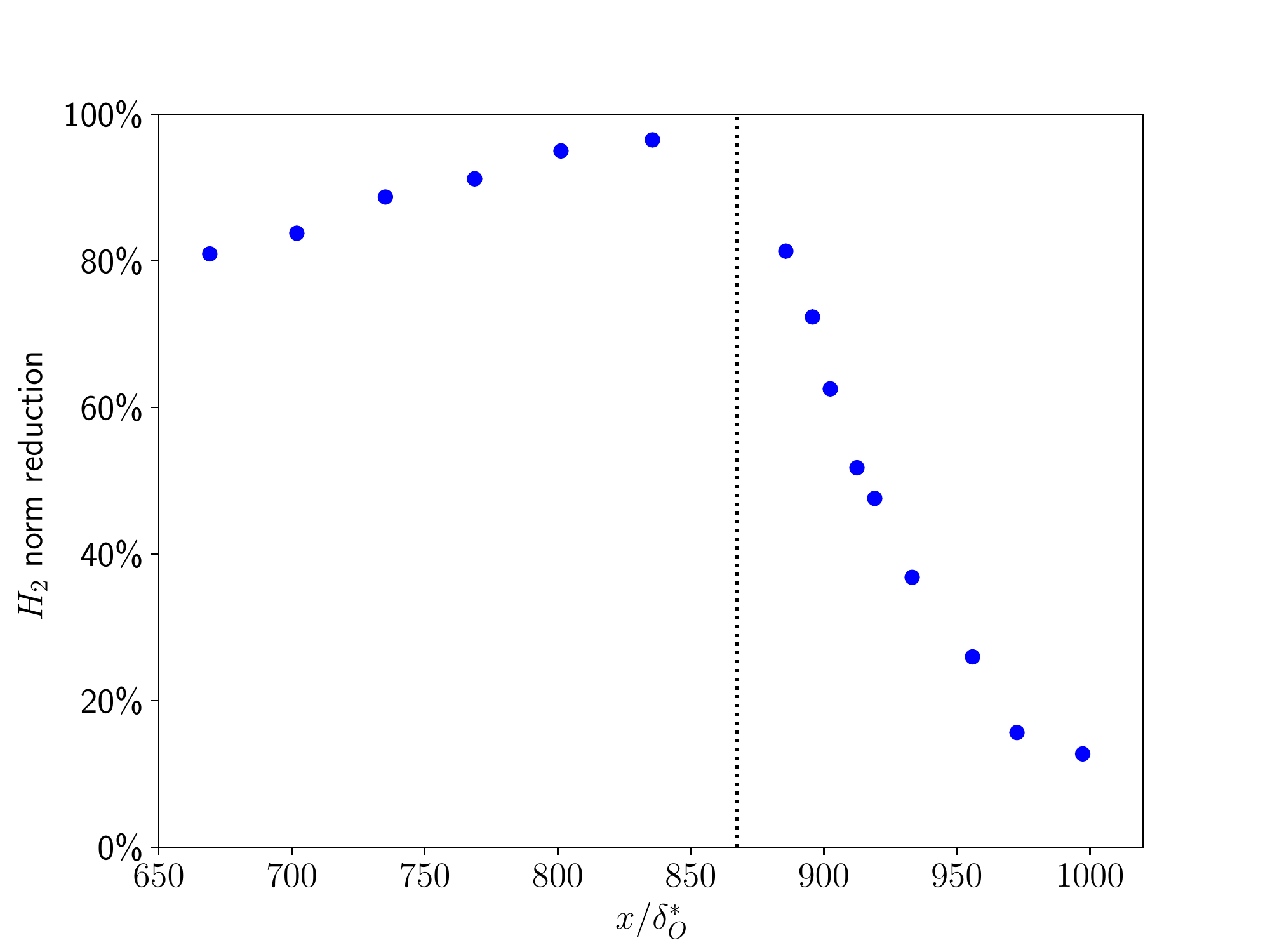}
\caption{Evolution of the maximum performance achievable on the ROM of the sensor \(z_{6}\) as a function of the position of the measurement sensor \(y\). The dotted line represents the actuator position; feedforward and feedback designs are respectively to the left and right of this dotted line.}
\label{fig:influence_y_max_perf}
\end{figure}

For the controller structure developed \S \ref{sec:synthesis_description_algo}, the constraint minimization problem (\ref{eq:min_problem}) is therefore written as
\begin{equation}
\label{eq:min_problem_bis}
\begin{split}
    &\text{minimize}\  ||T_{z_{6}w}^{c}||_{2}\\
    & \text{subject to}\ ||W_{S}S||_{\infty}<1.
\end{split}
\end{equation}
Figure \ref{fig:influence_y_max_perf} shows the evolution of the maximum performance achievable on the ROM of the performance sensor \(z_{6}\) as a function of the actuator/measurement sensor distance. The \(H_{2}\) norm reduction represents the quantity \((||T_{z_{6}w}^{c}||_{2}-||T_{z_{6}w}||_{2})/||T_{z_{6}w}||_{2}\). On the one hand, the actuator/measurement sensor distance influences very strongly the maximum performance achievable for feedback designs (to the right of the dotted line). On the other hand, feedforward designs (to the left of the dotted line) are relatively unaffected by this distance over a certain range and they perform better than feedback ones, which is consistent with the results of the incompressible literature \citep{Belson_art,Juillet_art,Freire_art}. The rapid drop in performance in the feedback cases is largely due to the delay in \(T_{yu}\) \citep{Skogestad_art,Skogestad_art_Best, Belson_art}, which is the time it takes for the wave generated by the actuator to arrive at the estimation sensor. Intuitively, to counteract efficiently disturbances of a wavelength \(2\pi/\widetilde{\alpha}_{r}\), the actuator/measurement sensor distance must be less than \(2\pi/\widetilde{\alpha}_{r}\). Consecutively, the frequency spectrum of the performance sensor \(z_{6}\) containing a significant amount of energy up to \(F\approx0.282\), this requires actuator/measurement sensor distance of less than \(2\pi \delta_{0}^{*}/F\sim20\delta_{0}^{*}\) in this case. Then, to obtain significant performance in terms of amplitude reduction, it is decided to place the sensor \(y_{\textit{fb}}\) at \(x_{\textit{fb}}=885.7\delta_{0}^{*}\) for the feedback configuration (at a distance of \(18.5\delta_{0}^{*}\) from the streamwise position of the actuator). With regard to the feedforward design, the sensor \(y_{\textit{ff}}\) is placed at \(x_{\textit{ff}}=801.2\delta_{0}^{*}\) (at a distance of \(66\delta_{0}^{*}\) from the streamwise position of the actuator), in such a way as to ensure that it is possible to disregard \(T_{yu}\) in the synthesis while having an optimal performance. 

\section{Evolution of performance as a function of the number of sensors \(z_{i}\) used in the synthesis}
\label{sec:number_perfo_sensor}
For the two sensor positions \(x_{\textit{fb}}\) and \(x_{\textit{ff}}\) determined previously, the following minimization problem is solved:
\begin{equation}
\label{eq:annex_min_problem}
\begin{split}
     \text{minimize} &\underset{z_{i} \in z_{used}}{\mathrm{max}}(||T_{z_{i}w}^{c}||_{2})\\
     &\text{subject to}\ ||W_{S}S||_{\infty}<1 \text{ and } ||W_{KS}KS||_{\infty}<1, 
\end{split}
\end{equation} 
with the controller structure developed in \S \ref{sec:synthesis_description_algo}. Not all the six performance sensors are necessarily used for the minimisation problem and the evolution of performance as a function of the number of sensors \(z_{i}\) (and by extension their positions) employed in the synthesis is assessed. The set of sensors \(z_{i}\) used in the synthesis is denoted \(z_{used}\). In table \ref{tab:labe_perfo_zsensor}, the different configurations tested are listed: the cases labelled 'Fb\(k\)\(z\)' (respectively 'Ff\(k\)\(z\)')  stand for feedback designs (respectively feedforward designs) with \(k\) performance sensors used in the synthesis; the  performance sensors used for each case are also given. Assuming that the transition to turbulence process begins shortly after the streamwise position of the actuator, it is chosen to scale the results by the local \(H_{2}\) norm of the uncontrolled system at the performance sensor \(z_{1}\), which is the closest performance sensor to the actuator. The maximum local \(H_{2}\) norm between the position of the sensors \(z_{1}\) and \(z_{6}\) (respectively the most upstream and the most downstream performance sensors used in some syntheses) for the controlled system is denoted \(\underset{x_{1}<x<x_{6}}{max}||T_{z(x)w}^ {c}||_{2}\).
\begin{table}
    \centering
 \begin{tabular}{cccc}
      Case
      &
      \makecell{Sensors z \\ used for synthesis}
      &
      \(\underset{z_{i} \in z_{used}}{max}\left(\frac{||T_{z_{i}w}^{c}||_{2}}{||T_{z_{1}w}||_{2}}\right)\)
      &
      \(\underset{x_{1}<x<x_{6}}{max}\left(\frac{||T_{z(x)w}^{c}||_{2}}{||T_{z_{1}w}||_{2}}\right)\) \\
      \midrule
      Without control
      &
      -
      &
      1.96
      &
      1.96\\
      \hline
      Fb1\(z\)
      &
      \(z_{used}=\{z_{6}\}\)
      &
      0.41
      &
      9.03\\
     Fb3\(z\)
      &
      \(z_{used}=\{z_{1}, z_{4}, z_{6}\}\)
      &
      0.67
      &
      1.98\\
    Fb4\(z\)
      &
      \(z_{used}=\{z_{1}, z_{2}, z_{3}, z_{6}\}\)
      &
      0.90
      &
      1.06\\
      Fb6\(z\)
      &
      \(z_{used}=\{z_{1}, z_{2}, z_{3}, z_{4}, z_{5}, z_{6}\}\)
      &
      0.92
      &
      0.92\\
	\hline      
     % \midrule
    Ff1\(z\)
      &
      \(z_{used}=\{z_{6}\}\)
      &
      0.14
      &
      2.48\\
    Ff6\(z\)
      &
      \(z_{used}=\{z_{1}, z_{2}, z_{3}, z_{4}, z_{5}, z_{6}\}\)
      &
      0.67
      &
      0.73\\
 \end{tabular}
    \caption{Evolution of the performance after the controllers are implemented in \textit{elsA} as a function of the number of sensors \(z_{i}\) used in the synthesis step. Cases labelled 'Fb\(k\)\(z\)' (respectively 'Ff\(k\)\(z\)') stand for feedback designs (respectively feedforward designs) with \(k\) performance sensors used in the synthesis. The results are normalized by the local \(H_{2}\) norm of the uncontrolled system at the position \(x_{1}\).}
    \label{tab:labe_perfo_zsensor}
\end{table}
\begin{figure}
    \centering
     \subfloat[\label{fig:evolution_z_fb}]{\includegraphics[scale=0.33]{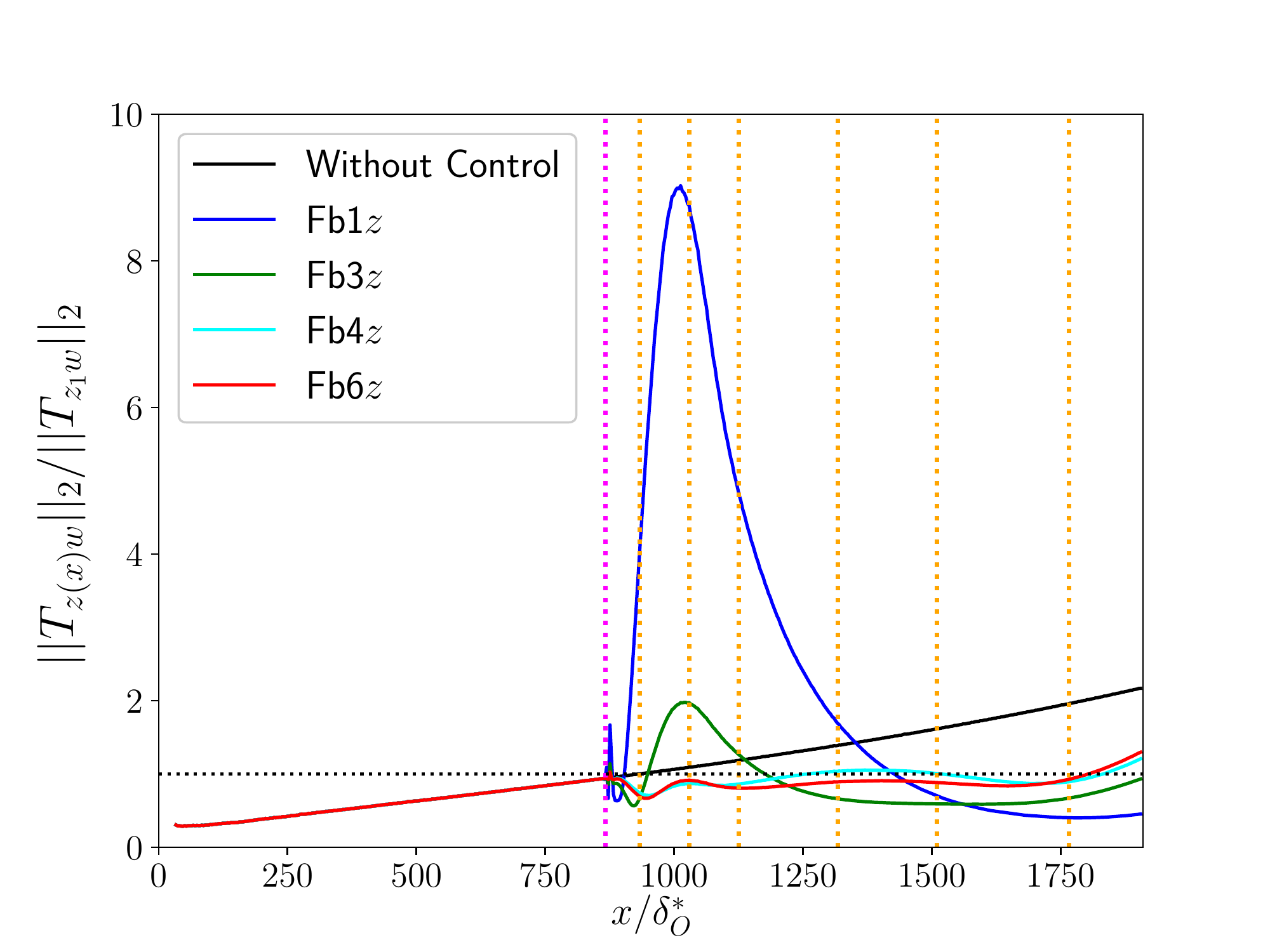}}
      \subfloat[\label{fig:Tzw_1sensorz}]{\includegraphics[scale=0.33]{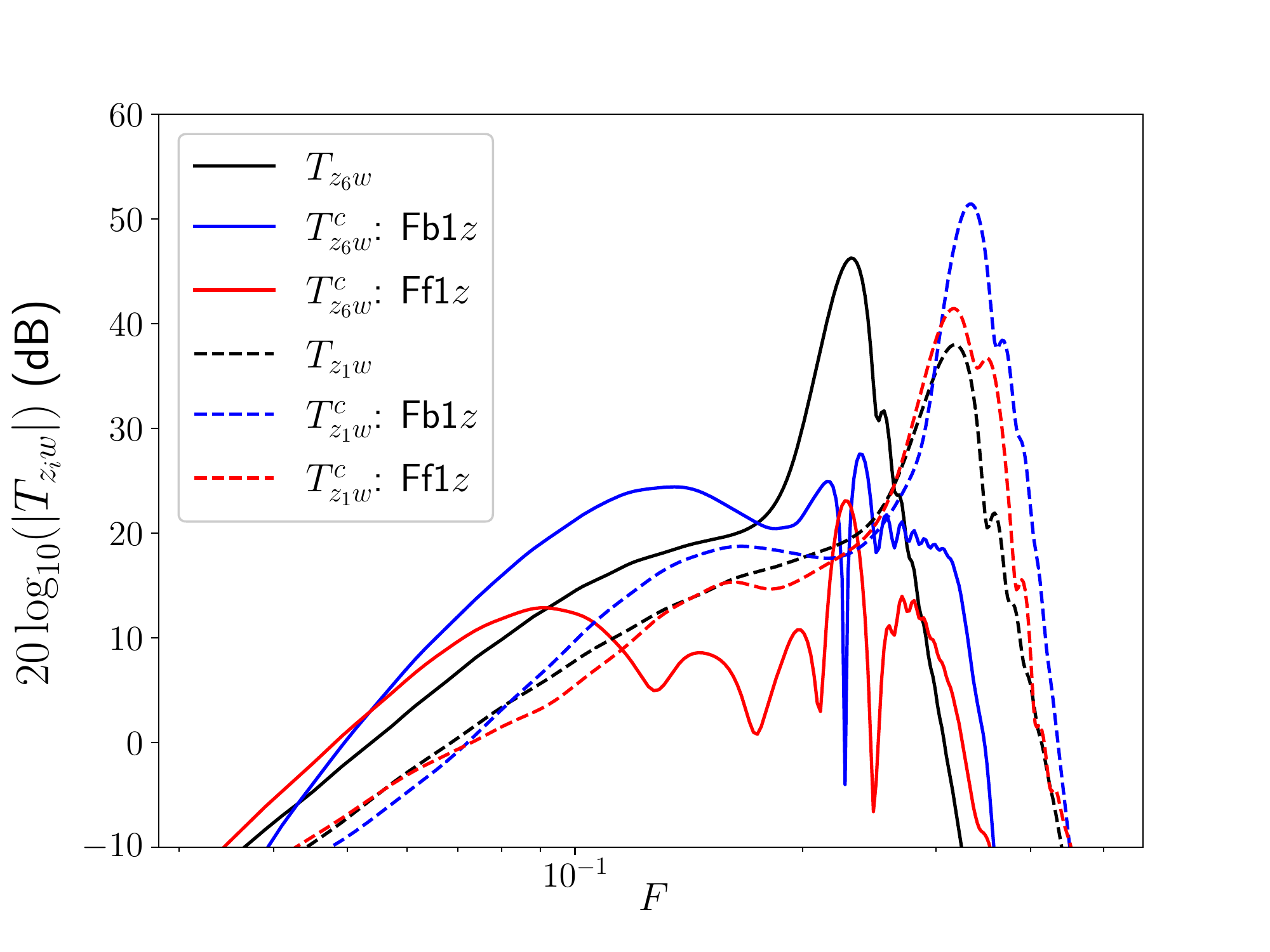}} \\
      \subfloat[\label{fig:sensitivity_compare}]{\includegraphics[scale=0.33]{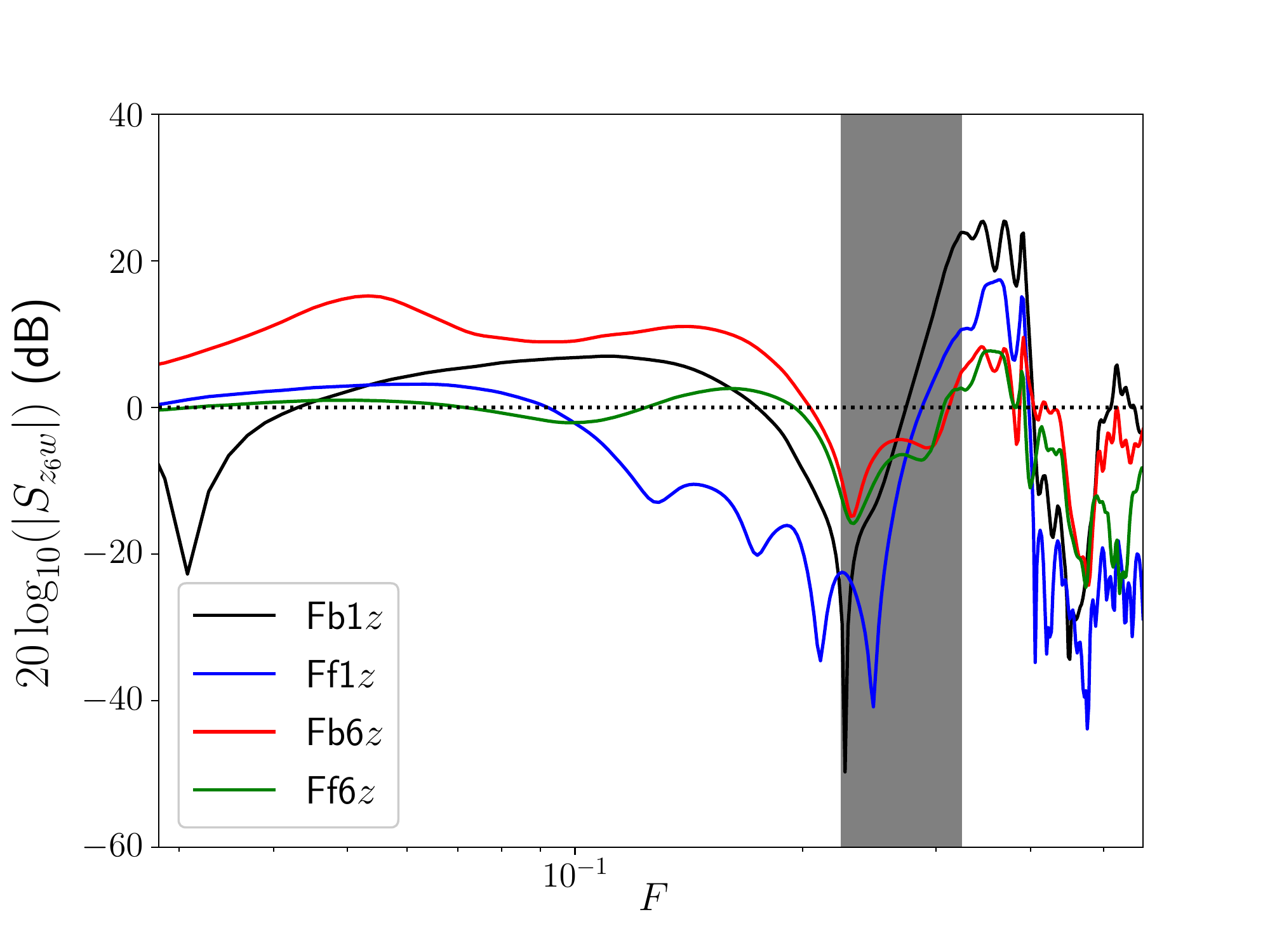}}
    \caption{a) Evolution of the local \(H_{2}\) norm of the transfer \(T_{z(x)w}\) from upstream noise \(w\) to wall-pressure fluctuation probes \(z(x)\). The vertical magenta and orange dotted lines represent, respectively, the position of the actuator (with the sensors \(y_{\textit{fb}}\) and \(y_{\textit{ff}}\) nearby) and the performance sensors \(z_{i}\) that can be used for synthesis. The values are normalized by \(||T_{z_{1}w}||_{2}\). b) Comparison of \(T_{z_{6}w}\) (solid lines) and \(T_{z_{1}w}\) (dashed lines) for the uncontrolled (black lines), Fb1\(z\) (blue lines) and Ff1\(z\) (red lines) cases. c) Comparison of the disturbance to performance attenuation criterion \(S_{z_{6}w}\) for different control cases. Disturbance rejection is improved (respectively degraded) below (respectively above) the dotted line. Grey shaded area represents the frequency bandwidth to be controlled from the actuator to the end of the domain.}
    \label{fig:Performance_z_evolution}
\end{figure}

The resulting controllers are then implemented in \textit{elsA} and we focus on the evolution of the local \(H_{2}\) norm of the transfers \(T_{z(x)w}\) at each abscissa of the plate. The evolution of the local \(H_{2}\) norm of the transfers \(T_{z(x)w}\) for the case without control and the  different feedback cases is depicted in figure \ref{fig:evolution_z_fb} (for feedforward cases, these results are summarised in table \ref{tab:labe_perfo_zsensor}). For the Fb1\(z\) and Ff1\(z\) cases, where the controller is designed to minimise the energy of the performance sensor \(z_{6}\), this results in a strong reduction of the local \(H_{2}\) norm at the end of the domain; in the feedback (respectively feedforward) configuration, \(||T_{z_{6}w}^{c}||_{2}\) is even about \(4.78\) (respectively \(14.\)) times lower than \(||T_{z_{6}w}||_{2}\). However, this significant decrease in energy downstream of the domain was accompanied by a strong increase in the local \(H_{2}\) norm upstream in the domain (blue solid line in figure \ref{fig:evolution_z_fb} for the feedback case). The quantity \(\underset{x_{1}<x<x_{6}}{max}||T_{z(x)w}^ {c}||_{2}\) for both feedforward and feedback configurations appears greater than the uncontrolled case; this increase of the local \(H_{2}\) norm may then lead to a faster transition to turbulence in a 3D setup, which is the opposite of the desired objective.

This increase of the local \(H_{2}\) norm can be explained from figure \ref{fig:Tzw_1sensorz}, which represents the module of \(T_{z_{6}w}\) (solid lines) and \(T_{z_{1}w}\) (dashed lines) for the uncontrolled (black lines), Fb1\(z\) (blue lines) and Ff1\(z\) (red lines) cases. On the one hand, the amplitudes of the dominant frequencies of the uncontrolled system for the sensor \(z_{6}\) (which is the only one used in the synthesis for these cases) are significantly reduced both in feedback and feedforward cases, which partly explains the significant reduction in the \(H_{2}\) norm for this transfer. On the other hand, the amplitudes of the dominant frequencies for the sensor \(z_{1}\) are amplified by both feedback and feedforward designs, leading to an increase of the \(H_{2}\) norm for this transfer and thus an amplification upstream of the domain.

Indeed, reducing the amplitude of disturbances in one part of the frequency spectrum can lead to increasing it in the other part, which could predominate in other abscissas of the domain. Figure \ref{fig:sensitivity_compare} shows the frequency spectrum of the disturbance to performance attenuation criterion, defined as \(S_{z_{i}w}=\frac{T_{z_{i}w}^c}{T_{z_{i}w}}\). For a sensor \(z_{i}\), disturbance rejection is achieved at frequencies where \(|S_{z_{i}w}|<1\). We can see from this figure that an effect similar to the waterbed effect \citep{Skogestad_art} appears: for the Fb1\(z\) and Ff1\(z\) cases, the significant disturbance rejection at frequencies around \(F=0.225\) is accompanied by an amplification for higher and lower frequencies. The frequency bandwidth to be controlled being around \(F \in [0.225,0.324]\) (see figure \ref{fig:Tzwx_FactorN}), amplifying lower frequencies is not a problem in our case as these will be found further downstream of \(z_{6}\) and therefore not taken into account in the computational domain. However, amplifying frequencies around \(F=0.324\) will directly impact performance on the sensor \(z_{1}\) which is dominated by these frequencies. This translates into the need to use several sensors \(z_{i}\) in the synthesis to obtain a suitable frequency representation in different abscissas of the domain to avoid an unwanted waterbed effect. Both Fb6\(z\) and Ff6\(z\) cases have lower disturbance rejection at frequencies around \(F=0.225\) but the waterbed effect on high frequencies is mitigated compared to Fb1\(z\) and Ff1\(z\) cases (see figure \ref{fig:sensitivity_compare}). By taking more and more performance sensors along the plate for the synthesis, we cover a wider spectrum of amplified frequencies. The larger the frequency bandwidth to be rejected, the more complicated it is to obtain very high attenuation on the spectrum. This is why the quantity \(\underset{z_{i} \in z_{used}}{max}||T_{z_{i}w}^{c}||_{2}\) increases with the number of performance sensors used in the synthesis (see table \ref{tab:labe_perfo_zsensor}). Nevertheless, due to the better coverage of amplified frequencies by increasing the number of \(z_{i}\) used in the synthesis, a more uniform performance along the plate is obtained (see table \ref{tab:labe_perfo_zsensor} and figure \ref{fig:evolution_z_fb}).

By taking three performance sensors (one near the actuator, one near the end of the domain and an other in between) and thus covering a wider frequency spectrum, the Fb3\(z\) case (green line in figure \ref{fig:evolution_z_fb}) allows to significantly reduce the local \(H_{2}\) norm increase near the actuator compared to the Fb1\(z\). However, immediately after the position of the sensor \(z_{1}\) (first vertical orange line), yet taken into account in this synthesis, the local \(H_{2}\) norm increases and a slight bump appears in \(x\approx1020\delta_{0}^{*}\). It is associated with strong non-modal effects in the vicinity of the actuator (see \S \ref{sec:Actuator_description}). For frequencies around \(F=0.296\), those dominant in the vicinity of the actuator, the modal behaviour is only found for \(x\gtrapprox1136.4\delta_{0}^{*}\) (see figure \ref{fig:non_modal_TzuTzw}). Therefore, we need to discretise the area from the actuator to the end of the transient non-modal region with several performance sensors as in the Fb4\(z\) and Fb6\(z\) cases. Because \(\underset{x_{1}<x<x_{6}}{max}||T_{z(x)w}^{c}||_{2}\) is lower in the Fb6\(z\) case than in the Fb4\(z\) one due to a better coverage of the amplified frequency spectrum along the plate, six performance sensors are therefore used in the syntheses of \S \ref{sec:Result_ff_vs_fb}.

\section{Impact of \(W_{KS}\) on the performance with noisy estimation sensors}
\label{sec:Impact_WKS_noisy_sensors}
\begin{figure}
    \centering
     %% cc pour dire deux colonnes CENTRE ll pour deux colonnes alignes a gauche er rr pour droite
     %\begin{tabular}{cc}
      %\hspace*{-1cm}
      \subfloat[\label{fig:KS_old}]{\includegraphics[scale=0.44]{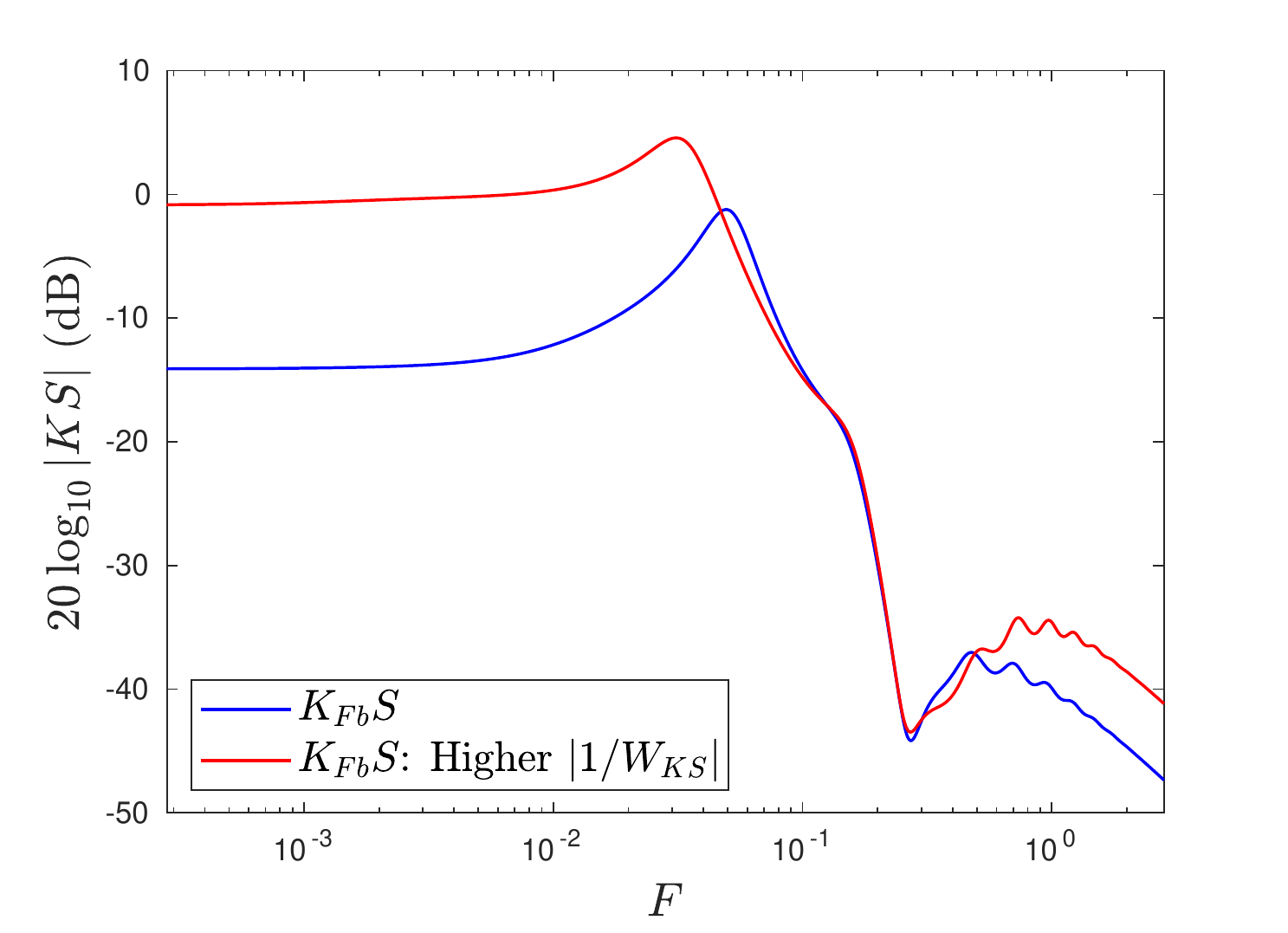}}
      %&
      %\hspace*{-1cm}
     \subfloat[\label{fig:PSDu_Old}]{\includegraphics[scale=0.33]{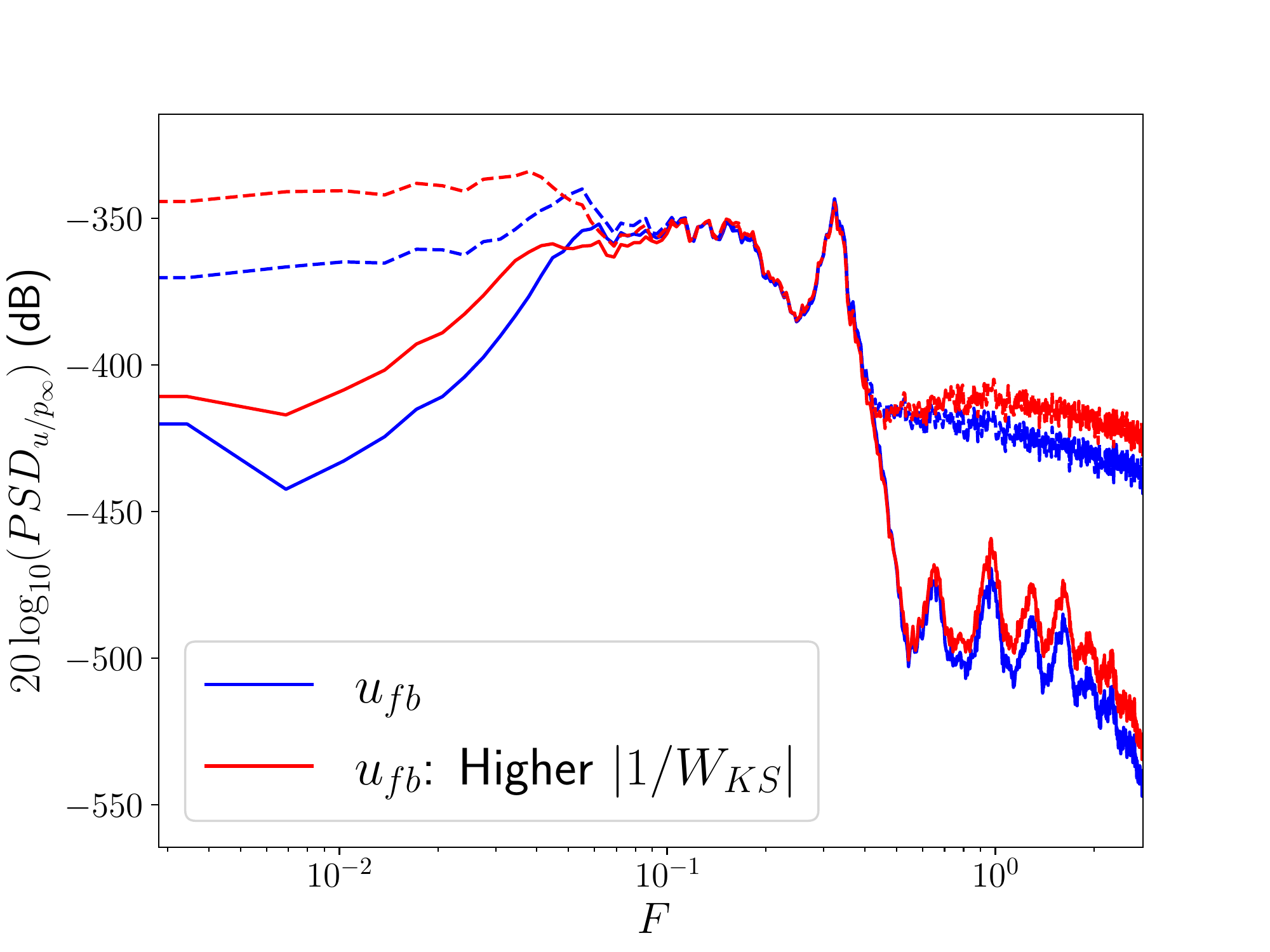}} \\
      %\subfloat[\label{fig:Champ_u_Old}]{\includegraphics[scale=0.21]{IMAGE/Urms_JFM_Noise_old-tiff-converted-to.png}} 
      %&
    \subfloat[\label{fig:maxu_Old}]{\includegraphics[scale=0.33]{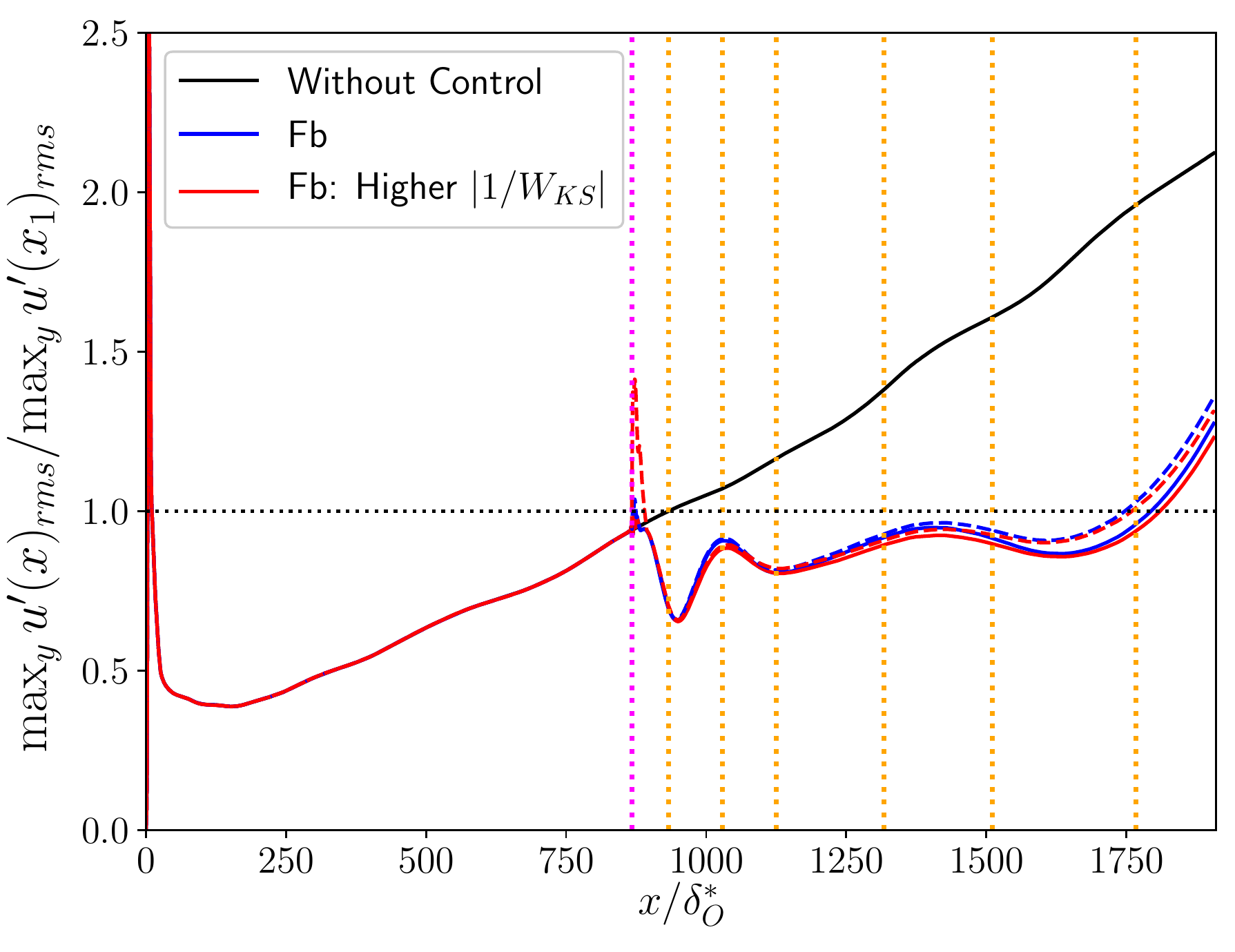}}
     %\end{tabular}
    \caption{Comparison of \(KS\) (a), \(PSD_{u}\) (b) and \(\mathrm{max}_{y}\ u'_{rms}\) (c) for a feedback controller resulting from a synthesis with an higher \(|1/W_{KS}|\) (red lines) than for the one used in \S \ref{sec:Result_ff_vs_fb} (blue lines). Solid lines and dashed lines in (b) and (c) represent the cases with ideal and noisy estimation sensors, respectively. The legend in (c) is the same as in figure \ref{fig:h2_final}.}
    \label{fig:urms_noise_sensor_old}
\end{figure}

To illustrate the impact of the weighting function \(W_{KS}\) on the performance, the constraint minimization problem (\ref{eq:min_problem}) is solved but with an higher \(|1/W_{KS}|\) compared to the one use all along \S \ref{sec:Result_ff_vs_fb}. The feedback controller resulting from this synthesis (red lines) is shown in figure \ref{fig:KS_old} and is compared to the previous one used in \S \ref{sec:Result_ff_vs_fb} (blue lines). The two controllers have globally the same behaviour in the frequency bandwidth of the second Mack mode, but the new controller has higher gain in low frequency bandwidth. In the case where the estimation sensors are corrupted by the same amount of white Gaussian noise as in \S \ref{sec:perfo_robustness} (\(50\%\) of the r.m.s. value without control action of \(y_{\textit{fb}}\)), it follows that the $u$-PSD for a corrupted signal \(y\) becomes more important in low frequencies for the controller resulting from the synthesis with an higher \(|1/W_{KS}|\) than for the previous controller (see figure \ref{fig:PSDu_Old}). For the noisy estimation sensor case and contrary to the controller used all along \S \ref{sec:Result_ff_vs_fb}, the new controller leads to a strong energy injection in the vicinity of the actuator (see red dashed line in figure \ref{fig:maxu_Old}). As these injected low frequencies are convectively stable, they attenuate very quickly but the maximum along the wall-normal direction of \(u'_{rms}\) clearly exceeds the energy threshold before the last performance sensor \(z_{i}\) used in the synthesis, which could trigger the transition to turbulence in a 3D configuration. It should be noted that in the case of ideal estimation sensors, the controller resulting from the synthesis with an higher \(|1/W_{KS}|\) minimizes slightly more the velocity fluctuations compared to the previous feedback controller used in \S \ref{sec:Result_ff_vs_fb} (see solid lines in figure \ref{fig:maxu_Old}) because the constraint on \(W_{KS}\) is less important. There is therefore a trade-off between minimizing \(H_{2}\) norms and desensitize the controller in low frequency range during the synthesis. 
  
\bibliographystyle{jfm}
\bibliography{jfm-instructions}

\begin{thebibliography}{76}
\expandafter\ifx\csname natexlab\endcsname\relax\def\natexlab#1{#1}\fi
\def\au#1{#1} \def\ed#1{#1} \def\yr#1{#1}\def\at#1{#1}\def\jt#1{\textit{#1}}
  \def\bt#1{#1}\def\bvol#1{\textbf{#1}} \def\vol#1{#1} \def\pg#1{#1}
  \def\publ#1{#1}\def\arxiv#1{#1}\def\org#1{#1}\def\st#1{\textit{#1}}

\bibitem[Amestoy {\em et~al.\/}(2001)Amestoy, Duff, L'Excellent \&
  Koster]{MUMPS_art}
{\sc \au{Amestoy, P.R.}, \au{Duff, I.S.}, \au{L'Excellent, J.-Y.} \&
  \au{Koster, J.}} \yr{2001}  \at{A fully asynchronous multifrontal solver
  using distributed dynamic scheduling}.  \jt{SIAM}  \bvol{23}.

\bibitem[Apkarian {\em et~al.\/}(2014)Apkarian, Gahinet \&
  Buhr]{apkarian_systune_art}
{\sc \au{Apkarian, P.}, \au{Gahinet, P.} \& \au{Buhr, C.}} \yr{2014}
  Multi-model, multi-objective tuning of fixed-structure controllers.  \bt{In
  {\em 2014 European Control Conference (ECC)\/}},  \pg{pp. 856--861}.

\bibitem[Apkarian \& Noll(2006)]{apkarian_hinfstruct}
{\sc \au{Apkarian, P.} \& \au{Noll, D.}} \yr{2006}  \at{Nonsmooth
  \({H}_{\infty}\) synthesis}.  \jt{IEEE Transactions on Automatic Control}
  \bvol{51},  \pg{71--86}.

\bibitem[Apkarian {\em et~al.\/}(2010)Apkarian, Noll \&
  Rondepierre]{apkarian_systune_art2}
{\sc \au{Apkarian, P.}, \au{Noll, D.} \& \au{Rondepierre, A.}} \yr{2010} Mixed
  \({H}_{2}\)/\({H}_{\infty}\) control via nonsmooth optimization.  \bt{In {\em
  Proceedings of the IEEE Conference on Decision and Control\/}}, ,
  \vol{vol.~47},  \pg{pp. 6460 -- 6465}.

\bibitem[Bagheri {\em et~al.\/}(2009)Bagheri, Brandt \&
  Henningson]{Bagheri_art}
{\sc \au{Bagheri, S.}, \au{Brandt, L.} \& \au{Henningson, D.S.}} \yr{2009}
  \at{Input-output analysis, model reduction and control of the flat-plate
  boundary layer}.  \jt{J. Fluid Mech.}  \bvol{620},  \pg{263--298}.

\bibitem[Barbagallo {\em et~al.\/}(2012)Barbagallo, Dergham, Sipp, Schmid \&
  Robinet]{Barbagallo_art}
{\sc \au{Barbagallo, A.}, \au{Dergham, G.}, \au{Sipp, D.}, \au{Schmid, P.J.} \&
  \au{Robinet, J.-C.}} \yr{2012}  \at{Closed-loop control of unsteadiness over
  a rounded backward-facing step}.  \jt{J.~Fluid Mech.}  \bvol{703},
  \pg{326--362}.

\bibitem[Barbagallo {\em et~al.\/}(2009)Barbagallo, Sipp \&
  Schmid]{Barbagallo_citation_art}
{\sc \au{Barbagallo, A.}, \au{Sipp, D.} \& \au{Schmid, P.J.}} \yr{2009}
  \at{Closed-loop control of an open cavity flow using reduced-order models}.
  \jt{J.~Fluid Mech.}  \bvol{641},  \pg{1--50}.

\bibitem[Belson {\em et~al.\/}(2013)Belson, Semeraro, Rowley \&
  Henningson]{Belson_art}
{\sc \au{Belson, B.A.}, \au{Semeraro, O.}, \au{Rowley, C.W.} \& \au{Henningson,
  D.S.}} \yr{2013}  \at{Feedback control of instabilities in the
  two-dimensional blasius boundary layer: The role of sensors and actuators}.
  \jt{Physics of Fluids}  \bvol{25},  \pg{054106}.

\bibitem[Beneddine {\em et~al.\/}(2015)Beneddine, Mettot \&
  Sipp]{Beneddine_art}
{\sc \au{Beneddine, S.}, \au{Mettot, C.} \& \au{Sipp, D.}} \yr{2015}
  \at{Global stability analysis of underexpanded screeching jets}.
  \jt{European Journal of Mechanics B/Fluids}  \bvol{49},  \pg{392--399}.

\bibitem[Bugeat {\em et~al.\/}(2019)Bugeat, Chassaing, Robinet \&
  Sagaut]{Bugeat_art}
{\sc \au{Bugeat, B.}, \au{Chassaing, J.-C.}, \au{Robinet, J.-C.} \& \au{Sagaut,
  P.}} \yr{2019}  \at{{3D} global optimal forcing and response of the
  supersonic boundary layer}.  \jt{J. Comput. Phys.}  \bvol{398},  \pg{108888}.

\bibitem[Cambier {\em et~al.\/}(2013)Cambier, Heib \& Plot]{Elsa_art}
{\sc \au{Cambier, L.}, \au{Heib, S.} \& \au{Plot, S.}} \yr{2013}  \at{The
  {Onera} {elsA} {CFD} sofware: input from research and feedback from
  industry}.  \jt{Mechanics \& Industry}  \bvol{14},  \pg{159--174}.

\bibitem[Chen {\em et~al.\/}(1994)Chen, Zhou \& Chang]{Reduction_Chen}
{\sc \au{Chen, J.}, \au{Zhou, K.} \& \au{Chang, B.-C.}} \yr{1994} Closed-loop
  controller reduction by a structured truncation approach.  \bt{In {\em
  Proceedings of 1994 33rd IEEE Conference on Decision and Control\/}}, ,
  \vol{vol.~3},  \pg{pp. 2726--2731}.

\bibitem[Dadfar {\em et~al.\/}(2014)Dadfar, Fabbiane, Bagheri \&
  Henningson]{Dadfar_centralized_art}
{\sc \au{Dadfar, R.}, \au{Fabbiane, N.}, \au{Bagheri, S.} \& \au{Henningson,
  D.S.}} \yr{2014}  \at{Centralised versus decentralised active control of
  boundary layer instabilities}.  \jt{Flow, Turbulence and Combustion}
  \bvol{93},  \pg{537--553}.

\bibitem[Dadfar {\em et~al.\/}(2013)Dadfar, Semeraro, Hanifi \&
  Henningson]{Dadfar2D_art}
{\sc \au{Dadfar, R.}, \au{Semeraro, O.}, \au{Hanifi, A.} \& \au{Henningson,
  D.S.}} \yr{2013}  \at{Output feedback control of blasius flow with leading
  edge using plasma actuator}.  \jt{AIAA Journal}  \bvol{51},  \pg{2192--2207}.

\bibitem[Doyle(1978)]{DoyleLQG_art}
{\sc \au{Doyle, J.}} \yr{1978}  \at{Guaranteed margins for {LQG} regulators}.
  \jt{IEEE Transactions on Automatic Control}  \bvol{23},  \pg{756--757}.

\bibitem[Doyle {\em et~al.\/}(1989)Doyle, Glover, Khargonekar \&
  Francis]{Doyle2_art}
{\sc \au{Doyle, J.}, \au{Glover, K.}, \au{Khargonekar, P.P.} \& \au{Francis,
  B.A.}} \yr{1989}  \at{State-space solutions to standard \({H}_{2}\) and
  \({H}_{\infty}\) control problems}.  \jt{IEEE Transactions on Automatic
  Control}  \bvol{34},  \pg{831--847}.

\bibitem[Doyle \& Stein(1981)]{LTR_Doyle}
{\sc \au{Doyle, J.} \& \au{Stein, G.}} \yr{1981}  \at{Multivariable feedback
  design: Concepts for a classical/modern synthesis}.  \jt{IEEE Transactions on
  Automatic Control}  \bvol{26},  \pg{4--16}.

\bibitem[Drmac {\em et~al.\/}(2015)Drmac, Gugercin \& Beattie]{tfest}
{\sc \au{Drmac, Z.}, \au{Gugercin, S.} \& \au{Beattie, C.}} \yr{2015}
  \at{Quadrature-based vector fitting for discretized \({H}_{2}\)
  approximation.}  \jt{SIAM Journal on Scientific Computing}  \bvol{2},
  \pg{A625--52}.

\bibitem[Erdmann {\em et~al.\/}(2011)Erdmann, P{\"a}tzold, Engert, Peltzer \&
  Nitsche]{erdmann2011active}
{\sc \au{Erdmann, R.}, \au{P{\"a}tzold, A.}, \au{Engert, M.}, \au{Peltzer, I.}
  \& \au{Nitsche, W.}} \yr{2011}  \at{On active control of laminar--turbulent
  transition on two-dimensional wings}.  \jt{Philosophical Transactions of the
  Royal Society A: Mathematical, Physical and Engineering Sciences}
  \bvol{369}~(1940),  \pg{1382--1395}.

\bibitem[Fabbiane {\em et~al.\/}(2014)Fabbiane, Semeraro, Bagheri \&
  Henningson]{Fabbiane_review_art}
{\sc \au{Fabbiane, N.}, \au{Semeraro, O.}, \au{Bagheri, S.} \& \au{Henningson,
  D.S.}} \yr{2014}  \at{Adaptive and model-based control theory applied to
  convectively unstable flows}.  \jt{Applied Mechanics Reviews}  \bvol{66},
  \pg{60801}.

\bibitem[Fabbiane {\em et~al.\/}(2015)Fabbiane, Simon, Fischer, Grundmann,
  Bagheri \& Henningson]{Fabbiane_JFMrapids_art}
{\sc \au{Fabbiane, N.}, \au{Simon, B.}, \au{Fischer, F.}, \au{Grundmann, S.},
  \au{Bagheri, S.} \& \au{Henningson, D.s.}} \yr{2015}  \at{On the role of
  adaptivity for robust laminar flow control}.  \jt{J.~Fluid Mechanics}
  \bvol{767}.

\bibitem[Fedorov(2011)]{Fedorov_art}
{\sc \au{Fedorov, A.}} \yr{2011}  \at{Transition and stability of high-speed
  boundary layers}.  \jt{Annu. Rev. Fluid Mech.}  \bvol{43},  \pg{79--95}.

\bibitem[Fedorov \& Tumin(2022)]{Fedo_Mack_H2}
{\sc \au{Fedorov, A.} \& \au{Tumin, A.}} \yr{2022}  \at{The mack's amplitude
  method revisited}.  \jt{Theor. Comput. Fluid Dyn.}  \bvol{36},  \pg{9--24}.

\bibitem[Flinois \& Morgans(2016)]{Flinois_art}
{\sc \au{Flinois, T.L.B.} \& \au{Morgans, A.S.}} \yr{2016}  \at{Feedback
  control of unstable flows: a direct modelling approach using the eigensystem
  realisation algorithm}.  \jt{J.~Fluid Mech.}  \bvol{793},  \pg{41--78}.

\bibitem[Franklin {\em et~al.\/}(1997)Franklin, Powell \& Workman]{book_foh}
{\sc \au{Franklin, G.F.}, \au{Powell, J.D.} \& \au{Workman, M.L.}} \yr{1997}
  {\em Digital Control of Dynamic Systems-Third Edition\/}.  \publ{Prentice
  Hall}.

\bibitem[Freire {\em et~al.\/}(2020)Freire, Cavalieri, Silvestre, Hanifi \&
  Henningson]{Freire_art}
{\sc \au{Freire, G.A.}, \au{Cavalieri, A.V.G}, \au{Silvestre, F.J.},
  \au{Hanifi, A.} \& \au{Henningson, D.S.}} \yr{2020}  \at{Actuator and sensor
  placement for closed-loop control of convective instabilities}.  \jt{Theor.
  Comput. Fluid Dyn.}  \bvol{34},  \pg{619--641}.

\bibitem[{Gad-el-Hak}(2000)]{gad_flow_book}
{\sc \au{{Gad-el-Hak}, M.}} \yr{2000} {\em Flow Control: Passive, Active, and
  Reactive Flow Management\/}.  \publ{Cambridge University Press}.

\bibitem[Gaponov \& Smorodsky(2016)]{Gaponov_art}
{\sc \au{Gaponov, S.A.} \& \au{Smorodsky, B.V.}} \yr{2016}  \at{Supersonic
  turbulent boundary layer drag control using spanwise wall oscillation}.
  \jt{International Journal of Theoretical and Applied Mechanics}  \bvol{1},
  \pg{97--103}.

\bibitem[Gear(1971)]{gear_book}
{\sc \au{Gear, C.W.}} \yr{1971} {\em Numerical Initial Value Problems in
  Ordinary Differential Equations\/}.  \publ{Prentice-Hall}.

\bibitem[Goddard \& Glover(1995)]{Reduction_perfo_Goddard}
{\sc \au{Goddard, P.~J.} \& \au{Glover, K.}} \yr{1995} Performance-preserving
  controller approximation.

\bibitem[Hanifi {\em et~al.\/}(1996)Hanifi, Schmid \& Henningson]{Hanifi_art}
{\sc \au{Hanifi, A.}, \au{Schmid, P.J.} \& \au{Henningson, D.S.}} \yr{1996}
  \at{Transient growth in compressible boundary layer flow}.  \jt{Physics of
  Fluids}  \bvol{8},  \pg{826}.

\bibitem[Herv{\'e} {\em et~al.\/}(2012)Herv{\'e}, Sipp, Schmid \&
  Samuelides]{Herve_art}
{\sc \au{Herv{\'e}, A.}, \au{Sipp, D.}, \au{Schmid, P.J.} \& \au{Samuelides,
  M.}} \yr{2012}  \at{A physics-based approach to flow control using system
  identification}.  \jt{J.~Fluid Mech.}  \bvol{702},  \pg{26--58}.

\bibitem[Huerre \& Monkewitz(1990)]{Huerre_art}
{\sc \au{Huerre, P.} \& \au{Monkewitz, P.~A.}} \yr{1990}  \at{Local and global
  instabilities in spatially developing flows}.  \jt{Annu. Rev. Fluid Mech.}
  \bvol{22},  \pg{473--537}.

\bibitem[Jahanbakhshi \& Zaki(2021)]{Jahanbakhshi_control_art}
{\sc \au{Jahanbakhshi, R.} \& \au{Zaki, T.A}} \yr{2021}  \at{Optimal heat flux
  for delaying transition to turbulence in a high-speed boundary layer}.
  \jt{J.~Fluid Mech.}  \bvol{916}.

\bibitem[Juang \& Pappa(1985)]{juang_pappa_art}
{\sc \au{Juang, J.-N.} \& \au{Pappa, R.S.}} \yr{1985}  \at{An eigensystem
  realization algorithm for modal parameter identification and model
  reduction}.  \jt{J. Guid. Control Dyn.}  \bvol{8},  \pg{620--627}.

\bibitem[Juillet {\em et~al.\/}(2013)Juillet, Schmid \& Huerre]{Juillet_art}
{\sc \au{Juillet, F.}, \au{Schmid, P.J.} \& \au{Huerre, P.}} \yr{2013}
  \at{Control of amplifier flows using subspace identification techniques}.
  \jt{J.~Fluid Mech.}  \bvol{725},  \pg{522--565}.

\bibitem[Juliano \& Borg(2015)]{Juliano_art}
{\sc \au{Juliano, T.J.} \& \au{Borg, M.P.}} \yr{2015}  \at{Quiet tunnel
  measurements of hifire-5 boundary-layer transition}.  \jt{AIAA Journal}
  \bvol{53},  \pg{1980--1993}.

\bibitem[Kalman(1964)]{Kalman64}
{\sc \au{Kalman, R.}} \yr{1964}  \at{When is a linear control system optimal}.
  \jt{Journal of Basic Engineering}  \bvol{86},  \pg{51--60}.

\bibitem[Kendall(1975)]{Kendall_art}
{\sc \au{Kendall, J.M.}} \yr{1975}  \at{Wind tunnel experiments relating to
  supersonic and hypersonic boundary-layer transition}.  \jt{AIAA Journal}
  \bvol{13},  \pg{290}.

\bibitem[Kwakernaak(1969)]{LTR_Kwakernaak}
{\sc \au{Kwakernaak, H.}} \yr{1969}  \at{Optimal low-sensitivity linear
  feedback systems}.  \jt{Automatica}  \bvol{5},  \pg{279–285}.

\bibitem[Leclercq {\em et~al.\/}(2019)Leclercq, Demourant, Poussot-Vassal \&
  Sipp]{Leclercq_art}
{\sc \au{Leclercq, C.}, \au{Demourant, F.}, \au{Poussot-Vassal, C.} \&
  \au{Sipp, D.}} \yr{2019}  \at{Linear iterative method for closed-loop control
  of quasiperiodic flows}.  \jt{J.~Fluid Mech.}  \bvol{868},  \pg{22--65}.

\bibitem[Leer(1979)]{MUSCL_art}
{\sc \au{Leer, B.~Van}} \yr{1979}  \at{Towards the ultimate conservative
  difference scheme. {V. A} second-order sequel to godunov's method}.  \jt{J.
  Comput. Phys.}  \bvol{32},  \pg{101--136}.

\bibitem[Lehoucq {\em et~al.\/}(1998)Lehoucq, Sorensen \& Yang]{ARPACK_art}
{\sc \au{Lehoucq, R.}, \au{Sorensen, D.} \& \au{Yang, C.}} \yr{1998}
  \at{Arpack users' guide: Solution of large scale eigenvalue problems with
  implicitly restarted arnoldi methods}.  \jt{SIAM}  \bvol{6}.

\bibitem[Liou(2006)]{AUSM_art}
{\sc \au{Liou, M.-S.}} \yr{2006}  \at{A sequel to {AUSM}, {Part II}:
  Ausm\(^{+}\)-up for all speeds}.  \jt{J. Comput. Phys.}  \bvol{214},
  \pg{137--170}.

\bibitem[Lugrin {\em et~al.\/}(2022)Lugrin, Nicolas, Severac, Tobeli,
  Beneddine, Garnier, Esquieu \& Bur]{Lugrin_expe}
{\sc \au{Lugrin, M.}, \au{Nicolas, F.}, \au{Severac, N.}, \au{Tobeli, J.-P.},
  \au{Beneddine, S.}, \au{Garnier, E.}, \au{Esquieu, S.} \& \au{Bur, R.}}
  \yr{2022}  \at{Transitional shockwave/boundary layer interaction experiments
  in the r2ch blowdown wind tunnel}.  \jt{Experiments in Fluids}  \bvol{63}.

\bibitem[Ma \& Zhong(2003)]{Ma_Zhong_art}
{\sc \au{Ma, Y.} \& \au{Zhong, X.}} \yr{2003}  \at{Receptivity of a supersonic
  boundary layer over a flat plate. part 1. wave structures and interactions}.
  \jt{J.~Fluid Mech.}  \bvol{488},  \pg{31--78}.

\bibitem[Mack(1977)]{mack_1977_N}
{\sc \au{Mack, L.M.}} \yr{1977} {\em Transition and laminar instability\/}.
  NASA-CP-153203.

\bibitem[Mack(1984)]{mack_1984_art}
{\sc \au{Mack, L.M.}} \yr{1984} {\em Boundary-Layer Linear Stability Theory\/}.
  AGARD Report No.709.

\bibitem[Malik(1989)]{malick_89_art}
{\sc \au{Malik, M.R.}} \yr{1989}  \at{Prediction and control of transition in
  supersonic and hypersonic boundary layers}.  \jt{AIAA Journal}  \bvol{27},
  \pg{1487--1493}.

\bibitem[McKelvey \& Helmersson(1996)]{tridiagonal}
{\sc \au{McKelvey, T.} \& \au{Helmersson, A.}} \yr{1996} State-space
  parametrizations of multivariable linear systems using tridiagonal matrix
  forms.  \bt{In {\em Proceedings of 35th IEEE Conference on Decision and
  Control\/}}, ,  \vol{vol.~4},  \pg{pp. 3654--3659}.

\bibitem[Morkovin(1969)]{Morkovin_art}
{\sc \au{Morkovin, M.V.}} \yr{1969}  \at{{On the Many Faces of Transition}}.
  \bt{In {\em Viscous Drag Reduction\/} (ed. \ed{C.~Sinclair Wells})}.
  \publ{Springer US}.

\bibitem[Morra {\em et~al.\/}(2020)Morra, Sasaki, Hanifi, Cavalieri \&
  Henningson]{Morra_art}
{\sc \au{Morra, P.}, \au{Sasaki, K.}, \au{Hanifi, A.}, \au{Cavalieri, A.V.G.}
  \& \au{Henningson, D.S.}} \yr{2020}  \at{A realizable data-driven approach to
  delay bypass transition with control theory}.  \jt{J.~Fluid Mech.}
  \bvol{883},  \pg{A33}.

\bibitem[Olazabal-Loume {\em et~al.\/}(2017)Olazabal-Loume, Danvin, Mathiaud \&
  Aupoix]{Clicet_art}
{\sc \au{Olazabal-Loume, M.}, \au{Danvin, F.}, \au{Mathiaud, J.} \& \au{Aupoix,
  B.}} \yr{2017} Study on k-$\omega$ shear stress transport model corrections
  applied to rough wall turbulent hypersonic boundary layers.  \bt{In {\em
  $7^{Th}$ European Conference for Aeronautics and Space Sciences\/}}.

\bibitem[Orr(1907)]{art_Orr}
{\sc \au{Orr, W.F.}} \yr{1907}  \at{The stability or instability of the steady
  motions of a perfect liquid and of a viscous liquid. part ii: A viscous
  liquid}.  \jt{Proceedings of the Royal Irish Academy. Section A: Mathematical
  and Physical Sciences}  \bvol{27},  \pg{69--138}.

\bibitem[Saint-James(2020)]{thesis_saintjames}
{\sc \au{Saint-James, J.}} \yr{2020}  \at{{Pr{\'e}vision de la transition
  laminaire-turbulent dans le code elsA. Extension de la m{\'e}thode des
  paraboles aux parois chauff{\'e}es}}. PhD thesis, Institut Sup{\'e}rieur de
  l'A{\'e}ronautique et de l'Espace (ISAE).

\bibitem[Sasaki {\em et~al.\/}(2020)Sasaki, Morra, Cavalieri, Hanifi \&
  Henningson]{Sasaki3D_art}
{\sc \au{Sasaki, K.}, \au{Morra, P.}, \au{Cavalieri, A.V.G}, \au{Hanifi, A.} \&
  \au{Henningson, D.S.}} \yr{2020}  \at{On the role of actuation for the
  control of streaky structures in boundary layers}.  \jt{J.~Fluid Mech.}
  \bvol{883},  \pg{A34}.

\bibitem[Sasaki {\em et~al.\/}(2018{\natexlab{{\em a\/}}})Sasaki, Morra,
  Fabbiane, Cavalieri, Hanifi \& Henningson]{Sasaki3D_wavecancelling_art}
{\sc \au{Sasaki, K.}, \au{Morra, P.}, \au{Fabbiane, N.}, \au{Cavalieri, A.V.G},
  \au{Hanifi, A.} \& \au{Henningson, D.S.}} \yr{2018{\natexlab{{\em a\/}}}}
  \at{On the wave-cancelling nature of boundary layer flow control}.
  \jt{Theor. Comput. Fluid Dyn.}  \bvol{32},  \pg{593--616}.

\bibitem[Sasaki {\em et~al.\/}(2018{\natexlab{{\em b\/}}})Sasaki, Tissot,
  Cavalieri, Silvestre, Jordan \& Biau]{SasakiPSE_art}
{\sc \au{Sasaki, K.}, \au{Tissot, G.}, \au{Cavalieri, A.V.G.}, \au{Silvestre,
  F.J.}, \au{Jordan, P.} \& \au{Biau, D.}} \yr{2018{\natexlab{{\em b\/}}}}
  \at{Closed-loop control of a free shear flow: a framework using the
  parabolized stability equations}.  \jt{Theor. Comput. Fluid Dyn.}  \bvol{32},
   \pg{765--788}.

\bibitem[Schmid(2007)]{Schmidt_nonmodal_art}
{\sc \au{Schmid, P.J.}} \yr{2007}  \at{Nonmodal stability theory}.  \jt{Annu.
  Rev. Fluid Mech.}  \bvol{39},  \pg{129--162}.

\bibitem[Schmid \& Sipp(2016)]{Sipp_control_art}
{\sc \au{Schmid, P.~J.} \& \au{Sipp, D.}} \yr{2016}  \at{Linear control of
  oscillator and amplifier flows}.  \jt{Phys. Rev. Fluids}  \bvol{1},
  \pg{040501}.

\bibitem[Semeraro {\em et~al.\/}(2011)Semeraro, Bagheri, Brandt \&
  Henningson]{Semeraro_art}
{\sc \au{Semeraro, O.}, \au{Bagheri, S.}, \au{Brandt, L.} \& \au{Henningson,
  D.S.}} \yr{2011}  \at{Feedback control of three-dimensional optimal
  disturbances using reduced-order models}.  \jt{J.~Fluid Mech.}  \bvol{677},
  \pg{63--102}.

\bibitem[Semeraro {\em et~al.\/}(2013{\natexlab{{\em a\/}}})Semeraro, Bagheri,
  Brandt \& Henningson]{Semeraro2_art}
{\sc \au{Semeraro, O.}, \au{Bagheri, S.}, \au{Brandt, L.} \& \au{Henningson,
  D.S.}} \yr{2013{\natexlab{{\em a\/}}}}  \at{Transition delay in a boundary
  layer flow using active control}.  \jt{J.~Fluid Mech.}  \bvol{731},
  \pg{288--311}.

\bibitem[Semeraro {\em et~al.\/}(2013{\natexlab{{\em b\/}}})Semeraro, Pralits,
  Rowley \& Henningson]{Semeraro_fb_art}
{\sc \au{Semeraro, O.}, \au{Pralits, J.O.}, \au{Rowley, C.W.} \&
  \au{Henningson, D.S.}} \yr{2013{\natexlab{{\em b\/}}}}  \at{Riccati-less
  approach for optimal control and estimation: an application to
  two-dimensional boundary layers}.  \jt{J.~Fluid Mech.}  \bvol{731},
  \pg{394--417}.

\bibitem[Shaqarin {\em et~al.\/}(2021)Shaqarin, Oswald, Noack \&
  Semaan]{Shaqarin_art}
{\sc \au{Shaqarin, T.}, \au{Oswald, P.}, \au{Noack, B.R.} \& \au{Semaan, R.}}
  \yr{2021}  \at{Drag reduction of a d-shaped bluff-body using linear parameter
  varying control}.  \jt{Physics of Fluids}  \bvol{33},  \pg{077108}.

\bibitem[Sharma {\em et~al.\/}(2019)Sharma, Shadloo, Hadjadj \&
  Kloker]{Sharma_art}
{\sc \au{Sharma, S.}, \au{Shadloo, M.S.}, \au{Hadjadj, A.} \& \au{Kloker,
  M.J.}} \yr{2019}  \at{Control of oblique-type breakdown in a supersonic
  boundary layer employing streaks}.  \jt{J.~Fluid Mech.}  \bvol{873},
  \pg{1072--1089}.

\bibitem[Sipp {\em et~al.\/}(2010)Sipp, Marquet, Meliga \&
  Barbagallo]{Sipp_Marquet_art}
{\sc \au{Sipp, D.}, \au{Marquet, O.}, \au{Meliga, P.} \& \au{Barbagallo, A.}}
  \yr{2010}  \at{Dynamics and control of global instabilities in open-flows: A
  linearized approach}.  \jt{Appl. Mech. Rev.}  \bvol{63},  \pg{030801}.

\bibitem[Sipp \& Schmid(2016)]{Sipp_Important_art}
{\sc \au{Sipp, D.} \& \au{Schmid, P.J.}} \yr{2016}  \at{Linear closed-loop
  control of fluid instabilities and noise-induced perturbations: A review of
  approaches and tools}.  \jt{Appl. Mech. Rev.}  \bvol{68},  \pg{020801}.

\bibitem[Skogestad(2009)]{Skogestad_art_Best}
{\sc \au{Skogestad, S.}} \yr{2009}  \at{Feedback: Still the simplest and best
  solution}.  \jt{Modeling, Identification and Control}  \bvol{30},
  \pg{149--155}.

\bibitem[Skogestad \& Postlethwaite(2005)]{Skogestad_art}
{\sc \au{Skogestad, S.} \& \au{Postlethwaite, I.}} \yr{2005} {\em Multivariable
  Feedback Control: Analysis and Design\/}.  \publ{John Wiley \& Son}.

\bibitem[Smith \& Gamberoni(1956)]{smith1956transition}
{\sc \au{Smith, A.M.O.} \& \au{Gamberoni, N.}} \yr{1956} {\em Transition,
  Pressure Gradient and Stability Theory\/}.  \publ{Douglas Aircraft Company}.

\bibitem[Tol {\em et~al.\/}(2019)Tol, Kotsonis \& Visser]{Tol_fbff_art}
{\sc \au{Tol, H.J.}, \au{Kotsonis, M.} \& \au{Visser, C.C.~De}} \yr{2019}
  \at{Pressure output feedback control of tollmien-schlichting waves in
  falkner--skan boundary layers}.  \jt{AIAA Journal}  \bvol{57},  \pg{1--14}.

\bibitem[Tol {\em et~al.\/}(2017)Tol, Kotsonis, Visser \& Bamieh]{Tol_channel}
{\sc \au{Tol, H.J.}, \au{Kotsonis, M.}, \au{Visser, C.C.~De} \& \au{Bamieh,
  B.}} \yr{2017}  \at{Localised estimation and control of linear instabilities
  in two-dimensional wall-bounded shear flows}.  \jt{J.~Fluid Mech.}
  \bvol{824},  \pg{818–865}.

\bibitem[Utku \& Garba(1989)]{Riccati_complexity}
{\sc \au{Utku, A. V. Ramesh~S.} \& \au{Garba, J.~A.}} \yr{1989}
  \at{Computational complexities and storage requirements of some riccati
  equation solvers}.  \jt{Journal of Guidance Control and Dynamics}  \bvol{12},
   \pg{469--479}.

\bibitem[Vemuri {\em et~al.\/}(2018)Vemuri, Bosworth, Morrison \&
  Kerrigan]{Vemuri_fb}
{\sc \au{Vemuri, SH.~S.}, \au{Bosworth, R.}, \au{Morrison, J.~F.} \&
  \au{Kerrigan, E.~C.}} \yr{2018}  \at{Real-time feedback control of
  three-dimensional tollmien-schlichting waves using a dual-slot actuator
  geometry}.  \jt{Phys. Rev. Fluids}  \bvol{3},  \pg{053903}.

\bibitem[Yao \& Hussain(2019)]{Yao_wall_osci}
{\sc \au{Yao, J.} \& \au{Hussain, F.}} \yr{2019}  \at{Supersonic turbulent
  boundary layer drag control using spanwise wall oscillation}.  \jt{J.~Fluid
  Mech.}  \bvol{880},  \pg{388–429}.

\bibitem[Zhang \& Freudenberg(1987)]{LTR_Delay}
{\sc \au{Zhang, Z.} \& \au{Freudenberg, J.~S.}} \yr{1987} Loop transfer
  recovery with non-minimum phase zeros.  \bt{In {\em 26th IEEE Conference on
  Decision and Control\/}}, ,  \vol{vol.~26},  \pg{pp. 956--957}.

\end{thebibliography}

\end{document}